\UseRawInputEncoding
\documentclass[%
 reprint,
 amsmath,amssymb,
 aps,
]{revtex4-2}
\usepackage{xcolor} 
\usepackage{textgreek}

\usepackage{graphicx}
\usepackage{dcolumn}
\usepackage{bm}

\usepackage[normalem]{ulem} 
\usepackage{amssymb,amsfonts,latexsym,cancel}

\usepackage{amsmath,amssymb}
\usepackage{amsmath}

\newcommand{\beq}{\begin{equation}}
\newcommand{\eeq}{  \end{equation}}
\newcommand{\beqa}{\begin{eqnarray}}
\newcommand{\eeqa}{  \end{eqnarray}}
\newcommand{\beas}{\begin{eqnarray*}}
\newcommand{\eeas}{\end{eqnarray*}}

\def\R{\mathrm{Re}}

\def\strutdepth{\dp\strutbox}
\def\nw#1{\strut\vadjust{\kern-\strutdepth\vtop to0pt{\vss\hbox to\hsize
{\hskip\hsize\hskip5pt$\leftarrow$\hss\strut}}}{\em #1}}

\usepackage{graphicx,floatrow}
\usepackage{dcolumn}
\usepackage{bm}

\newcommand{\force} { F}

\newcommand{\gdot}[0] {\dot{\gamma}}

\newcommand{\bi}{\begin{itemize}}
\newcommand{\ei}{\end{itemize}}

\newcommand{\be}{\begin{equation}}
\newcommand{\ee}{\end{equation}}
\newcommand{\bea}{\begin{eqnarray}}
\newcommand{\eea}{\end{eqnarray}}

\newcommand{\edot}{\dot{\epsilon}}

\def\strutdepth{\dp\strutbox}
\def\nw#1{\strut\vadjust{\kern-\strutdepth\vtop to0pt{\vss\hbox to\hsize
{\hskip\hsize\hskip5pt$\leftarrow$\hss\strut}}}{\em #1}}

\begin{document}

\preprint{APS/123-QED}
\title{Perspectives on viscoelastic flow instabilities and elastic turbulence}


\author{Sujit S. Datta}
\email{ssdatta@princeton.edu}
 \affiliation{Department of Chemical and Biological Engineering, Princeton University, Princeton, NJ 08544, USA}
\author{Arezoo M. Ardekani}%
 \affiliation{School of Mechanical Engineering, Purdue University, West Lafayette, IN 47907, USA}
 
  \author{Paulo E. Arratia}%
 \affiliation{Department of Mechanical Engineering and Applied Mechanics, University of Pennsylvania, Philadelphia, PA 19104, USA}

  \author{Antony N. Beris}%
 \affiliation{Center for Research in Soft Matter and Polymers, Department of Chemical and Biomolecular Engineering, University of Delaware, Newark, DE 19716, USA}
 
  \author{Irmgard Bischofberger}%
 \affiliation{Department of Mechanical Engineering, Massachusetts Institute of Technology, Cambridge, MA 02139, USA}

  \author{Jens G. Eggers}%
 \affiliation{School of Mathematics, University of Bristol, Fry Building, Woodland Road, Bristol BS8 1UG, UK}

  \author{J. Esteban L\'{o}pez-Aguilar}%
 \affiliation{Facultad de Qu\'{i}mica, Departamento de Ingenier\'{i}a Qu\'{i}mica, Universidad Nacional Aut\'{o}noma de M\'{e}xico (UNAM), Ciudad Universitaria, Coyoac\'{a}n, Ciudad de M\'{e}xico 04510, Mexico}

\author{Suzanne M. Fielding}
\affiliation{Department of Physics, Durham University, Science Laboratories, South Road, Durham DH1 3LE, UK}

 \author{Anna Frishman}%
 \affiliation{Department of Physics, Technion Israel Institute of Technology, 32000 Haifa, Israel}

  \author{Michael D. Graham}%
 \affiliation{Department of Chemical and Biological Engineering, University of Wisconsin - Madison, Madison, WI 53706, USA}

  \author{Jeffrey S. Guasto}%
 \affiliation{Department of Mechanical Engineering, Tufts University, Medford, MA 02155, USA}

\author{Simon J. Haward}%
 \affiliation{Okinawa Institute of Science and Technology, Onna-son, Okinawa 904-0495, Japan}

\author{Sarah Hormozi}%
 \affiliation{Robert Frederick Smith School of Chemical and Biomolecular Engineering, Cornell University, Ithaca, NY 14853, USA}

\author{Gareth H. McKinley}%
 \affiliation{Department of Mechanical Engineering, Massachusetts Institute of Technology, Cambridge, MA 02139, USA}
 \author{Robert J. Poole}%
 \affiliation{School of Engineering, University of Liverpool, Liverpool, L69 3GH, UK}
  \author{Alexander Morozov}%
 \affiliation{SUPA, School of Physics and Astronomy, The University of Edinburgh, James Clerk Maxwell Building, Peter Guthrie Tait Road, Edinburgh, UK}

 \author{V. Shankar}%
 \affiliation{Department of Chemical Engineering, Indian Institute of Technology, Kanpur 208016, India}

 \author{Eric S. G. Shaqfeh}%
 \affiliation{Department of Chemical Engineering, Stanford University, Stanford, CA 94305, USA}
 \affiliation{Department of Mechanical Engineering, Stanford University, Stanford, CA 94305, USA}

\author{Amy Q. Shen}%
 \affiliation{Okinawa Institute of Science and Technology, Onna-son, Okinawa 904-0495, Japan}

\author{Holger Stark}%
 \affiliation{Institute of Theoretical Physics, Technische Universität Berlin, Germany}

\author{Victor Steinberg}%
 \affiliation{Department of Physics of Complex Systems, Weizmann Institute of Science, Rehovot 76100, Israel}
  \affiliation{The Racah Institute of Physics, Hebrew University of Jerusalem, Jerusalem 91904, Israel}
  
   \author{Ganesh Subramanian}%
 \affiliation{Engineering Mechanics Unit, Jawaharlal Nehru Centre for Advanced Scientific Research, Bangalore 560064, India}

  \author{Howard A. Stone}%
 \email{hastone@princeton.edu}
 \affiliation{Department of Mechanical and Aerospace Engineering, Princeton University, Princeton, NJ 08544, USA}

\date{\today}

\begin{abstract}
Viscoelastic fluids are a common subclass of rheologically complex materials that are encountered in diverse fields from biology to polymer processing. Often the flows of viscoelastic fluids are unstable in situations where ordinary Newtonian fluids are stable, owing to the nonlinear coupling of the elastic and viscous stresses. Perhaps more surprisingly, the instabilities produce flows with the hallmarks of turbulence---even though the effective Reynolds numbers may be $O(1)$ or smaller. We provide perspectives on viscoelastic flow instabilities by integrating the input from speakers at a recent international workshop: historical remarks, characterization of fluids and flows, discussion of experimental and simulation tools, and modern questions and puzzles that motivate further studies of this fascinating subject. The materials here will be useful for researchers and educators alike, especially as the subject continues to evolve in both fundamental understanding and applications in engineering and the sciences.

\end{abstract}

\maketitle


\section{\label{sec:level1} Introduction}
\begin{figure*}[!t]
    \centering
    \includegraphics[width=0.75\textwidth]{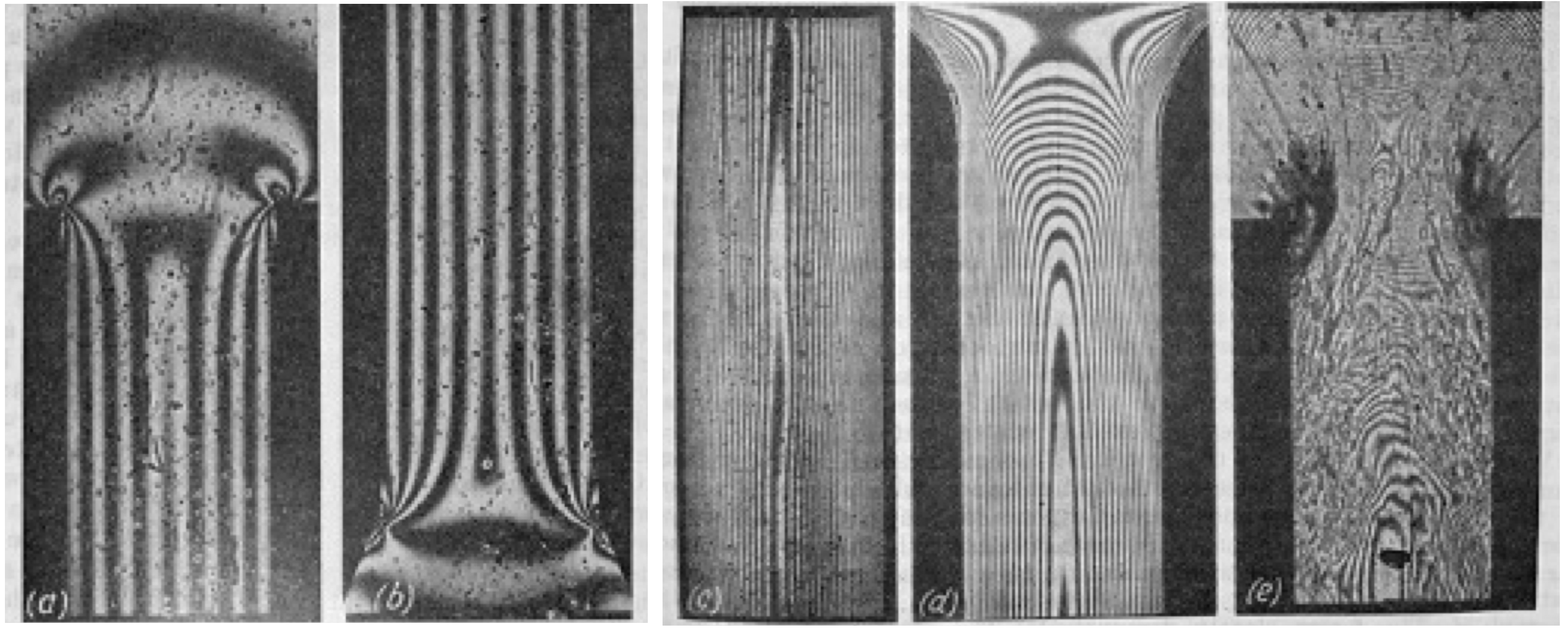}
    \caption{Flow induced birefringence of a viscoelastic polymer melt flowing into (a) and exiting (b) a channel die; note that the shapes of the corners are slightly different between the two. The fringe patterns show lines of constant retardance and constant principal stress difference in the flowing melt. As the flow rate into the die increases, the elastic stress differences increase rapidly and fringes become increasingly numerous and closely spaced (c), (d). Beyond a critical imposed pressure drop the flow becomes unsteady and time-dependent (e) as reflected in the chaotic fringe pattern, which is an apparent signature of a phenomenon named \textit{elastic turbulence}. Adapted from \cite{Vinogradov1980}. }
    \label{vinogradov}
\end{figure*}

Viscoelastic instabilities often occur during the flow, at sufficiently strong forcing,  
of polymer solutions and other
viscoelastic fluids---driven by the 
strong coupling between the (viscous) fluid flow and the material's
elasticity. A classic example of a molten polymer entering a planar contraction is shown in Fig.~\ref{vinogradov}; beyond a critical flow rate, the flow field is dramatically disrupted, even though inertial effects are negligible. The dynamics of 
these complex fluids is both
fundamentally interesting and technologically  important, and 
continues to be studied by
researchers around the world. In some cases, such 
flow instabilities lead to \textit{elastic
turbulence} -- a chaotic, strongly fluctuating regime of fluid flow, such as Fig.~\ref{vinogradov}(e) -- which, amazingly, occurs at low Reynolds number. The statistical features of the flow in this regime have been suggested to
be universal, insensitive to the details of the viscoelastic fluid. Although some flow configurations are well studied, perhaps surprisingly there remain poorly understood aspects of the flows, and these questions lead to many open fundamental and applied problems in the dynamics of complex fluids.

S. Datta and H. Stone organized a virtual workshop of the Princeton Center for Theoretical Sciences in January
2021 to bring together researchers to discuss problems related to viscoelastic flow instabilities, 
assess successes as well as examples of the lack of predictability in current theory, models and
simulations,  identify theoretical pathways linking tools of statistical and polymer physics to
mean field models of the flows, and highlight applications of these instabilities. The ultimate
goal was to bring this community together and clarify, as well as identify, unifying/open questions for future
research to address. We had nearly 500 registered participants from institutions in academia and
industry from all over the world. This Perspective, which includes contributions from the invited speakers, summarizes some of the research presented
and discussions generated at the workshop. Indeed, the participants expressed the viewpoint that the discussions were particularly enlightening as they crystallized poorly understood topics, offered ideas where theory and experimental findings diverged, and highlighted where mechanistic understanding was poor or even lacking.

\subsection{This Article: A Summary and Perspective}
The goal of this article is to provide a short record of the main themes of the workshop, including, where possible, some of the spirit of the discussions that occurred between the talks.
A wide range of topics are discussed briefly, with the hope that a researcher new to the field, or even an established researcher in one corner of the subject, will find introductory ideas that  can launch them into a new research topic if they are so enthused.

This article is organized similar to the structure of the talks in the workshop. Any study of the subject of the flow of complex fluids necessarily touches on important ideas in fluid mechanics, from the geometry of a wide range of (steady) base flows that are possible to the nature of constitutive equations that are needed to close the equations of motion. Even though the latter treats the fluids as continua, it must include the fact that the microstructure of the fluid is deformable, which changes the mechanical, i.e., elastic, response. In addition to reversible deformations, various irreversible effects take place in the form of relaxation and viscous dissipation. Experimental observations have been crucial to identifying the rich nature of the dynamics that can occur, and it should not be forgotten that often the experts were surprised when many of the observations were first made. Furthermore, in some cases, elementary models have been helpful in rationalizing observations, at least qualitatively, in the field of complex fluids generally and polymer solutions in particular, e.g., the bead-spring model suggested in 1953 by Rouse~\cite{Rouse} and improved three years later by Zimm~\cite{Zimm} and then used by de Gennes~\cite{deGennes1974} and Hinch~\cite{Hinch1977} to propose the coil-stretch transition as responsible for some viscoelastic flow responses.

Hence, by means of introduction to this field, we provide a discussion of these complex fluid dynamics in \S~\ref{ViscoelasticIntro}, with emphasis on the dimensionless parameters needed to characterize the flow, and with historical developments in the field highlighted. We introduce viscoelastic flow instabilities and provide a discussion of the different kinds of kinematics that both characterize different flow configurations and indicate their potential as a trigger for flow instability. Also, we highlight how in the case of instabilities in common flows with curved streamlines,  Pakdel and McKinley~\cite{Pakdel1996} provided an insightful characterization that has proven to be helpful quite broadly. Most of the discussion here and elsewhere is thinking about polymeric fluids (but see \S~\ref{Nonpolymericsystems} below).

Numerical simulations using macroscopic, necessarily approximate, constitutive equations linking the state of stress to the strain and rate of strain in these viscoelastic materials have proven to be increasingly insightful in unraveling (no pun intended) the dynamics of these flows with deformable microstructure. Hence, \S~\ref{ModelsAndNumericalSimulations} provides background on constitutive models and numerical tools, including open-source code, for studying viscoelastic fluid flows. This discussion can also serve as an introduction to flow modeling more generally.

The subject of viscoelastic flow instabilities in model (canonical)  geometries and the connections to turbulent dynamics is introduced in \S~\ref{FlowTransitionSection}. The discussion includes a review of the main observations as well as new ideas, related to mechanisms, that have come from two-dimensional (2D) and three-dimensional (3D) numerical simulations. 
Elastic flow instabilities in more complex geometries, such as the flow between a pair of cylinders, or flow in ordered or disordered porous media, are discussed in \S~\ref{ElasticInstability}.
Section~\ref{EITsection} deals with the combined effects of elasticity and inertia in engendering novel instabilities in rectilinear shearing flows. The role of these instabilities \textit{vis-a-vis} transition from the laminar state, and in reducing drag in the fully developed turbulent regime (the maximum drag-reduced regime, in particular), is discussed. Connections between the different turbulent regimes -- elastic, inertial and elastoinertial -- are highlighted.  Flow instabilities in simplified free-surface flows are discussed in \S~\ref{SectionFreeSurface}. Some of the flow instabilities that are observed in other non-polymeric complex fluids are indicated in \S~\ref{Nonpolymericsystems} and the article closes in \S~\ref{outlook} with future outlooks across this fascinating subject. We hope the reader enjoys this tour of instabilities in the flow of complex fluids.

\section{Viscoelastic fluids and flows}
\label{ViscoelasticIntro}

As introduced in undergraduate classes, ordinary small-molecule liquids are {\em viscous} and Newtonian: the stress is a linear function of the shear rate (also known as the strain rate), with the coefficient of proportionality given by the dynamic shear viscosity. By contrast, ordinary solid materials are {\em elastic}: the stress is a linear function of the strain, and the coefficient of proportionality is the elastic modulus. Conceptually it is useful to imagine an elastic material as made up of simple springs with a restoring force linear in displacement, e.g., Hooke's law derives from Robert Hooke's observation, in latin, in the mid-late 1600s that ``ut tensio, sic vis" or ``as the extension, so the force". The materials we are concerned with have elements of both: they are viscoelastic, so that the state of stress depends on both the strain and the strain rate. Conceptually, such a material has some responses expected of a matrix of damped linear springs, e.g., the Maxwell or Voigt linear viscoelastic models have this character and can be used to characterize such fluids in the small amplitude limit. 

Polymer solutions are a common and industrially relevant example of a viscoelastic fluid. Thus, a wide variety of industrial processes such as molding, extrusion, coating, spraying etc. that involve polymer solutions give rise to the challenge of modeling and controlling viscoelastic flows. The polymer solutions can be dilute or concentrated (and in the latter case they also share many properties with polymer melts), and have very different constitutive relations~\cite{BAHvolume1,Larson1999}. Other viscoelastic fluids include micellar surfactant solutions, emulsions, liquid crystals, etc.

\subsection{Typical Flows and Kinematics} 
As in all areas of fluid dynamics, flows can be driven by the motion of boundaries or by a pressure difference.  Thus, the two prototype flows that are characterized most are Taylor-Couette flow, which refers to the wall-driven flow in the annular gap between two concentric cylinders, and pressure-driven (or Poiseuille) flow in a channel or circular pipe. 
The former has curved streamlines, whereas the latter flow is rectilinear far from the inlet when the flow is fully-developed---although streamline curvature can also be introduced when the channel or pipe has a curving centerline (often termed ``Dean" flow) or when a boundary-driven flow is forced to recirculate in a closed cavity. 

Early in the 20th century, in an experimental and theoretical study of a Newtonian fluid in a concentric cylinder device, G.I. Taylor characterized the instability experimentally and numerically: above a critical rotation rate (Reynolds number) the flow is unstable when the inner boundary is rotated with the outer boundary fixed, but the opposite case (inner boundary fixed and outer boundary rotating) is always linearly stable~\cite{Taylor1923}. Over the next few decades there  were hints that viscoelastic fluids had a qualitatively different response---but the definitive work on the topic, clarifying {\em elastic instabilities}, and the fact that they could occur for low-Reynolds-number flows when the inner boundary was fixed and outer boundary rotating would not occur until the late 1980s and early 1990s.

In addition, to understand the motion of fluids that contain a deformable microstructure, it is important to recognize the distinction between {\em shear-dominated} and {\em extension-dominated} flows. For steady flows, in the former case, because of the presence of a finite rate of rotation, material points separate  algebraically in time, and orientable objects tumble at a rate nominally tied to the vorticity in the flow. By contrast in the latter flow type, because of the absence  of rotation, material points separate exponentially in time. It should not be surprising that exponential stretching can cause large changes in the stress in a viscoelastic material.

\begin{table}[t]
\centering
\begin{tabular}{|c|l|}\hline
 Symbol&  Meaning \\ \hline\hline
  $\rho$ & Density \\ \hline
    $\sigma$ & Interfacial tension \\ \hline
  
    $\boldsymbol{v}$ & Velocity vector \\ \hline

  $V$ & Velocity scale \\ \hline
  $t$ & Time \\ \hline
  $\ell$ & Flow length scale \\ \hline
  $\gamma$ & Shear strain \\ \hline
 $\tau$ & Stress (shear or extensional, as defined in the text) \\ \hline
   $\eta$ & Shear viscosity (dynamic) \\ \hline
      $\nu$ & Shear viscosity (kinematic) \\ \hline
   $\lambda$ & Stress relaxation time \\ \hline
   $\gamma_y$ & Yield strain \\ \hline
   $\tau_y$ & Yield stress \\ \hline
      $N_1$ & First normal stress difference \\ \hline
      $N_2$ & Second normal stress difference \\ \hline
      $\Psi_1$ & First normal stress coefficient \\ \hline
      $\Psi_2$ & Second normal stress coefficient \\ \hline
      $G$ & Shear modulus \\ \hline
$G^*$ & Complex shear modulus \\ \hline
$G'$ & Storage modulus \\ \hline
$G''$ & Loss modulus \\ \hline
  $\epsilon$ & Extensional Hencky strain \\ \hline
$\eta_E$ &Extensional viscosity\\ \hline
\end{tabular}
\caption{Due to the use of different symbolic conventions by different communities, here we summarize the choices of the symbols used most commonly in this article along with their meanings. Where relevant, the mathematical definitions are presented in the text as the symbols are introduced. Other symbols are also introduced in the text as needed for specific other quantities. Note that an overlying dot represents a time derivative, e.g., $\dot{\gamma}$ denotes the shear rate.}
\label{table0}
\end{table}

\subsection{Rheology and Rheological Parameters}
The field of non-Newtonian fluid mechanics, with its  unfamiliar notation and specific terminology/jargon, can be initially bewildering to newcomers. This is, in some sense, unavoidable because of the vast range of fluids that fall into the class of what used to be called generically {\em non-Newtonian fluids} and that are now increasingly defined by the catchall phrase {\em complex fluids}. These materials  may range from dilute polymer solutions and melts to dense suspensions with high volume fractions of particles, surfactant solutions that self-assemble into long `wormlike' micellar structures to soft swollen polymer microgels, liquid crystalline dispersions and beyond.  The key feature of interest in understanding viscoelastic flow instabilities is the presence of an underlying deformable microstructure in the fluid that can be affected by the flow, and which, in turn, can modify the underlying equations of motion -- as a consequence of the generation of additional non-Newtonian contributions to the total stress field that arise from changes in the microstructural configurational distribution. It is the nonlinear feedback between these two features of flow and fluid that give rise to entirely new instabilities not present in Newtonian fluids that are characterized by a linear relationship between stress and deformation rate. 

It is not possible in a short perspective article to cover this entire zoology of fluids (for more details the reader is directed to \cite{Larson1999}), but it is possible to summarize five of the key phenomena and the corresponding material properties that are displayed by prototypical complex fluids. These involve:
\begin{enumerate}
\item Fluid viscoelasticity, as parameterized by a stress relaxation time, commonly denoted $\lambda$ (but also often defined as $\tau$ in the physics literature) and a complex (shear) modulus $G^* (\omega)=
G^\prime (\omega)+iG^{\prime\prime} (\omega)$, where $\omega$ is the frequency of the time-varying strain field in small amplitude oscillatory shear (SAOS). Here the storage modulus $G^\prime$ characterizes the elastic response of the material and the loss modulus $G^{\prime\prime}$ characterizes the viscous response of the material. The elastic and viscous stresses resulting from this deformation grow linearly with the strain amplitude and are, respectively, in phase and 90 degrees out of phase with the imposed sinusoidal strain oscillation.

\item The development in steady simple shear of large normal stresses along the principal axes of the flow characterized by two nonzero {\em normal stress differences}, commonly denoted by material functions $N_1$ and $N_2$.

\item 	Shear rate-dependence in the viscometric material functions that are measured in steady shearing flow with shear rate ${\dot\gamma}$; e.g., a shear-thinning viscosity, $\eta \left ({\dot\gamma}\right ) = \tau/\dot{\gamma}$ and normal stress coefficients $\Psi_1\left ({\dot\gamma}\right )=N_1/{\dot\gamma}^2$ and $\Psi_2\left ({\dot\gamma}\right )=N_2/{\dot\gamma}^2$. 

\item 	Time and rate-dependence, often corresponding to {\em strain-hardening} and {\em tension-thickening} (e.g., flexible linear polymers in dilute solution) respectively, in the time-dependent extensional viscosity function, $\eta_E^+ \left ({\dot\epsilon}, t\right )$, where ${\dot\epsilon}$ is the local extension rate in an extensional flow.   

\item 	The possible appearance of a yield stress $\tau_y$ (and a corresponding yield strain $\gamma_y$) at sufficiently high concentrations of the microstructural constituents. 
\end{enumerate}

Just as in the case of a communicable disease, such as the flu (or COVID-19!), the list of symptoms described above may be present or absent to different extents in a particular fluid, or constitutive model, and care must be taken to understand these limitations.  For example, the Oldroyd-B model, which is  discussed extensively in the rest of this paper, predicts some of the phenomena in the above list; specifically items \#1, 2
(but only partially as $N_2 = 0$), and 4, but does not predict rate-dependency of the viscometric functions.  To compare experimental observations and theoretical predictions, the rheological material response of a given fluid needs to be carefully characterized in several different flow fields, e.g., at minimum, SAOS, as well as a large deformation shearing flow such as steady simple shear flow at large shear rates (i.e. $\dot{\gamma}\gg1/\lambda$) or large amplitude oscillatory shear (LAOS) as well as an extensional flow of some kind, so that accurate model parameters can be extracted from experimental data.  Highly elastic dilute polymer solutions, or `Boger fluids'~\cite{Boger1977,James2009} were formulated to exhibit an approximately constant shear viscosity but significant elasticity. They thus correspond quite closely to the predicted rheological response of the Oldroyd-B model so that quantitative comparisons between the predictions of linear stability analysis and careful experimental observations could be performed without the complications of the interplay between viscoelasticity and shear thinning.

Additional, more complex rheological phenomena arising from complications such as finite extensibility of the polymer chains or coil-overlap in semi-dilute/concentrated solutions can be incorporated by including additional physics in the microstructural description of the complex fluid. This invariably gives rise to additional (often dimensionless) model parameters, such as the finite extensible, nonlinear elastic (FENE), parameter $L$ that describes the finite extensibility of solvated polymer chains~\cite{Rallison1988} or the Giesekus mobility parameter $\alpha$~\cite{Giesekus1982} describing anisotropic drag on the underlying flow-aligned microstructural elements, detailed further in \S~\ref{ModelsAndNumericalSimulations}. Incorporating these additional microscopic effects into the constitutive model typically lead to nonlinear responses such as shear-thinning in the viscometric material functions as well as a bounded stress growth in steady homogeneous elongational flows, which agree even more closely with experimental data. An extensive review of the current state of the art in the constitutive modeling of polymer melts and solutions is given by Larson and Desai~\cite{LarsonDesai2015}.  For more complex materials, such as particle-filled viscoelastic fluids, in which a yield stress also appears, suitable frame-invariant tensorial constitutive models are just beginning to appear~\cite{Saramito2009, Dimitriou2019}, but very few stability analyses have yet been performed for this class of materials, which can be conveniently described as elastoviscoplastic (EVP) materials.  

\begin{figure*}
    \includegraphics[width=0.9\textwidth]{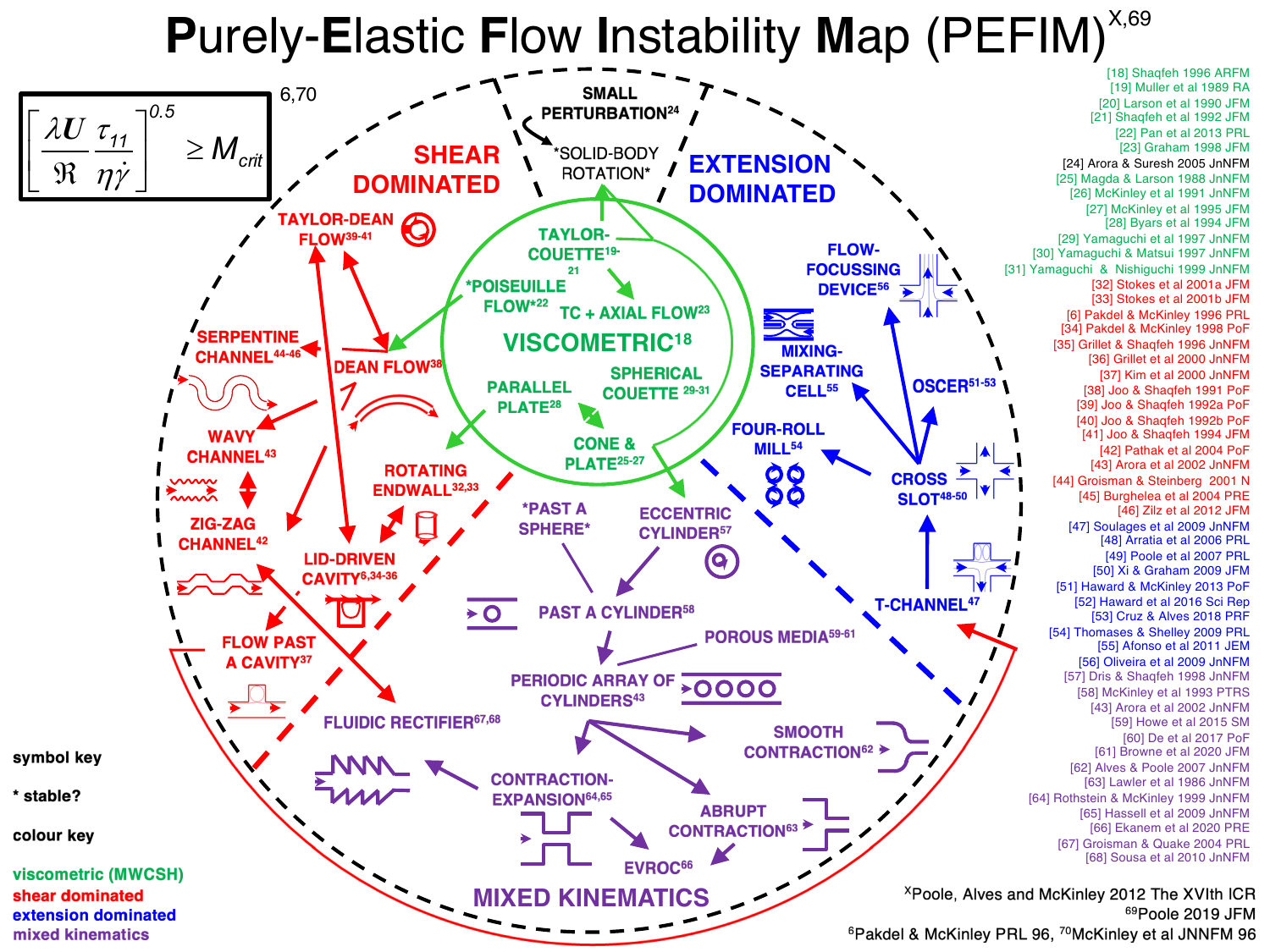}
    \caption{A map of the different viscoelastic flow instabilities that have been documented to date, with references to selected works exemplifying the different kinematics.}
    \label{pefim}
\end{figure*}

\subsection{Dimensionless Parameters for Complex Fluids}
\label{subsec:Dimlessparameters}

For fluid mechanicians it is natural to aim to quantitatively compare experimental observations and numerical computations of steady flows and flow stability in terms of appropriate dimensionless quantities. However the number of material parameters or functions required to describe a specific complex fluid can lead to a rapid increase in the dimensionality of the problem as well as some non-uniqueness in the definitions of material parameters.

For a fluid of density $\rho$ and shear viscosity $\eta$, in a canonical flow with a typical velocity scale $V$ and length scale $\ell$, in addition to a Reynolds number $\textrm{Re}=\frac{\rho V \ell}{\eta}$ parameterizing the relative importance of inertial stresses $O\left (\rho V^2\right )$ to viscous stresses $O\left (\eta V/\ell\right )$, it is essential to also quantify the level of non-Newtonian effects in the flow.  This is commonly done through a Deborah number, $\textrm{De}$, or a Weissenberg number, $\textrm{Wi}$. The preferred usage of each term has evolved over time and formal definitions of these dimensionless parameters are subtly different from each other (see~\cite{Dealy2010} for additional discussion). Nevertheless, they both represent a dimensionless relaxation time with respect to a characteristic time for the flow.  In the Deborah number, the characteristic time for the flow is represented directly by an estimate of the time over which the flow changes, whereas in the Weissenberg number, the time-scale is parameterized indirectly using an inverse shear rate.  Alternatively, a stress ratio, defined as the ratio of the first normal stress difference to the shear stress, can also be used to quantify the level of non-Newtonian effects in the flow (as seen using the Oldroyd-B model, described further in \S\ref{ModelsAndNumericalSimulations}, in the limit of no solvent viscosity).

Careful inspection of any particular research paper is required to ensure that one understands clearly the definition being used; but a simple example can suffice here. In the steady viscometric flow of a viscoelastic fluid in a cone-and-plate rheometer, where $\Omega$ is the steady rotation rate of the conical fixture and $\theta_0\ll 1$ is the (very small) conical angle of the domain within which the fluid is confined,  the conical fixture generates a homogeneous deformation rate ${\dot\gamma}_{\phi\theta}=\Omega/\theta_0$.  The Deborah number would be properly defined in terms of a {\em ratio of time scales} between the stress relaxation timescale and the characteristic flow time scale, $T_{flow}\approx 1/\Omega$, so that $\textrm{De}=\lambda \Omega$.  However the magnitude of the normal stress differences and the shear stress developed in the viscometric shearing flow established by the rotating cone (typically) have magnitudes that scale with ${\dot\gamma}^2$ and ${\dot\gamma}$, respectively, so that the dimensionless Weissenberg number parameterizing this stress ratio is properly defined as $\rm{Wi} = \lambda {\dot\gamma}=\lambda\Omega/\theta_0$. 

These two non-Newtonian dimensionless groups, De and Wi, are of course not completely independent and their ratio is  given simply by a dimensionless geometric factor characterizing the flow; for example, in the cone-and-plate example above $\textrm{Wi/De}=1/\theta_0$. Similar definitions and distinctions apply in all flow domains, from viscometric flows such as a Taylor-Couette geometry -- with inner and outer radii $R_i$ and $R_o$, respectively, where the gap ratio $\left (R_o-R_i\right )/R_i$ plays the corresponding role to $\theta_0$ -- to more complex geometries such as viscoelastic flow past a cylinder or sphere, or through a contraction/expansion.  One possible taxonomy, based on kinematic distinctions, of the different viscoelastic flow instabilities that have been documented to date is suggested in Fig.~\ref{pefim}.

Because of this indeterminacy and variability between different approaches and geometries it is good practice -- particularly in stability analyses where one seeks to compare experiments and predictions -- to use only one dimensionless group as a dynamical parameter measuring the flow strength and then report dimensionless ratios of material and geometric parameters that are independent of kinematics to fully specify the flow  situation being considered.  For example, a dimensionless elasticity number $\textrm{El = De/Re} = \lambda\eta/\left (\rho \ell^2\right )$ conveniently represents the relative magnitude of viscoelastic and inertial effects in a flow (or the ratio of the time scale of the fluid to the time for vorticity to diffuse across a distance $\ell$) and is constant for a particular fluid and geometry. Unfortunately, in many computational studies the Deborah number and Reynolds number may be systematically varied independently to explore the dynamical response of a system; for an experimentalist this corresponds to having to perform experiments with a range of different fluids and/or flow geometries, which can be very challenging! Another point to keep in mind is that in numerical simulations, one can ``turn off" the influence of inertia, whereas experiments typically have some finite degree of inertia.

\begin{figure}
    \includegraphics[width=\columnwidth]{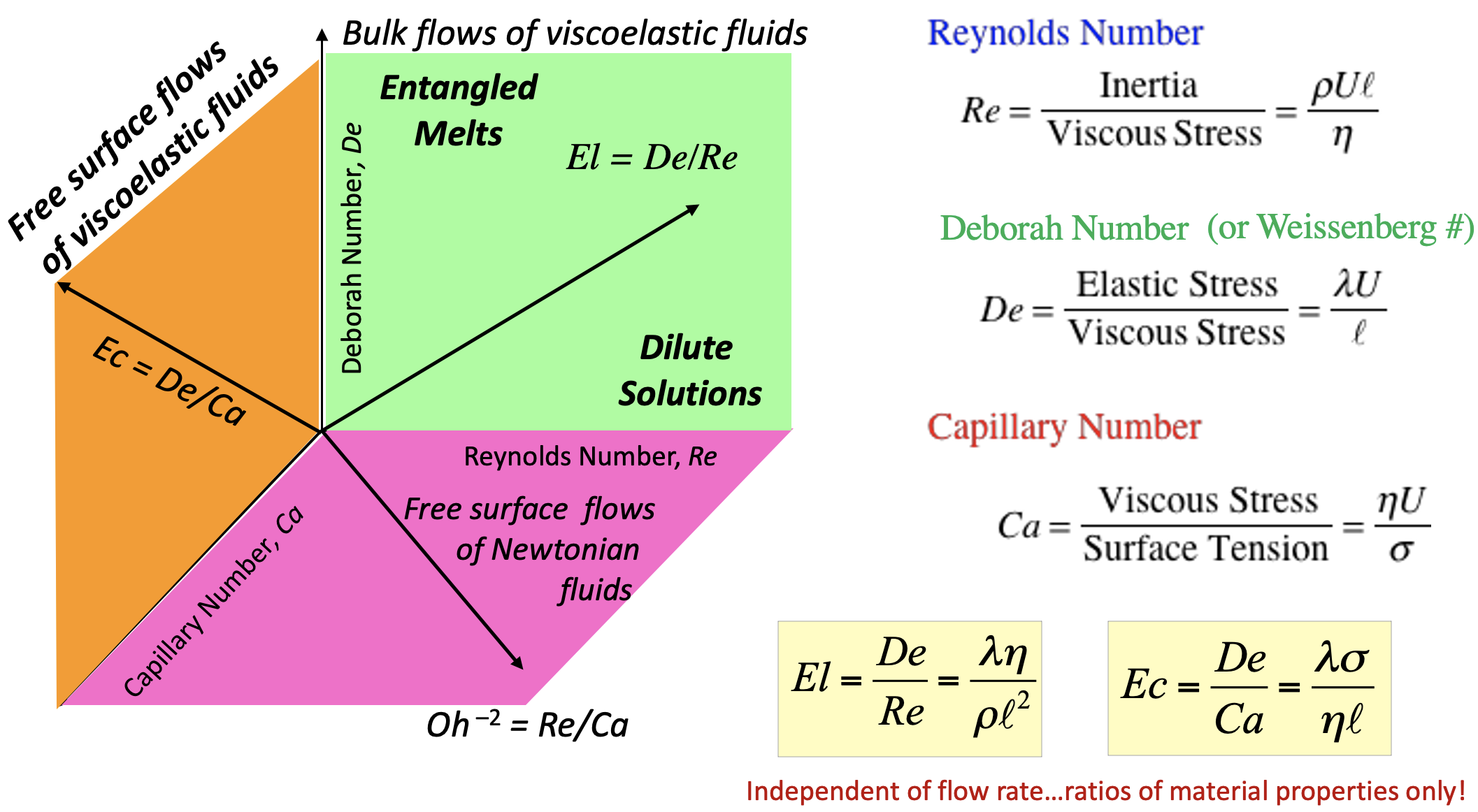}
    \caption{A simple three-dimensional representation of the parameter space accessed in processing flows of complex fluids. Taking ratios of each pair of axes results in new dimensionless material parameters that are independent of the kinematics of the flow \cite{McKinley2005}. Note that here we have defined De as a ratio of stresses instead of the ratio of time scales presented in the text. }
    \label{surfaceflows}
\end{figure}

This problem of flow characterization is further compounded when a free surface is present due to the additional introduction of a surface or interfacial tension coefficient (which we denote here by $\sigma$). We neglect more complex interfacial effects such as surface diffusivity or viscoelasticity, which would result in yet more dimensionless parameters. It is then natural to discuss a capillary number, $\textrm{Ca}=\eta V/\sigma$. In principle, for viscoelastic free surface flows the locus of a particular process, e.g., a fiber-spinning or inkjet printing or spraying/atomization operation, can then be represented in a three-dimensionless space constructed from the Deborah number, Reynolds number, and capillary number as sketched in Fig.~\ref{surfaceflows}. Taking ratios of the dimensionless parameters plotted on each axis gives rise, respectively, to the elasticity number $\textrm{El = De/Re}$ and the Ohnesorge number $\textrm{Oh}^2=\frac{\eta^2}{\rho\sigma\ell}=\textrm{Ca/Re}$.  Additional dimensionless parameters can be defined in specific problems, such as the Weber number $\textrm{We = ReCa}$, which is commonly encountered in analyses of jet stability. The (un-named) dimensionless ratio of the elastic stresses and capillary pressure is also of importance in the stability analysis of such problems and can be conveniently parameterized by the ratio of the Deborah number and capillary number, which we suggest should be referred to as an elastocapillary number, $\textrm{Ec = De/Ca} = \lambda \sigma/(\eta\ell)$.

With these ideas and scalings it becomes natural to represent experimental measurements and theoretical analyses of the critical conditions for onset of viscoelastic flow instabilities in terms of stability diagrams such as the one sketched generically in Fig.~\ref{dragons}; detailed results for pipe and channel flows, respectively, are provided below in Figs.~\ref{fig:bigpicturepipe} and \ref{fig:bigpicturechannel}. The characteristic Reynolds number of the flow is plotted on the abscissa and the corresponding Deborah number of the complex fluid, or alternatively the Weissenberg number, is represented on the ordinate. The stability boundaries for each given wavenumber or perturbation type correspond to bounding curves in this parameter space.  A set of exploratory experiments with a given fluid in a fixed geometry correspond to a sequence of points that traverse along a line of constant slope $\textrm{El = De/Re}$ that eventually intersects with a stability boundary marking the critical conditions for onset of an observable instability.  All experiments with Newtonian fluids can explore only the abscissa of this plot and a purely inertial instability corresponds to a critical point along this horizontal line. In macroscale flows, weakly-elastic fluids, such as dilute aqueous polymer solutions, move along lines of small slope and explore the lower right of this parameter space. Highly elastic materials, such as gels and polymer melts, explore the upper left portions of the plot. 

Although every flow geometry and instability mode studied represents a unique stability locus in a stability diagram such as Fig.~\ref{dragons}, some general remarks are possible. Elasticity can either strengthen or weaken inertial mechanisms of instability, with the precise results depending on the geometric details of the flow geometry as well as the magnitudes of rheological parameters such as the first and second normal stress differences (see for example the seminal studies by \cite{Beard1964,Ginn1969}). In Fig.~\ref{dragons} we show schematically a case in which weak elastic effects destabilize an inertial mode of instability (as for example in viscoelastic flow in the Taylor-Couette geometry) i.e., 
the critical \textrm{Re} required for the instability is reduced by increasing elasticity, thus inclining the stability locus, indicated schematically by the blue line, as shown. Conversely, small levels of inertia often stabilize viscoelastic base flows against {\em purely elastic instabilities}, which correspond to critical loci that intersect the ordinate axis, leading to trajectories as indicated by the red line; nevertheless, we note that 
Joo and Shaqfeh  have shown, for viscoleastic Taylor-Couette flow, that the purely elastic instability is destabilized (for nonzero inertia) when the inner cylinder is rotating, while it is stabilized when the outer cylinder is rotating~\cite{JooShaqfeh1992b}.  The spatiotemporal characteristics of these {\em purely elastic} and {\em inertioelastic modes} often differ very substantially.

\begin{figure}
    \includegraphics[width=\columnwidth]{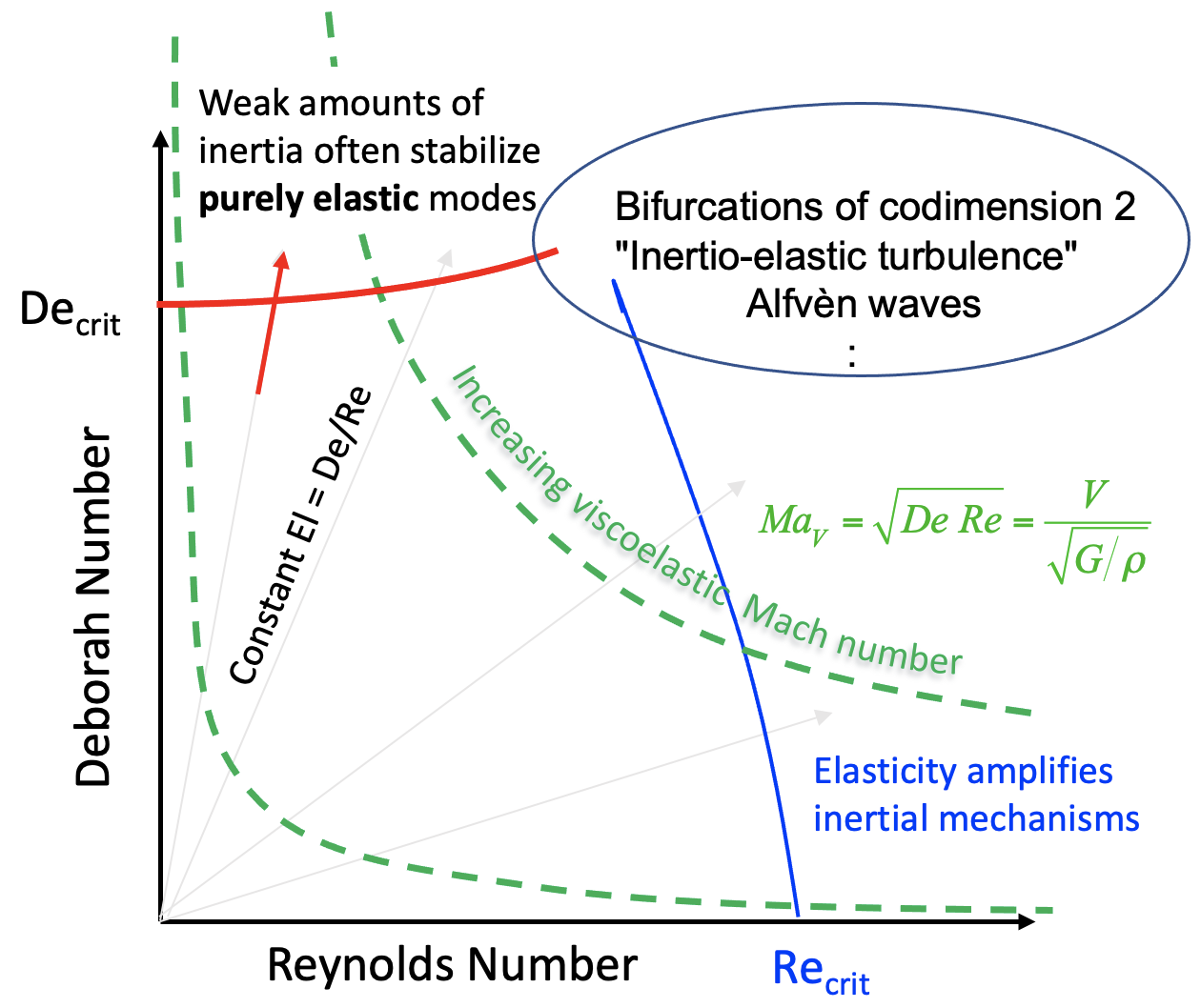}
    \caption{Sketch of a canonical stability diagram for representing the onset of viscoelastic flow instabilities in a complex fluid.}
    \label{dragons}
\end{figure}

At large levels of both flow inertia and fluid elasticity (corresponding to the upper right of this plot) more exotic `beasts' and complex dynamical modes of instability such as `€˜codimension two' bifurcations exist; see for example the seminal work of Renardy et al.~\cite{Renardy1996}. With the definitions introduced in this section it becomes self-evident that because the elasticity number $\textrm{El}=\lambda\eta/\left (\rho \ell^2\right )$ depends inversely on the length scale of the geometry, it is possible to experimentally access such regimes using microfluidic devices in which the characteristic length scale is very small. Indeed for microfluidic researchers handling complex microstructured fluids, such as blood, DNA suspensions or protein solutions, flow instabilities are to be expected in such devices whenever they have smallest dimensions $\ell \ll \left (\lambda \eta/\rho\right )^{1/2}$.  

Finally, with respect to this representation of parameter space, we note that in inertioelastic flow fields the speed of viscoelastic shear waves is given by $c_s=\left (G/\rho\right )^{1/2}$, where $G \equiv |G^*|$ is the magnitude of the complex shear modulus of the viscoelastic fluid~\cite{Joseph1985}. It is important to note that these viscoelastic shear waves are distinct from the sound modes associated with the bulk modulus of the fluid, and instead are associated with the transmission of perturbations through the entropic elasticity of the underlying microstructure in the complex fluid. Since the elastic modulus may only be of order $10^2 - 10^6$ Pa, the resulting viscoelastic shear wave speed may therefore be quite modest.  If we neglect the role of the solvent viscosity and identify $G\approx \eta/\lambda$ (in which case $\eta$ is dominated by the polymer contribution), then in dimensionless form we can construct a viscoelastic Mach number $\textrm{Ma}_V=V/c_s =\left (\textrm{Re De} \right )^{1/2}$, and lines of constant Mach number correspond to hyperbolae in Fig.~\ref{dragons}, as indicated by the green dashed lines.  Flows at high viscoelastic Mach number result in a \textit{change of type} in the underlying constitutive equations, from parabolic (similar to a diffusion equation) to hyperbolic (similar to a wave equation), and give rise to many anomalous phenomena in the inertioelastic flow of complex fluids~\cite{Delvaux1990}. Examples include finite upstream propagation of vortices ahead of a blockage such as a cylinder in a channel \cite{Qin2019,Zhao2016} as well as the development of traveling waves of elastic stress that are analogous to the Alfven waves observed in magnetohydrodynamic flows~\cite{Varshney2019}. We note that this topic was an active topic of discussion at the workshop, with multiple perspectives presented; another perspective put forward by V. Steinberg is presented in \S\ref{outlook}.

Analogous stability diagrams may also be constructed for free surface instabilities of complex fluids, for example in terms of the Deborah number and Ohnesorge number~\cite{Bhat2010} or a capillary and Weber number~\cite{Clasen2012}.  For complex problems, in which multiple dimensionless material parameters control the constitutive response of a complex fluid, the stability loci correspond to surfaces in three- or higher-dimensional diagrams, which can be difficult to represent graphically. However, simpler two-dimensional `slices' of this space are still useful graphically to represent the sensitivity of the stability diagrams to other effects, such as the magnitude of second normal stress differences, changes in the finite extensibility of the dissolved macromolecules, or sensitivity to the effects of viscous heating for example~\cite{AlMubaiyedh2000,Rothstein01}.

A particularly common, and indeed almost unavoidable, example of this kind is the role of fluid shear-thinning which becomes increasingly important at progressively higher shear rates (except for the case of carefully formulated highly-elastic constant viscosity fluids such as `Boger fluids'~\cite{James2009}). Understanding the central role of shear-thinning in viscoelastic flow instabilities is critical because both the fluid relaxation timescale and the viscosity typically decrease in most complex fluids (with the important exception of shear-thickening materials; see for example \cite{Denn2018}).  A convenient way to graphically represent these effects is by defining a (dimensionless) function $\mathcal{S}(\dot\gamma)=1-(d \ln \tau / d \ln \dot\gamma)$, which is evaluated from the flow curve measuring the shear stress $\tau$ at a steady shear rate $\dot\gamma$ ~\cite{Sharma2012,Haward2012a,Haward2020}; thus, $\mathcal{S}=0$ corresponds to no shear-thinning (i.e., the Oldroyd-B limit) and $\mathcal{S}\rightarrow 1$ corresponds to the upper limit of a strongly shear-thinning fluid such as an elastoviscoplastic material near its yield stress, or a shear-banding wormlike micellar solution.  It is clear from the definition of the elasticity number given above that the slope (given by $\textrm{El = De/Re}$) of a specific fluid'€™s trajectory through the $\left \{ \textrm{Re, De}\right \}$ stability diagram becomes progressively shallower under increasingly strong shearing deformations, and this can dramatically constrain the range of parameter space that can be explored.

\begin{figure}
    \includegraphics[width=\columnwidth]{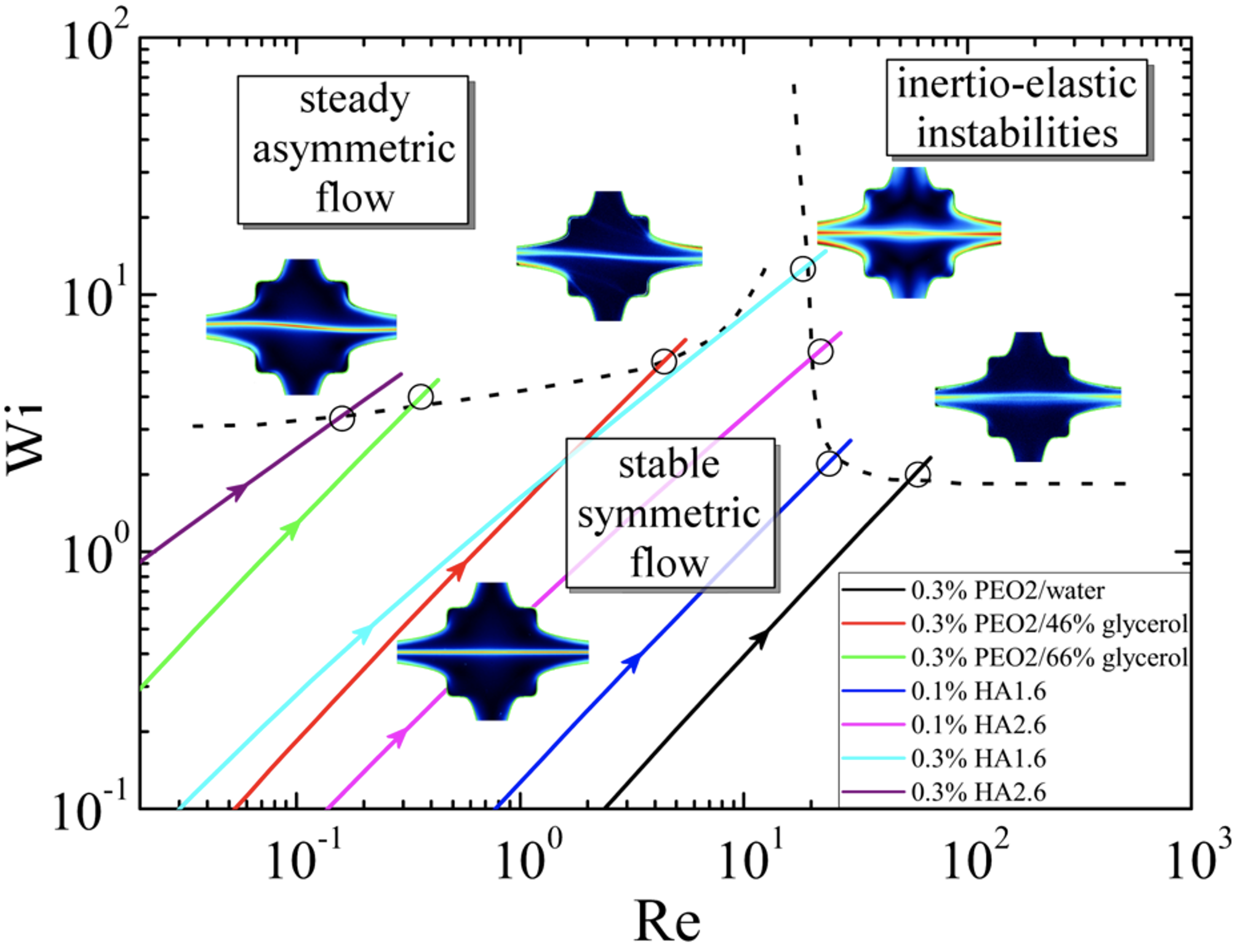}
    \caption{A stability diagram for viscoelastic flow of dilute solutions of poly(ethylene oxide) (PEO) and hyaluronic acid (HA) through an OSCER (Optimized Shape Cross-slot Extensional Rheometer) device. The elasticity number of each fluid (shown by the colored lines of constant Wi/Re are varied by changing the polymer concentration or solvent viscosity. Two distinct modes of instability can be observed at high Weissenberg and Reynolds numbers \cite{Haward2012a}.}
    \label{stability}
\end{figure}

To briefly illustrate these ideas we show in Fig. \ref{stability} the results of a detailed study of the types of viscoelastic flow instabilities observed in the flow of a range of dilute polymer solutions through a microfluidic cross-slot device that has been optimized to generate a strong and homogeneous extensional flow \cite{Haward2013}. The trajectories through $\{\textrm{Re,Wi}\}$ space followed by each fluid as the imposed flow rate is increased are shown by colored lines. These pathlines are almost straight (corresponding to the limit $\textrm{El = Wi/Re}= constant$) but the weakly varying effects of shear thinning in the fluid rheology slowly modulate this (corresponding to evolving values of the function $S(\dot{\gamma})$).  A fully three-dimensional representation of this kind of stability diagram in $\{\textrm{Re, Wi, }S\}$ space can also be constructed \cite{Haward2012a}. However, it is clear even in this 2D projection that two distinct modes of instability are observed: a steady two-dimensional symmetry-breaking purely-elastic mode (corresponding to a transcritical bifurcation) at high levels of fluid elasticity, as well as an oscillatory inertioelastic mode (corresponding to a Hopf bifurcation) that is dominant at higher levels of fluid inertia.  In each case, the onset of these viscoelastic instabilities disrupt the homogeneous elongational kinematics that are desired for an extensional rheometric device.  Understanding, exploring and predicting these kinds of instabilities was a major focus of this workshop.

\subsection{Some (Pre)History}
Although the term \textit{elastic turbulence} has grown into relatively common usage in the 21st century, the origins of the term date back to the very beginnings of research in complex fluids.  As early as 1926, Ostwald and Auerbach~\cite{Ostwald1926} remarked on the anomalously high pressure drops (and enhanced fluctuations) that were required to pump certain complex microstructured fluids through cylindrical tubes at low flow rates where laminar flow conditions were to be expected. As they noted ``... it is thus a peculiarity of ammonia oleat and similar sols to exhibit, {\em in addition to normal turbulence also a structural turbulence}''€.  Some 40 years later Hanswalter Giesekus~\cite{Giesekus1968}, in his pioneering studies on nonlinear effects of viscoelastic effects through converging nozzles and slits, carefully documented the apparently turbulent, i.e., strongly time-varying and highly chaotic flow conditions, that could be achieved in non-dilute polyacrylamide solutions even at moderate concentrations of 3-4 wt.\% polymer.  

In the years between these two papers the more applied polymer processing literature is full of many documented instances of unstable flow fields arising from the viscoelastic nature of the molten plastics being used in injection molding and other processing operations.  The systematic analysis of these empirical observations by Pearson and colleagues, as well as Denn and colleagues, is particularly impactful in this regard; see, for example, the extensive reviews by these authors~\cite{Pearson1976,DennARFM2001}. Much of the work on polymer processing instabilities is nicely captured in the path-breaking book by Vinogradov and Malkin~\cite{Vinogradov1980} published first in the Soviet Union in 1977 and subsequently in English in 1980. The term {\em elastic turbulence} appears explicitly in the index of this work multiple times, and beautiful flow visualization images of the stress field (using crossed polarizing optical elements to reveal flow-induced birefringence) in a molten polymer entering a planar contraction (reproduced in Fig.~\ref{vinogradov}) show the dramatic disruption in the flow field that occurs beyond a critical flow rate, even though the relevant Reynolds number is $\textrm{Re} \ll 1$.

Quantitative study of such flow instabilities and the term ``purely elastic instability" first appeared in 1989 through the work of Muller, Larson and Shaqfeh in a seminal paper on viscoelastic Taylor-Couette flow \cite{Muller1989} that was dedicated to Giesekus on his retirement as editor of {\em Rheologica Acta}. In particular, in the late 1980s and early 1990s, these researchers at Bell Labs, MIT and later at Stanford focused on discovering and understanding purely elastic instabilities, i.e., those elastic instabilities for which inertial forces play a negligible role, in viscometric flows. These flows were marked by curved streamlines and the associated elastic instabilities prevented rheological measurements in certain parameter regimes. In all instances there existed a separation of scales: a thin gap across which there was shear, characterized for elastic fluids by the shear Weissenberg number, Wi, and a significantly larger radius of curvature, where the ratio of gap to radius of curvature was denoted by $\epsilon$. These flows included Taylor-Couette flow, and torsional shearing flows between parallel plates, and in a cone-and-plate flow~\cite{LarsonShaqfehMuller1990,McKinleyByars1991,ByarsOztekin1994}. Measurements in Boger fluids  (where the viscosity remains approximately constant and the elasticity number is high)  demonstrated that beyond a certain critical Weissenberg number, the flows were unstable, thus bifurcating from axisymmetric shear flows to cellular three-dimensional flows. 

 The linear stability of a number of canonical shearing flows and the dependence of the spatiotemporal waveforms of the resulting three-dimensional flow fields was studied in the subsequent years and is summarized in a 1996 review by Shaqfeh~\cite{Shaqfeh1996}; see also Steinberg~\cite{Steinberg_AFM}. In particular, the linear stability analysis and resulting eigenvalue problems that were developed to describe the experiments demonstrated that such instabilities could be driven by the nonlinearities associated with the upper-convected Oldroyd derivative acting on the stress and the rate-of-strain tensors. In a small $\epsilon$  expansion, these critical conditions generally scaled with $\epsilon \textrm{Wi}^2$, which is the elastic equivalent of the Taylor number to use the language common to the Taylor-Couette literature, or equivalently, $\textrm{DeWi}$, where De is a Deborah number based on the time it takes a fluid element to be advected a distance corresponding to the radius of curvature of the flow.

Although the initial Taylor-Couette studies probing the conditions for instability exhibited qualitative agreement between experiments and the first linear stability calculations, and had recognized the physical mechanism underlying the elastic instability, there was quantitative disagreement, which took some time to explore and understand. In particular, the initial linear stability theory assumed that the observed unstable mode was axisymmetric and resulted in a time-dependent response, but reality proved more complex: Beris and colleagues showed numerically that, in fact, the largest growth rate corresponds to a non-axisymmetric mode~\cite{AvgoustiBeris,SureshkumarBeris1994}. To probe the quantitative details, Al-Mubaiyedh et al.~\cite{Khomami2000} showed theoretically and by numerical simulations that viscous heating can also play a significant role influencing the nature of the instability, which affected conclusions regarding flow stability based on axisymmetric versus non-axisymmetric modes. The relative magnitude of non-isothermal effects in a highly elastic fluid depend on the magnitude of the viscosity and relaxation time, as well as their (typically exponential) sensitivity to temperature, through a {\em thermoelastic number}~\cite{Rothstein01}. In detail, an isothermal analysis of the Oldroyd-B model yields non-axisymmetric and time dependent modes, while a non-isothermal (energetic) analysis, yields a time-independent (stationary) but axisymmetric instability.  While these details are inconsistent with the interpretation offered for the original experimental results of Larson, Muller and Shaqfeh, they are consistent with later more detailed experimental work by
Baumert and Muller~\cite{Baumert95,Baumert97,Baumert99} and Groisman and Steinberg \cite{Groisman96,Groisman97,Groisman98}, who found good agreement with isothermal theory based on the most unstable non-axisymmetric mode and consistent trends with the non-isothermal theoretical predictions; see also~\cite{Steinberg_AFM}. This understanding then sheds light on the detailed ``polymer-scale" mechanism coupling radial perturbations and polymer stretch in driving instability in the Taylor-Couette geometry, as discussed more below.

This class of elastic instabilities was  broadened by other researchers to include the Taylor-Dean (cylindrical Couette flow with an additional pressure-driven flow in the axial and azimuthal directions) and Dean flows (flows in curved channels)~\cite{JooShaqfeh1992} as well as lubrication bearing flows~\cite{DrisShaqfeh1995}, which are generally not used for viscometric measurements but contain the same kinematic elements, i.e., shear-dominated flow along curved streamline, as those in the original Taylor-Couette studies. Researchers again demonstrated through experiment and eigenvalue analysis that the instabilities were characterized by a critical value of the product \textrm{(De Wi)}. An examination of the mechanisms of all of these instabilities demonstrated at least three separate modes and mechanisms of instability, all of which involved the interaction of the base state (either the kinematics or existing hoop stress field) with a velocity fluctuation to locally enhance hoop stresses and further drive the fluctuation. 

All mechanisms leading to unstable conditions scaled, in the small gap limit, with \textrm{DeWi}. Thus, McKinley et al.~\cite{McKinley1996} suggested a scaling approach to the critical conditions, now known as the Pakdel-McKinley scaling, namely that if one writes the Weissenberg number more generally as $\textrm{Wi} = \tau_{11}/\left (\eta {\dot\gamma}\right )$, where $\tau_{11}$ is the primary normal stress component along the streamlines and $\eta$ the total fluid viscosity, then the dimensionless magnitude $M^2\textrm{= DeWi}$ could be developed into a general criterion whose critical value signaled the onset of elastic instabilities in many curvilinear shear flows. In particular, the onset of elastic instability is related to the characteristic curvature of the flow and stress along the streamlines, and is given by
			\begin{equation}
			   \frac{\lambda U}{\cal R}\frac{N_1}{\left | \tau\right |}\ge M^2,
			   \label{mparameter}
			\end{equation}
where $\lambda$ is the relaxation time of the fluid, $U$ is the characteristic streamwise fluid velocity, ${\cal R}$ is the characteristic radius of curvature of the streamline, $N_1$ is the first normal stress difference of the fluid, and $\tau$ is the total shear stress in the fluid. 

As summarized by Poole in Fig.~\ref{pefim}, now an enormous plethora of flows have been demonstrated to be elastically unstable, primarily by experimental observations and measurements. Most of these flows involve curved streamlines and thus their instability is attributed to hoop-stress driven instabilities. As such, they are typically characterized, in some manner, by the $M$ parameter. The geometric scaling has been enormously successful, as, for example, it has been shown that instability during flow in a  serpentine channel is directly related to the Dean instability~\cite{PooleLindner2012}. Remarkably, the dependence of the instability threshold in Taylor-Couette flow predicted by the Pakdel-McKinley criterion is in a better agreement with the experimental values than the results of linear stability analysis carefully tailored to the fluid's rheology~\cite{Schafer2018}.

As the instability develops in time, or conditions beyond the critical conditions are considered, the dynamics of these purely-elastic instabilities become increasingly complex, even at very small Reynolds numbers. In the late 1990s, Baumert and Muller~\cite{Baumert95,Baumert97,Baumert99} and Groisman and Steinberg \cite{Groisman96,Groisman97,Groisman98} reported a series of experimentally observed transitions in Taylor-Couette flow involving axial vortices developing into localized ``diwhirls" and ``flame" patterns followed by oscillating states and finally disordered oscillations. Kumar and Graham \cite{Kumar:2000ut,Kumar:2001tj} studied Taylor-Dean flow  and computed stationary nontrivial solutions with the FENE-P model that strongly resemble some of the experimentally observed diwhirls, showing that they arise in a nonlinear transition scenario.  The self-sustaining mechanism is related to the mechanism of instability in viscoelastic Dean flow \cite{JooShaqfeh1992}, arising from a finite-amplitude perturbation giving rise to a locally parabolic profile of the azimuthal velocity near the upper wall. The more complex time-dependent states were later simulated, at least qualitatively, by Khomami and coworkers \cite{Thomas06,Liu13,Ghanbari14}. 
These ideas of self-sustaining nonlinear interactions between the velocity field and the state of stress in the flow form a robust mechanistic basis for a transition to elastic turbulence \cite{Groisman00}. 

As described further in \S~\ref{FlowTransitionSection}, the first two decades of this century have focused on achieving a deeper understanding and progressively unraveling the complex viscoelastic dynamics for a range of different flow geometries and fluid rheologies. Furthermore, in more recent work, Khomami and coworkers have focused on examining through direct numerical simulations the connection between inertial and elastic turbulence, as well as the connection to curvature and curvature-induced elastic instabilities (of the type described by the Padel-McKinley criterion) in strengthening large-scale Taylor vortices at the expense of small-scale G{\"o}rtler vortices as the curvature in the flow is increased while keeping the same Reynolds and Weissenberg numbers \cite{Song2019}. Additional work exploring these connections is described in \S~\ref{FlowTransitionSection}-\ref{EITsection}.

\begin{table*}[t]
\centering
\begin{tabular}{|c|c|c|c|c|c|}\hline
 Model& $\xi$ &  $\lambda ( {\bf c})$&  $\alpha$&  $f( {\bf c})$&  Remarks/References \\ \hline\hline
 Maxwell & $0$&$\lambda_0$ &$0$ &$1$ &Hookean or linear \\ 
 & & & & & dumbbell model~\cite{BAHvolume1};\\
 Oldroyd-B& & & & & Oldroyd-B includes \\
 & & & & & viscous stress\\ \hline
 Johnson-& $0\le \xi \le 2$ & $\lambda_0$ & $0$ &$1$  & \cite{JohnsonSegalman1977}\\ 
 Segalman&  &  &  &  & \\ \hline
 FENE-P& $0$ & $\lambda_0$ & $0$ &$\frac{L^2-3}{L^2-tr({\bf c})}$  & $L$ represents the dimensionless maximum chain \\
 & & & & &extensibility; $3< L^2<\infty$~\cite{BAHvolume1} \\ 
\hline
 Giesekus & $0$ &$\lambda_0$  & $0<\alpha <1$ & $1$ & $\alpha$ is an anisotropic mobility parameter~\cite{Giesekus1982} \\ \hline
 Phan-Thien  & $0\le \xi \le 2$ & $\frac{\lambda_0}{1+\epsilon tr\left ({\bf c}-{\bf I}\right)}$ & $0$ & $1$ &This is the linear PTT model~\cite{PhanThienTanner1977};  \\
 \& Tanner & & & & & for a nonlinear version use $\lambda=\lambda_0 e^{-\epsilon tr \left ({\bf c}-{\bf I}\right )}$; $\epsilon>0$\\ \hline
 Extended &  &  &  &  & \\ 
 White-Metzner& $0$ & $\lambda_0 \left (\frac{1}{3} tr ({\bf c})^{-k}\right)$  & $0$ & $1$ & $k>0$ \cite{SouvaliotisBeris1992}\\ \hline
\end{tabular}
\caption{List of commonly used models along with the corresponding expressions with respect to the conformation tensor, $\bf c$, or values for materials parameters in Eq.~\ref{MicroStructureEvolutionEqn}. The chain length parameter $b= L^2/\left (L^2-3\right )$ is also used in some modeling studies.}
\label{table1}
\end{table*}

\section{Constitutive models and numerical simulations of elastic flow instabilities}
\label{ModelsAndNumericalSimulations}

In addition to experimental characterization, many researchers are seeking insight into this large class of instabilities \textit{via} large scale numerical simulation. As an example, during the workshop M. Alves presented results from {\em RheoTool} (a numerical library based on the open-source OpenFOAM\textsuperscript{\textregistered}) on cross slot~\cite{Alves2016} and contraction flow instabilities that are purely elastic. Even though the flows have a large region of extensional flow, there is evidence that these instabilities are again driven by elastic hoop stresses. However, that evidence comes from calculating local fields of the \textit{M} parameter in a flow and demonstrating that the flows break symmetry and/or become time dependent when the maximum value of \textit{M} becomes sufficiently large, e.g., $M> 4 - 5$. In this context, there is a lack of linear (or energy) stability analyses for these extension-dominated flows such as the cross-slot geometry.

What are the underlying models used in theory/numerical simulations? For the analysis of flow instabilities and/or simulation of highly elastic viscoelastic flows, differential models are typically used that connect the stress and its time and space derivatives to the velocity gradient and its time derivatives~\cite{BAHvolume1,TannerBook}.  The simplest of these models is the Upper Convected Maxwell (UCM), or Oldroyd-B model (when a Newtonian solvent viscosity contribution is added).  This model originates from a simple mechanical analog of polymer solutions corresponding to a spring and dashpot in series, with the upper convected time derivative expressing the second-order (contravariant) tensor generalization of the material time derivative of the stress tensor, as beautifully shown first in the pioneering work of Oldroyd~\cite{Oldroyd1950}.  Most importantly, some time later, a formal connection was made  to an idealized Hookean dumbbell polymer structure~\cite{BAHvolume2}.  This image of a solution's microstructure has allowed a number of considerable generalizations to be obtained, like the Finitely Extensible Non-linear Elastic dumbbell with the Peterlin approximation (FENE-P dumbbell) that allows for a finite polymer extensibility~\cite{BAHvolume2,Rallison1988}.  Other notable generalizations of the Oldroyd-B model are: the Johnson-Segalman model~\cite{JohnsonSegalman1977} involving a non-affine correction to the upper convected time derivative, first proposed by Gordon and Schowalter~\cite{GordonSchowalter1972}; the Giesekus model involving a nonisotropic drag controlled by a mobility parameter~\cite{Giesekus1982}; and the Phan-Thien and Tanner (PTT) model involving a dependence of the relaxation time on the stress~\cite{PhanThienTanner1977}.  

All of these models can be  described conveniently using a time-evolution equation  in terms of the stress tensor,  $\boldsymbol{\tau}$.  However, given the connection of microstructural models to kinetic and network theories the stress is assumed to be related to an internal structural parameter, $\bf c$, which is typically identified as the second moment $\left\langle {\bf R}{\bf R}\right \rangle$  of the end-to-end distance ${\bf R}$ vector if macromolecular chains are involved~\cite{BAHvolume1};   an elastic deformation strain~\cite{LeonovProkuninBook,TruesdellNoll} can also be described in terms of $\bf c$.  An advantage of this representation is that it allows for a connection to the theories of nonlinear elasticity~\cite{TruesdellNoll}, while providing for a nonlinear thermodynamics foundation~\cite{GrmelaCarreau,BerisEdwards} that allows for both a straightforward extension/mixing of models (like the extended White-Metzner model~\cite{SouvaliotisBeris1992}) and an evaluation for their thermodynamic consistency and Hadamard-type instabilities~\cite{LeonovProkuninBook}.   

Indeed, all of the above-mentioned models can be concisely represented as~\cite{BerisEdwards}:
\begin{eqnarray}
   \frac{D{\bf c}}{Dt}  &-& \left (\nabla {\bf v}\right )^T \cdot {\bf c} -{\bf c}\cdot\nabla{\bf v}+{\xi} 
   \left (\textbf{D}\cdot{\bf c}+ {\bf c}\cdot{\textbf{D}}\right )=\nonumber \\
   &-&\frac{1}{\lambda({\bf c})}\left( {\boldsymbol\tau}^+ +\alpha {\boldsymbol\tau}^+\cdot{\boldsymbol\tau}^+\right ); \quad{\boldsymbol\tau}^+=f({\bf c}){\bf c}-{\bf I},
   \label{MicroStructureEvolutionEqn}
\end{eqnarray}where $D/Dt$ is the material derivative, $\xi$ is a dimensionless non-affine motion  parameter with $0\le \xi \le 2$, $\textbf{D}=\frac{1}{2}\left (\nabla{\bf v}+\left (\nabla{\bf v}\right )^T\right )$  represents the rate of deformation tensor, $\lambda(\bf c)$ represents the relaxation time (which may be a function of $\bf c$), ${\boldsymbol\tau}^+$ is the dimensionless polymer extra stress,
${\boldsymbol\tau}^+={\boldsymbol\tau}/\left (G_0 (1-\xi )\right )$, with $G_0$  a characteristic elastic modulus, and  $f({\bf c})$ is a model-dependent parameter representing finite polymer extensibility effects.  The left-hand side of Eq. (\ref{MicroStructureEvolutionEqn}) corresponds to the Johnson-Segalman derivative~\cite{JohnsonSegalman1977}.  For most polymeric systems $\xi=0$, in which case the left-hand side of Eq. (\ref{MicroStructureEvolutionEqn}) reduces to the Oldroyd upper-convected time derivative, as is appropriate for a structural material parameter connected to the Cauchy elastic strain tensor~\cite{TruesdellNoll}.  

As discussed, Eq. (\ref{MicroStructureEvolutionEqn}) can represent all the above-mentioned constitutive models, with suitable choice of the model parameters, as shown in Table \ref{table1}.  We note that the highest elasticity (as for example determined by the magnitude of the normal stresses in shear flows) is obtained with the Maxwell/Oldroyd-B model. Alternatively,  $\xi\rightarrow 0$,  $L\rightarrow\infty$,  $\alpha\rightarrow 0$ (see Eq. (\ref{MicroStructureEvolutionEqn}) and Table~\ref{table1}), limits for which we recover the Maxwell/Oldroyd-B models,
are values often selected in numerical simulations/analyses seeking to maximize the effects of elasticity, such as, for example, simulations of highly elastic, viscoelastic turbulent flow~\cite{SureshkumarBeris1997,HousiadasBeris2005,GrahamFloryan2021,ZhuXi2021}.

Another advantage of the conformation tensor representation is that it allows checking for numerically induced instabilities as, from theory and its physical interpretation, $\bf c$ is a positive definite tensor~\cite{BerisEdwards}.  Consequently, numerical schemes have been devised so that they guarantee that $\bf c$ always remains  positive definite, such as the log-conformation tensor methodology proposed by Fattal and Kupperman \cite{Fattal1,Fattal2} or the matrix decomposition proposed by Vaithianathan and Collins \cite{Vaithianathan}. As such, when used, the approach significantly improves the numerical stability and allows reaching substantially higher values of elasticity (i.e., higher $\rm{Wi}$ or $\rm{De}$ numbers)~\cite{HulsenFattal2005}.  Different types of stabilization techniques commonly used in computational rheology were reviewed recently by Alves and co-workers~\cite{Alves}.

How successful have studies using these rather idealized single relaxation mode viscoelastic models been in describing highly elastic, viscoelastic flows?  
One could say fairly successful, judging from several important accomplishments.  First and foremost, these include the capability of reproducing the polymer-induced drag reduction phenomenon in direct numerical simulations (DNS) of turbulent flows using the above-mentioned single relaxation viscoelastic models, when the elasticity in the flow was high enough~\cite{SureshkumarBeris1997,HousiadasBeris2005,GrahamFloryan2021,ZhuXi2021}, e.g., see the discussion of elastoinertial turbulence in \S~\ref{EITsection}.  The drag reduction is observed experimentally when high molecular weight polymers are added, usually in small concentrations, to a Newtonian solvent~\cite{Toms1949}.  The  DNS results, typically carried out in a channel, showed, in addition to the drag reduction that increased with increasing elasticity in the flow, all the main kinematic effects accompanying it, such as the increase in the extent of the buffer layer, the widening of the streaky structure, and the enhancement of the larger, more coherent turbulent structures at the expense of the smaller scales, etc.~\cite{SureshkumarBeris1997,HousiadasBeris2005,GrahamFloryan2021,ZhuXi2021}.  

Second, the detailed results and sensitivity analysis to the model parameters allowed one to deduce the main mechanism behind the drag reduction, namely the weakening of the vortical structures (eddies) due to the enhancement of the resistance to extensional deformations induced by viscoelasticity~\cite{SureshkumarBeris1997,HousiadasBeris2005}, as originally proposed by Seyer and Metzner~\cite{SeyerMetzner1969} and Lumley~\cite{Lumley1969}.  

Third, in the most recent of these works~\cite{GrahamFloryan2021,ZhuXi2021} the analysis of the underlying mechanism revealed further details of the reasons for the experimentally observed maximum drag reduction~\cite{Virk}. In this respect, worth mentioning are the results of a recent linear stability analysis of a highly elastic viscoelastic pipe flow that showed it to be linearly unstable for a certain range of the parameters of the Oldroyd-B fluid model~\cite{garg2018,chaudharyetal_2021}.  Moreover, linear stability analysis and direct numerical simulations of a highly elastic viscoelastic channel flow demonstrated the presence of an ``arrowhead" two-dimensional wave instability for a certain range of parameters of a FENE-P model~\cite{PageDubiefKerswell2020,Dubief2021prerint} -- see the relevant discussion in \S~\ref{EITsection}A. 

The successes just described in extracting new physics out of highly elastic but rather idealized models raise the question as to how accurately the models, in the limit of high elasticity, can predict real polymer flow behavior.  Of course, this depends on the complexity of the polymeric system.  Research along these lines, involving much more accurate (but also much more highly time-consuming from a computational stand point) microscopic models, along with a comparison to dilute polymer solutions~\cite{DoyleShaqfeh1998,RadhakrishnanUnderhill2012}, has shown that it may be possible to make quantitative predictions if some modifications are implemented to the description of the friction drag on the individual beads in the microscopic multi-bead models used.  Thus, the fact remains that microscopic multi-bead models still need to be used, which is not very hopeful from a macroscopic, continuum mechanics, viewpoint. However, some early work on a modified FENE-P model (using two conformation tensor parameters instead of one) has shown that it may still be possible to capture those nonlinear effects, which arise primarily due to the non-Gaussian microscopic distribution of the deformation of the polymer chains and which are reflected in hysteresis phenomena~\cite{LielensKeunings1999}.  This is therefore an avenue that still remains to be exploited.  Still, for more complex physics, such as presented by concentrated polymer solutions and melts~\cite{Schroeder2018,SefiddashtiEdwards2019}, or micellar solutions~\cite{Fielding2007,GermannCookBeris2013} or under confinement~\cite{MavrantzasBeris1999} and especially in the presence of entanglements~\cite{SefiddashtiEdwards2019}, additional components in the models may be needed as outlined in the referenced papers. A particular outstanding challenge for these models are correct pressure-drop predictions for well-characterized dilute polymer solutions in mixed kinematics flows \cite{rothstein99}.

\section{Transition to elastic turbulence}
\label{FlowTransitionSection}

We have seen that even relatively ``simple" polymeric flows at low Reynolds numbers can become unstable. Beyond the instability we expect the flows to be more complicated; indeed, features consistent with ``turbulence" have been identified. Here, we describe some of the challenges and puzzles raised by recent research on elastic turbulence in two- and three-dimensional flows, especially parallel shear flows, such as pressure-driven flows in channels and pipes.

\subsection{Taylor-Couette Flow}
One archetypical geometry for studies of purely elastic instabilities in flows with curved streamlines is the Taylor-Couette geometry. In simulations of this configuration, what is especially surprising is that all unstable modes that have been computed so far rely on the three-dimensional character of the flow. This topic was addressed during the workshop by H. Stark, who presented  numerical simulations of two dimensional (i.e., zero axial wave number) elastic Taylor-Couette flow using {\em RheoTool}.  At large enough Wi, the elastic turbulence that had been reported by Steinberg and coworkers in serpentine channel flow and plate-plate flows  \cite{GroismanNJP,Rothstein01,Steinberg_AFM,PooleLindner2012}appeared .

\begin{figure}
\begin{centering}
\includegraphics[width=0.9\columnwidth]{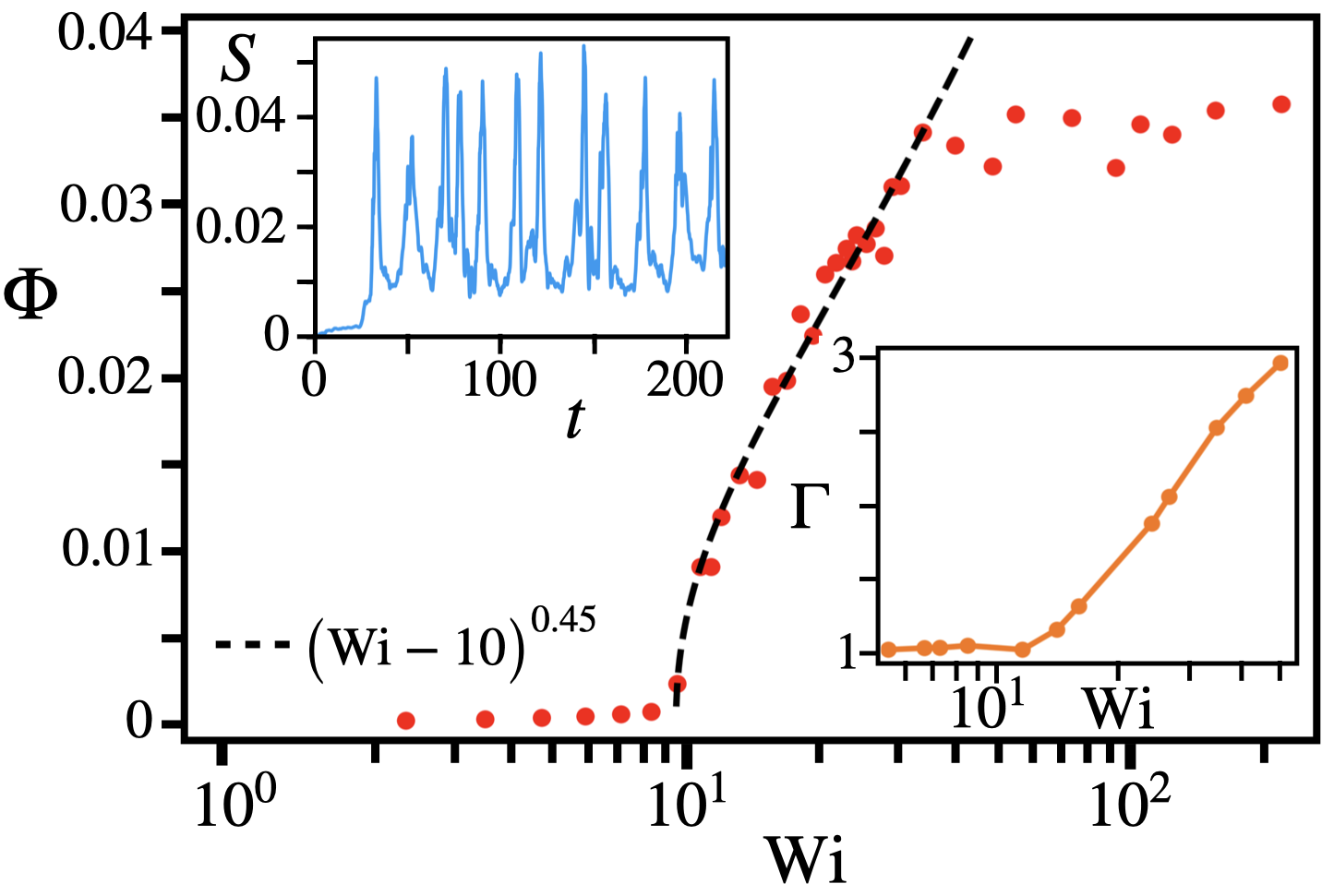}
\caption{Order parameter 
$\Phi = \overline{S(t)}$ 
versus Weissenberg number $\textrm{Wi}$. The dashed line indicates
the fitted scaling law beyond the elastic instability. Upper inset: Secondary flow strength 
$S(t)$
plotted versus $t$ for
$\textrm{Wi} = 16.3$. Lower inset: Flow resistance quantified by the azimuthal stress on the outer cylinder plotted versus
$\textrm{Wi}$. Adapted from Ref. \cite{Buel18} with permission from \textit{EPL}; copyright (2018).
}
\label{fig.2DTC}
\end{centering}
\end{figure}

In particular, using the Oldroyd-B model for numerical simulations, van Buel and Stark reported the
onset of the elastic instability towards elastic turbulence in the two-dimensional Taylor-Couette flow
with a wide gap~\cite{Buel18}. They locate the
instability by an order parameter 
$\Phi = \overline{S(t)}$, 
which is the time average of the secondary-flow strength
$S(t)$ 
that measures the overall deviation from the Taylor-Couette base flow. The upper inset of
Fig.~\ref{fig.2DTC} shows a typical  example for 
$S(t)$
beyond the critical Weissenberg number $\mathrm{Wi}_c\approx 10$
and reveals the irregular nature of the secondary flow. The continuous increase of the order parameter with Wi
in Fig.~\ref{fig.2DTC} indicates a supercritical instability that is
accompanied by an increase in the flow resistance.

\begin{figure}
\begin{centering}
\includegraphics[width=0.9\columnwidth]{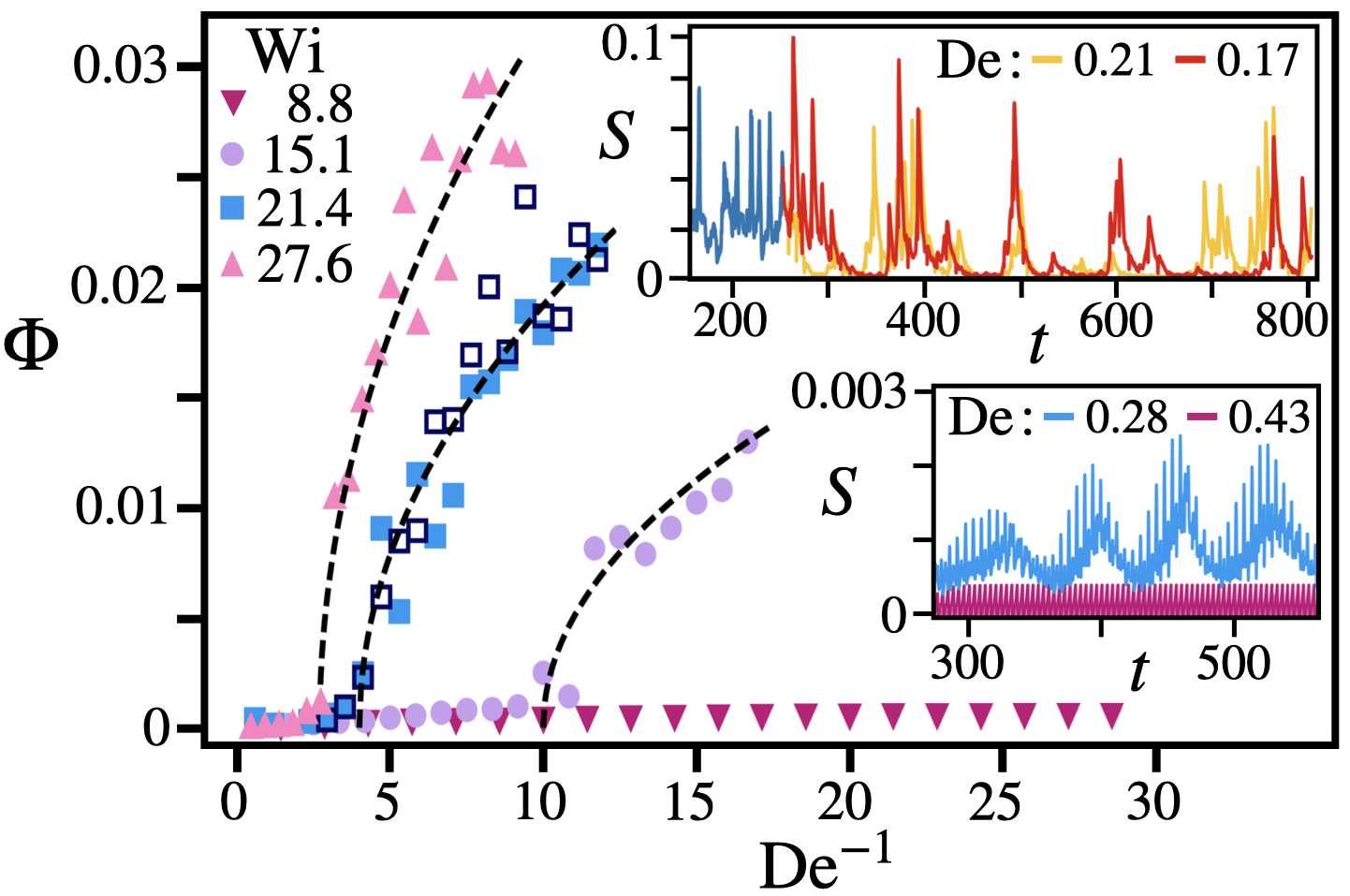}
\caption{Active control of elastic turbulence. Order parameter versus inverse Deborah number, $\textrm{De}^{-1}$, for
different $\text{Wi}$. Insets: Secondary-flow strength versus time for different $\textrm{De}$ at $\textrm{Wi} = 21.4$.
The time-modulated driving is switched on at $t = 250$. Adapted from Ref. \cite{Buel20}.}
\label{fig.control}
\end{centering}
\end{figure}

Furthermore, for the spatial power spectrum of the
secondary flow along the azimuthal direction, a power law decay $m^{-\zeta}$ with
exponent $\zeta > 3$ for all $\textrm{Wi} > \textrm{Wi}_c$ was reported~\cite{Buel18}. This result is consistent with a theoretical bound on the exponent $\zeta$~\cite{Fouxon03}. Note that we do not expect the Kolomogorov scaling $k^{-5/3}$ of inertial
turbulence since elastic turbulence is initiated by elastic stresses. Also,  the exponent $\alpha$
of the temporal power spectrum was found to be generally smaller than $\zeta$~\cite{Buel18}, hence, does
not obey Taylor's hypothesis for inertial turbulence that demands that the exponents are equal \cite{Taylor38}. Only for small $\text{Wi}$ was the exponent $\alpha> 3$, which is considered to be a signature for turbulent flow and was measured in several experiments~\cite{Groisman00,Sousa18}. Groisman and Steinberg  suggested that the reduced value of the exponent $\alpha$ found in three-dimensional Taylor Couette flow indicates that the flow is transitional~\cite{GroismanNJP}, and not fully in the elastic turbulence regime, due to the large shear strain component in Taylor-Couette flow with a small gap and a possible increase in the Wi needed to generate the coil-stretch transition of individual polymers~\cite{LiuSteinberg}.

Typically, changes to the flow geometry or boundary conditions are used to passively control the onset of the elastic instability and elastic
turbulence\ \cite{Rothstein01,Neelamegam13,Davoodi19,Walkama20}.
In this spirit, van Buel and Stark realized active open-loop control
in simulations of the two-dimensional Taylor-Couette flow~\cite{Buel20}. They apply a time-modulated shear stress by periodically reversing the
rotational velocity of the outer cylinder. The modulation frequency is quantified by the Deborah
number, the product of frequency times stress relaxation time. The insets of Fig.~\ref{fig.control}
show how the secondary-flow strength of the turbulent velocity field decreases with increasing $\textrm{De}$
until a modulated laminar flow remains. For high modulation frequencies the elastic stresses cannot fully build up
in order to generate turbulent flow. The transition from laminar to turbulent flow is again supercritical
(Fig.~\ref{fig.control}) and for larger Weissenberg numbers a larger critical frequency (or $\textrm{De}_c$) is needed
to suppress elastic turbulence. Note that the different curves in Fig.~\ref{fig.control} collapse onto a master curve
when plotting $\Phi / \textrm{Wi}^{3/2}$ versus $\textrm{De}^{-1} - \textrm{De}_c^{-1}$.

\subsection{Parallel Shear Flows}

In the sections that follow we discuss both experimental observations, including instabilities and later fully developed turbulence, and associated theoretical attempts to describe these flows in channels and pipes. The field has come a long way. In a prior version of this workshop that was held at the Princeton Center for Theoretical Science in 2018, the mere existence of sustained fluctuations in viscoelastic flows in straight channels was in doubt. Three years later, their existence is now well established, and researchers are now working on understanding the origins and mechanisms governing these observed instabilities. Even after much effort, however, there are different interpretations offered.


\subsubsection{Theoretical analyses}
\label{subsubsec:TheorylowRe}

\textbf{Modal linear stability analysis.} As we have discussed above, low Reynolds number, polymeric fluid flows with curvilinear streamlines are characterized by an elastic hoop stress that generates a bulk (body) force acting on the fluid in the direction of the center of curvature, which leads to an elastic instability and subsequently to elastic turbulence. This instability mechanism ceases to be effective at zero curvature in flows with straight streamlines, such as parallel channel shear flow, i.e.,
based on the criterion given in Eq. (\ref{mparameter}), purely elastic hoop stress-driven instabilities are not possible as the curvature of the streamlines decreases to zero (${\cal R}\rightarrow0$).
 Thus, a common assumption is that parallel or rectilinear shear flows of viscoelastic fluids, such as plane Couette and Poiseuille flows, are linearly stable in the absence of inertia~\cite{Leonov1967}. This form of stability is described using linear stability analysis, which decomposes a perturbation in the flow into normal modes, familiar from studies of Fourier series. For solutions with the time dependence assumed to be of the form $e^{i \omega t}$, eigenvalues $\omega$ with negative imaginary parts correspond to perturbations that grow exponentially in time, thus leading to a linear instability as we consider the limit $t\rightarrow \infty$.  
 

There are a number of directions that have been pursued to examine the possible linear stability of viscoelastic flows. Motivated by the polymer extrusion instability and the problem of ``melt fracture," Ho and Denn examined the stability of plane Poiseuille flow of a UCM fluid and concluded, based on an eigenvalue analysis, that the flow is stable to infinitesimal perturbations~\cite{HoDenn1977}. The authors did acknowledge the possibility of the flow becoming unstable to finite amplitude perturbations, but it was deemed unlikely. Similar results were found by Lee and Finlayson  for Poiseuille and planar Couette flows~\cite{LeeFinlayson1986}, by Renardy and Renardy for Couette flow using spectral methods~\cite{RenardyRenardy1986}, and by Gorodtsov and Leonov~\cite{Leonov1967} for plane Couette flow. A rigorous proof that such rectilinear viscoelastic flows are indeed linearly stable was provided by M. Renardy~\cite{Renardy1992}, who studied the stability of plane Couette flow of a UCM fluid. Importantly, the author cautioned that artificial instabilities could arise from numerical discretization in simulations of viscoelastic flows.


Taken together, these results suggested that parallel shear flows of model viscoelastic fluids are indeed linearly stable \cite{Larson1992,Shaqfeh1996}, with the exception of fluids with strongly shear-thinning material properties~\cite{WILSON199975,WILSON2015200}. So, it came as a surprise when Khalid et al.~\cite{khalid_creepingflow_2021} recently reported a linear instability in purely elastic channel flows of Oldroyd-B fluids --- described further in the next section. Although only found in the part of the parameter space that might be difficult to access experimentally ($\textrm{Wi}= O(10^3)$ and the ratio of the solvent to the total shear viscosity $\beta \gg 0.9$), these results suggest that the linear stability analysis of parallel shear flows of simple viscoelastic model fluids needs to be revisited.\\


\textbf{Transient non-modal growth of non-normal perturbations.} While the linear stability analyses discussed above rule out the existence of a linear instability for a broad class of viscoelastic parallel shear flows, they do not automatically imply that such flows remain laminar. Indeed, there are several mechanisms that can potentially lead to flows that are very different than the corresponding laminar ones, even in the absence of a linear instability. One of such mechanisms was uncovered in the early 1990s in the field of Newtonian hydrodynamics. It relies on the observation that the dynamics of infinitesimal perturbations introduced to a laminar flow are governed by linear equations that often involve non-self-adjoint (non-normal) operators~\cite{Trefethen1993,Grossmann2000,schmid2000stability,Schmid2007}.  Although the real parts of the associated eigenvalues can all be negative, the associated eigenmodes may not be `orthogonal' to each other, i.e., they do not represent unique, independent flow perturbations. Instead, some of the modes become almost parallel to each other, especially for sufficiently large Reynolds numbers~\cite{Grossmann2000}. For additional discussion of the transition to turbulence in inertially dominated flows, see \S~\ref{EITsection}.

This \textit{non-normality} has a profound implication for the short-time evolution of flow perturbations: an initial state, prepared as a combination of several such eigenmodes (in other words, a general random perturbation of the form expected to be experimentally relevant) will see its kinetic energy increase algebraically in time, reaching values that are many times larger than the initial value; in the framework of constant coefficient differential equations with repeated roots, there are solutions $te^{-at}$, where $a$ is (in general) a complex constant with positive but small in magnitude real part, which grow at early times $t$. The solutions then decrease exponentially in time, as predicted by the modal linear stability analysis of the previous subsection. It was shown that the maximum energy amplification that can be achieved through this mechanism in plane Couette and channel flows is $O(Re^2)$~\cite{Schmid2007}. For sufficiently large Reynolds numbers, such strong amplification can lead to perturbations becoming sufficiently large so that their dynamics are no longer described by the linearized equations on which the analysis is based. Thus, if a particular linearly stable flow is unstable to finite-amplitude perturbations (in other words, there exists a `bifurcation from infinity'), as is the case with plane Couette and pipe flows of Newtonian fluids, non-normal growth can amplify small experimental noise helping to tip the system over the instability threshold.

The corresponding theory for viscoelastic non-normal growth was developed by M. Jovanovich, S. Kumar, and colleagues during 2008-2018 \cite{jovanovic_kumar_2010,jovanovic_kumar_2011} and T. Zaki and colleagues \cite{Page2014,Page2015} in 2014-2018. Specifically, it was demonstrated that even in the absence of inertia, infinitesimal perturbations in plane Couette and channel flows can be significantly amplified \cite{jovanovic_kumar_2010}. Perturbations of the streamwise velocity achieve growth by a factor of $O(\textrm{Wi})$, while the streamwise component of the polymer stress tensor can be amplified up to $O(\textrm{Wi}^2)$ compared to its initial value; the time to reach the maximum values scales as $t_{max} \sim \lambda \textrm{Wi}$, where $\lambda$ is the relaxation time of the fluid. Similar to its Newtonian counterpart, the purely elastic non-normal growth theory predicts that the most amplified initial flow structures comprise (almost) streamwise-independent vortices, leading to streamwise-independent streaks \cite{jovanovic_kumar_2010}, although other forms of stress amplification have also been examined more recently \cite{hariharan2021}. 


The non-normal growth mechanism provides a powerful pathway to significantly amplify small-amplitude experimental noise until it becomes large enough to ignite some non-linear process that would sustain turbulent flow. However, the non-normal growth mechanism does not provide any insight into the non-linear process. As a linear theory, it cannot predict a critical Weissenberg number at which an instability might set in. We should also mention that although streamwise vortices and streaks are naturally produced by this theory, their experimental observation is not a proof that non-normal amplification is at play in that particular flow: as discussed by Waleffe \cite{Waleffe1995}, these flow structures may also be produced by other, non-linear mechanisms. \\

\textbf{Weakly nonlinear analysis.} As already discussed above, linear stability does not imply global stability. One of the classical examples of such behaviour is Newtonian pipe flow that is linearly stable for all Reynolds numbers but is unstable when a sufficiently large perturbation is added to the flow \cite{Eckhardt2007}. In the early 2000s, Bonn, van Saarloos, Morozov, and collaborators~\cite{MeulenbroekBonn2003,BertolaBonn2003} proposed that viscoelastic parallel shear flows exhibit analogous behaviour. Using weakly nonlinear analysis, the authors in~\cite{MeulenbroekBonn2003} tackled an interesting observation, namely that the fracture instability in polymer melts, which occurs when the solution flows out of a slit or ``die'', seems to occur at an approximately constant value of \textrm{Wi}. Using the UCM fluid model, it was then shown that viscoelastic Poiseuille flows could exhibit a nonlinear  ``subcritical''€™ instability due to normal stress effects; the flow was predicted to become unstable at $\hbox{Wi}_c \approx 5$. This analysis was followed by experiments \cite{BertolaBonn2003} that showed melt fracture instability at $\textrm{Wi}$ values that are quantitatively similar to those predicted by the nonlinear expansion theory \cite{MeulenbroekBonn2003}.

Subsequent analysis by  Morozov and van Saarloos~\cite{MorozovVanSaarloos2005,morozov_saarloos2007} for plane Couette and Poiseuille geometries showed that the viscoelastic flows could be unstable to finite-amplitude perturbations without curved streamlines or inertia. They developed a novel amplitude-equation technique that constructs a non-linear solution as power series in its amplitude (relative to the laminar flow), with the lowest-order term being the least-stable eigenmode of the linear stability analysis. Morozov and van Saarloos found that plane Couette and channel flows of Oldroyd-B fluids exhibit sub-critical instabilities for $\rm{Wi} \gtrapprox 3$ and $\rm{Wi} \gtrapprox 5$, respectively. The non-linear flow structures predicted by this analysis are travelling-wave solutions, similar to their Newtonian counterparts \cite{Eckhardt2007,Barkley2020}; for channel flows their spatial profiles are reported in \cite{morozov_saarloos2007}. Their origin can be understood as a two-step process. While the underlying laminar flow has straight streamlines, and is thus linearly stable according to the Pakdel-McKinley criterion \cite{Pakdel1996} (Eq. \ref{mparameter}), a slowly decaying perturbation with curvature in its streamlines can drive an instability. This \textit{perturbation of a perturbation} scenario then leads to a finite-amplitude threshold.

To understand the relevance of these solutions to purely elastic turbulence in parallel shear flows, Morozov and van Saarloos yet again drew an analogy with Newtonian turbulence in pipes and rectilinear channels \cite{Eckhardt2007,Barkley2020}. Our current understanding of the transition in these flows is centered on the exact solutions to the Navier-Stokes equations discovered by Nagata \cite{Nagata1990}, Waleffe \cite{Waleffe1997}, Hof \cite{Hof2004}, and others. These solutions, often referred to as \textit{coherent structures}, are either travelling waves or periodic orbits that comprise streamwise streaks and vortices, and instabilities connecting them; they are generated through a self-sustaining process uncovered by Waleffe \cite{Hamilton1995,Waleffe1997}. Importantly, coherent structures are linearly unstable: their vicinity in the phase space contains many attractive and a few repulsive directions \cite{Eckhardt2007,Barkley2020}, and a typical turbulent trajectory is performing a pin-ball-type motion among those coherent structures. While each of them is very regular in space (i.e. they do not look turbulent in any sense), an instantaneous snapshot of the flow caught in-between many coherent structures simultaneously appears turbulent. This scenario is at the frontier of the current research in Newtonian turbulence and there are strong early indications that it persists sufficiently far away from the transition \cite{Dennis2014}. Morozov and van Saarloos proposed \cite{morozov_saarloos2007} that the solutions found in \cite{MorozovVanSaarloos2005} are the viscoelastic counterparts to Newtonian coherent structures, and, while not being directly observable, they play a role in organising the phase space dynamics of purely elastic turbulence in parallel shear flows.

It is important to note that the weakly non-linear analyses presented above are the only theoretical results on non-linear structures in purely elastic channel and pipe flows currently available. Although suggestive, they are obtained by an approximate technique; also their linear stability is currently unknown. A significant amount of new research in this area is needed before the analogy with Newtonian turbulence, as proposed by Morozov and van Saarloos \cite{morozov_saarloos2007}, can be made more exact.

\subsubsection{Experimental results}

\textbf{Experiments of Arratia and colleagues.} Experimental evidence of nonlinear instability has been hard to come by. While the hysteretic behavior presented in \cite{BertolaBonn2003} is consistent with a nonlinear instability, it was unclear whether the instability originated inside or outside of the flow domain. A subsequent experimental study on the stability of viscoelastic flows inside a cylindrical straight pipe did find unusually large velocity fluctuations far downstream for the initial perturbation, but the subcritical nature of the instability was not established and no hysteretic behavior was reported~\cite{Bonn_Kellay_PRE2011}. 

\begin{figure}
    \includegraphics[width=\columnwidth]{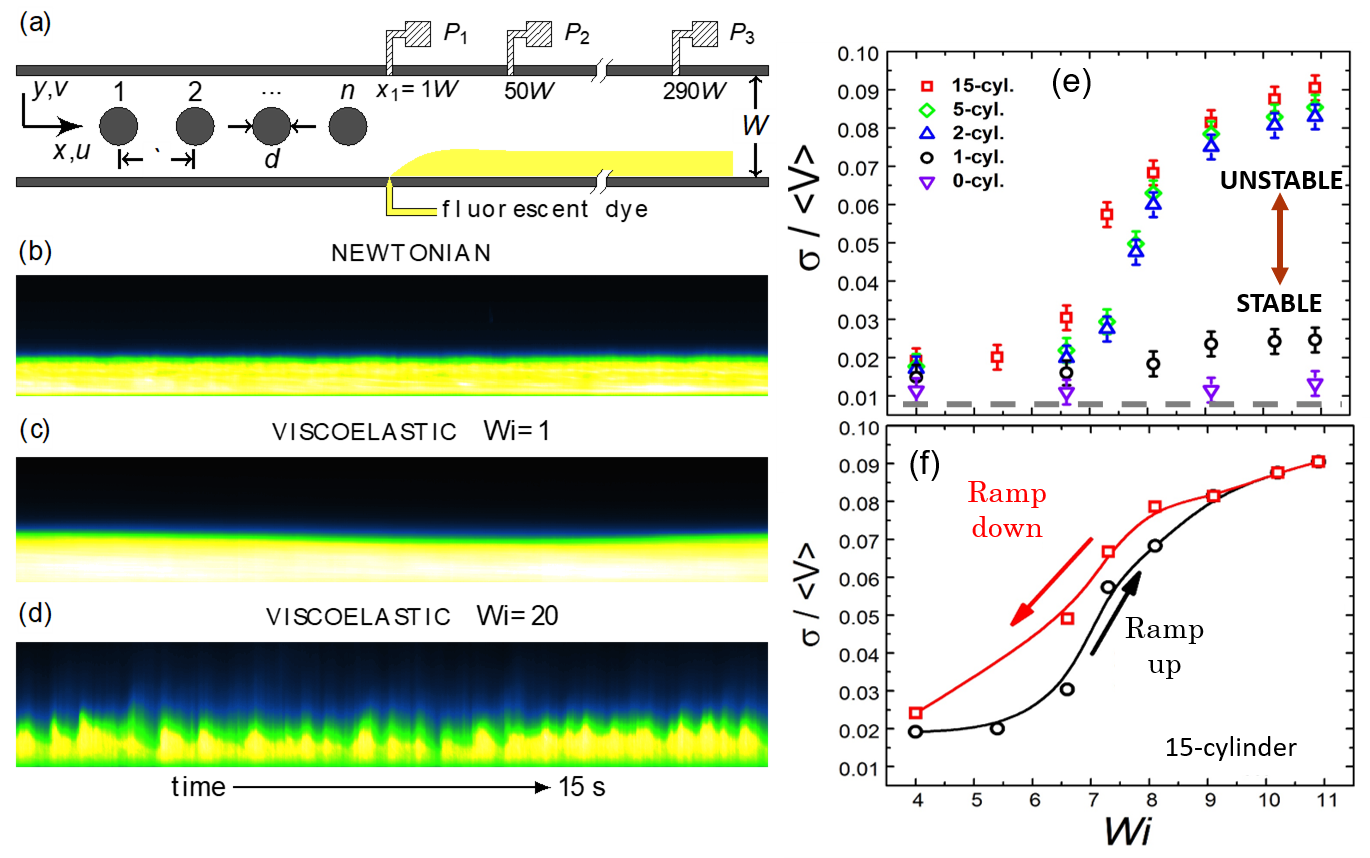}
    \caption{Nonlinear elastic instability in a microfluidic channel flow. (a) Experimental setup showing the initial linear array of cylinders followed by a long parallel shear flow region. (b,c,d) Space-time dye patterns for the case with 15-cylinders for Newtonian and polymeric fluids measured far downstream. (e) Normalized velocity fluctuations as a fucntion of initial perturbation (n) and Wi showing the appearance of two branches. (f) Hysteretic behavior, a hallmark of nonlinear sub-critical instabilities, found for polymeric fluids \cite{Pan_2012_PRL}.}
    \label{fig1}
\end{figure}

Thus, it was particularly notable when in 2012, Arratia and co-workers~\cite{Pan_2012_PRL} provided experimental evidence of such nonlinear subcritical instability in a straight microfluidic channel. A linear array of upstream posts provided the initial (finite amplitude) perturbation, and the researchers found large and sustained velocity fluctuations far downstream from the initial perturbation; no fluctuations were found without perturbations, indicating that viscoelastic flows are indeed linearly stable. In addition, the transition to this nonlinear state was found to be hysteretic upon the increase or decrease of the flow rate,  which is a typical behavior of a subcritical bifurcation. The flow became unstable at $\hbox{Wi}_c \approx 5$, in apparent agreement with the theory of Morozov and van Saarloos~\cite{MorozovVanSaarloos2005}. Subsequent work has shown that the nonlinear state possesses  features of elastic turbulence~\cite{Arratia_PRL_2017}, and a flow resistance law (pressure drop as a function of flow rate) that is nonlinear with \textrm{Wi}, followed by drag reduction~\cite{Arratia_PRL_2019} (which, intriguingly, seems to occur in other geometries as well~\cite{Kumar2020,JhaSteinberg1}). However, experiments have yet to report the existence of traveling wave
solutions predicted by Morozov and van Saarloos~\cite{MorozovVanSaarloos2005}.

An important question is whether the evidence provided by Arratia and colleagues results conform to the picture of non-normal transient growth. Such a scenario predicts that a non-modal perturbation should first grow algebraically before decaying exponentially in time. In the Lagrangian view, the scenario translates into a spatial region with large perturbations followed by a region where they decay. Experimental data, on the other hand, show that the velocity fluctuation levels remain essentially constant while moving downstream~\cite{Pan_2012_PRL}. Nevertheless, one expects non-normal growth to be a part of the mechanism that sustains elastic turbulence, but perhaps not the cause of it. This, however, is still an open question as the new results by Steinberg and colleagues suggest, as discussed next.\\

\begin{figure}
    \includegraphics[width=\columnwidth]{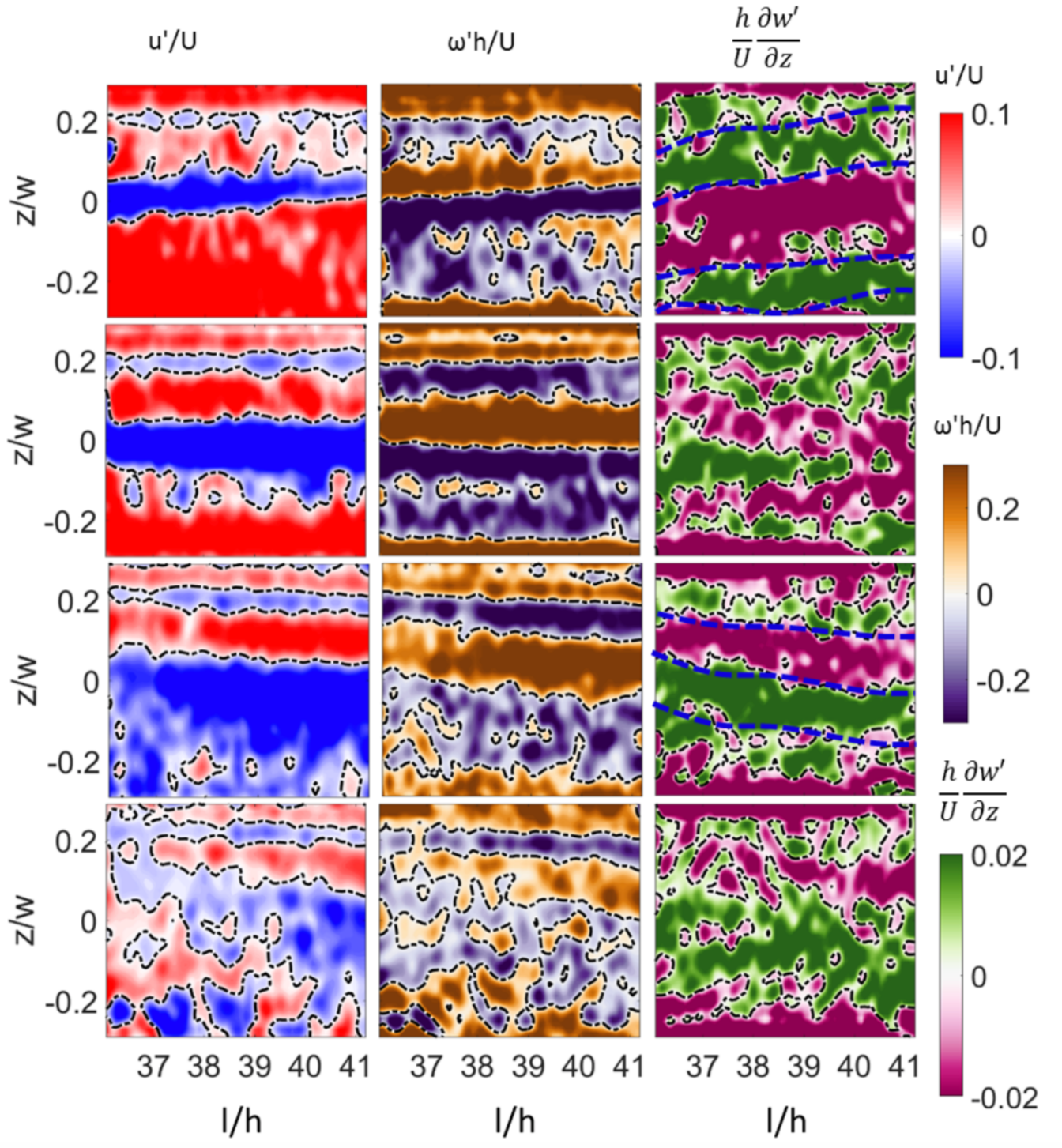}
    \caption{A cycle of coherent structures in ET at \textrm{Wi} = 185 and downstream distances $l/h = 36-41$. The normalized time $t^{*}=tf_{\rm{el}}$, where $f_{\rm{el}}$ is the elastic wave frequency. The  quantities reported here are from PIV measurements of velocity fluctuations reported in a reference frame moving with the average fluid velocity.
    Fluctuating streamwise velocity ($u'$), vertical vorticity ($\omega'$), and spanwise gradient of spanwise velocity ($\partial\omega'/\partial z$) are shown in each column with their scales shown on the right as marked in the plot~\cite{JhaSteinberg1}.}
    \label{SteinbergFig}
\end{figure}

\textbf{Experiments of Steinberg and colleagues.} N. Jha and V. Steinberg undertook experiments similar to the Arratia group but with a somewhat different geometry. The experiments of Steinberg and colleagues were conducted in a long channel with the width/height ratio = 7 and length/height ratio = 950 using channels with a height of 0.5 mm, i.e., of a large aspect ratio compared with a square channel cross-section used in the Arratia group's experiment. It is possible that this difference is one of the reasons for the difference in some of the unexpected flow states observed both at transition and beyond, such as in the observed coherent structures (CSs), such as
stream-wise rolls and streaks, and self-sustained cycling processes (SSPs). Also, elastic waves were found already at the onset to the elastically-driven flow transition and up to the highest Wi that varies from Wi = 7 up to Wi = 3500 defined as Wi = $\lambda U/h$ (instead of Wi = $U\lambda/w$, where $h$ and $w$ are the channel height and width, respectively) with a critical value $\textrm{Wi}_c = 140$. The CSs are evident in PIV measurements of velocity fluctuations reported in a reference frame moving with the average fluid velocity, as shown in Fig.~\ref{SteinbergFig}, which, remarkably, resemble those observed numerically and experimentally in Newtonian turbulence of a channel shear flow~\cite{JhaSteinberg1}. 

\begin{figure}
    \includegraphics[width=\columnwidth]{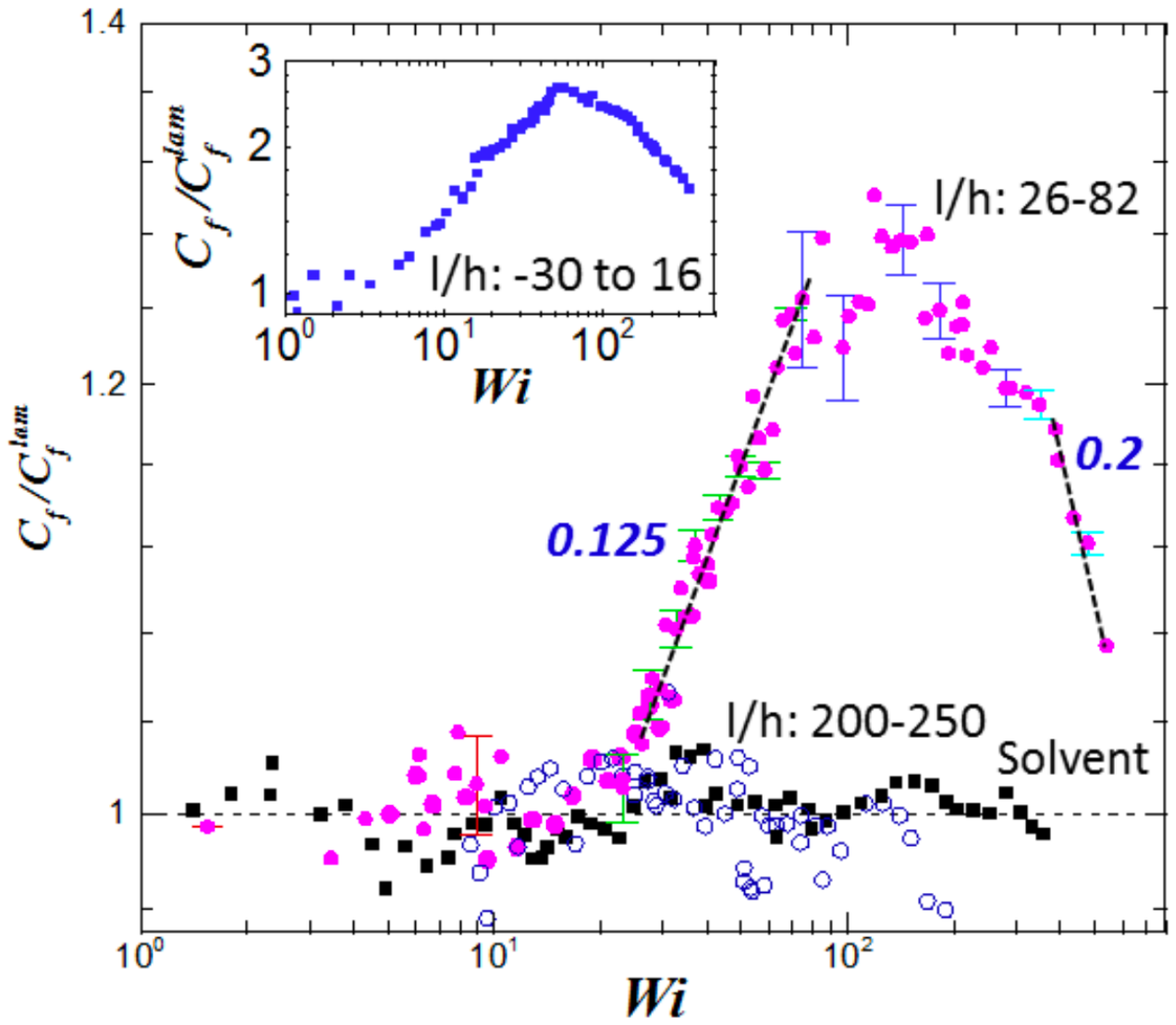}
    \caption{Dependence of the normalized friction factor $C_f /C_f^{lam}$ on \textrm{Wi} for three values of non-dimensional distance from the third layer of cylinders (at $l/h=0$) $l/h$: (i) -30 to 16 (inset), (ii) 26 to 82 (filled circle), (iii) 200 to 250 (open circle), and (iv) Newtonian solvent 26 to 82 (filled black square). Inset shows $C_f /C_f^{lam}$ versus \textrm{Wi} across three rows of cylinders ~\cite{JhaSteinberg1}.}
    \label{VictorFigure2}
\end{figure}

   We next make remarks about the continuous transition and the manner in which the friction factor for the pressure drop in pipe flow varies with the flow speed (here quantified by \textrm{Wi}). The first key observation at $\textrm{Wi}\gg 1$ and $\textrm{Re}\ll 1$ (i.e. $\textrm{El} \gg 1$) is the small magnitude of the exponent characterizing the power-law growth of the friction factor with the order parameter ($\textrm{Wi}-\textrm{Wi}_c$), which appears to have a value of 0.125 (in contrast to 0.5 for the normal mode instability) as shown in the data reported in Fig.~\ref{VictorFigure2}.
   This distinguishes the elastically driven transition in a straight channel flow from the continuous transition \textit{via} the most unstable normal mode. 
   
   Moreover, the velocity power spectra just above the instability threshold reveal the presence of a peak of elastic waves on the top of a continuous spectrum with the decay exponent -1.7~\cite{JhaSteinberg1}. 
   These new results indicate that the continuous transition cannot be described by the single, most unstable fastest growing normal mode, which would also support an instability mechanism based on the hoop stress picture~\cite{LarsonShaqfehMuller1990,Shaqfeh1996,Steinberg_AFM}.
   Thus, another possibility of a non-normal mode instability should be considered. 
    
    Steinberg and colleagues characterize their experimental observations as weakly unstable non-normal modes selected by the flow from strong perturbations at the inlet, which are further amplified due to nonlinear self-interactions to generate the coherent states (CSs0. The latter are self-organized into cycling SSP in particular in elastic turbulence, where CSs, namely stream-wise rolls and streaks, are clearly identified (Fig.~\ref{SteinbergFig}). The SSPs are synchronized by the elastic wave frequency, and consequently the SSP cycling frequency is equal to the elastic wave frequency. The synchronization is critical for the existence of CSs and SSP, which is interpreted as the pumping of energy into CSs and supporting the SSP~\cite{JhaSteinberg1}. This feature distinguishes the CSs and SSP from those found in Newtonian turbulence in shear flows.
    
   A surprising novel ingredient is the development of elastic waves above the flow transition, which are further amplified in ET and decay in the drag reduction regime. Moreover, elastic waves also pump energy into a secondary instability, whose dynamics destroy the counter-propagating streaks (compare the second and third rows in Fig.~\ref{SteinbergFig}) and bring to mind the Kelvin-Helmholtz instability [KHI] in the flow of Newtonian fluids. However, in spite of the similarity of this KH-like instability to the conventional KHI, the instability mechanism is strikingly different for the purely elastic case, where the main destabilizing factor results from interaction of transverse elastic waves with wall-normal vorticity generated by perturbations of the streaks~\cite{JhaSteinbergKHI}.
   
   Finally, it should be pointed out that CSs and cycling SSP are localized only in a finite spatial range inside the channel flow. Further downstream, only chaotic velocity power spectra with power-law decay in frequency were observed. The reason for the finite spatial range of the existence of these structures is the spatial attenuation of the elastic waves; estimates of the attenuation length of the elastic waves show an agreement with the observation~\cite{JhaSteinberg1}.\\

\begin{figure}
    \includegraphics[width=\columnwidth]{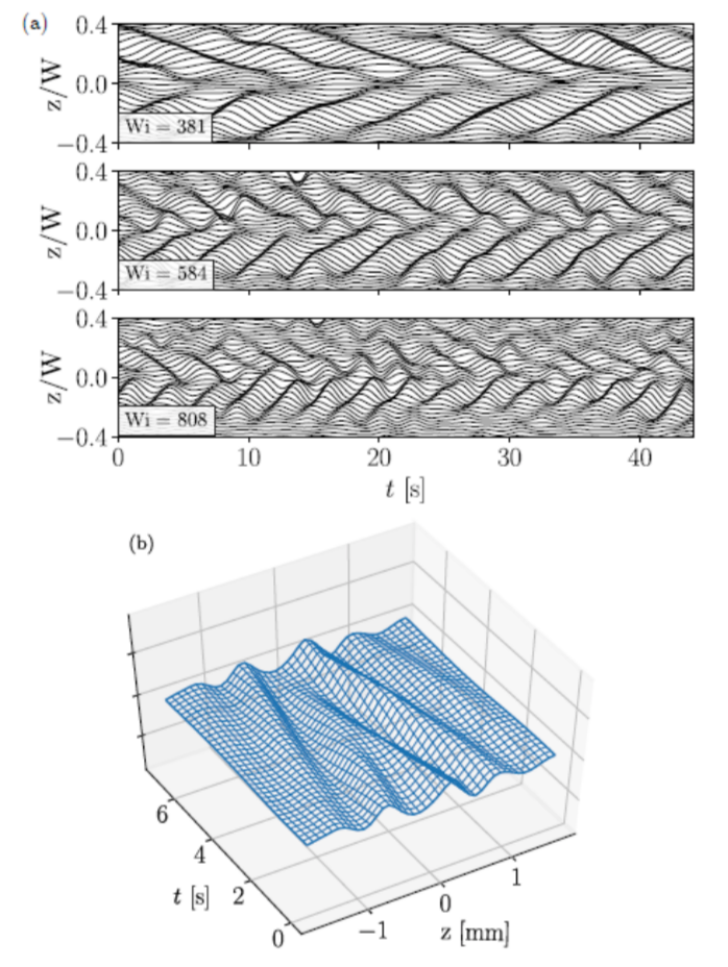}
    \caption{(a) Space-time plots at $-0.4 < z/W < 0.4$ of the stream-wise velocity fluctuations, $u^\prime(z, t)$ exhibiting elastic wave structures for three values of \textrm{Wi}. The time series are filtered via a band-pass Butterworth filter centered on the spectral peaks to remove background noise. (b) Stream-wise velocity, phase averaged at the elastic wave frequency for $\textrm{Wi = 407}$~\cite{ShnappSteinberg}.}
    \label{VictorFigure3}
\end{figure}

\textbf{Perspective.} One might think that understanding fluid flow in straight channels and pipes is easy. This is not the case at higher Reynolds numbers for Newtonian fluids, which produces inertial turbulence approximately when $\textrm{Re}>2000$. As described in this section, it is also not the case for low-Reynolds-number flows of highly elastic $(\textrm{Wi} \gg 1)$ polymeric fluids.

To summarize, the early experimental observations of Arratia and colleagues  demonstrate the existence of an elastically driven transition in a straight channel flow due to strong perturbations at the inlet. Furthermore, their results suggest the existence of a sub-critical instability, which can be viewed as the viscoelastic analogue of turbulence in classical Newtonian pipe flows, except that it is controlled by the elasticity of the fluid and not by inertia. 
On the other hand, the experimental results of Steinberg and colleagues, also for a rectilinear channel flow -- but with different cross-sectional dimensions and distinct form of imposed perturbations -- suggest that strong perturbations at the inlet are not a necessary condition to generate the elastic instability and subsequent ET in a straight channel flow. Instead, their experiments provide evidence for an elastic instability even in a straight channel with a smoothed inlet and a small hole on the top plate at the middle of the channel for pressure measurements. 
As a result, the transition at $\textrm{Wi}_c=125$ is
observed with well-characterized elastic waves at the onset, which are continued further into the ET and drag reduction regions with the same elastic wave velocity dependence on $\textrm{Wi}-\textrm{Wi}_c$ on the top of continuous velocity and pressure spectra in the transition; the latter include ET and drag reduction regions with decay of the velocity spectra having an exponent with a magnitude smaller than 3. Moreover, in this case they were able to visualize elastic waves propagating in the span-wise direction towards the center by presenting them in spatio-temporal plots for three \textrm{Wi} values (Fig.~\ref{VictorFigure3}). These features are consistent with the elastic instability occurring due to non-normal modes similar to the channel flow with the strong perturbations at the inlet, though the critical Weissenberg number $\textrm{Wi}_c$ is about twice as large (taking into account the approximately 2$\times$ smaller value of the longest relaxation time). These results once more suggest a similarity to dynamics of Newtonian parallel shear flow.



With reference to the generic flow stability diagram sketched in Fig.\ref{dragons}, it is worth noting that the Weissenberg number values are quite different between these two experimental studies; while Arratia and co-workers focused on a regime where $\hbox{Wi} = O(10)$, Steinberg and co-workers, using the same polymeric materials, focused on much higher values, $\textrm{Wi} = O(10^3)$; though some part of the discrepancy may be due to significant differences in the reported polymer relaxation times, as different characterization methods were used (again highlighting challenges with understanding even relatively simple flows of complex fluids such as dilute polymer solutions). Taking these results together, as well as the results of the theoretical analyses described in the previous sub-sections, it is possible that viscoelastic flows in pipes and channels are non-linearly unstable  at low \textrm{Wi}, as suggested by Morozov and van Saarloos, but linearly unstable at moderate to high \textrm{Wi}. Recent linear stability analysis by V. Shankar and collaborators~\cite{khalid_creepingflow_2021} seem to suggest such  a possibility --- but as summarized in this section, this question (and many others) remains unresolved.

\section{Elastic instabilities in more complex geometries}
\label{ElasticInstability}

\begin{figure*}[!t]
    \centering
    \includegraphics[width=\textwidth]{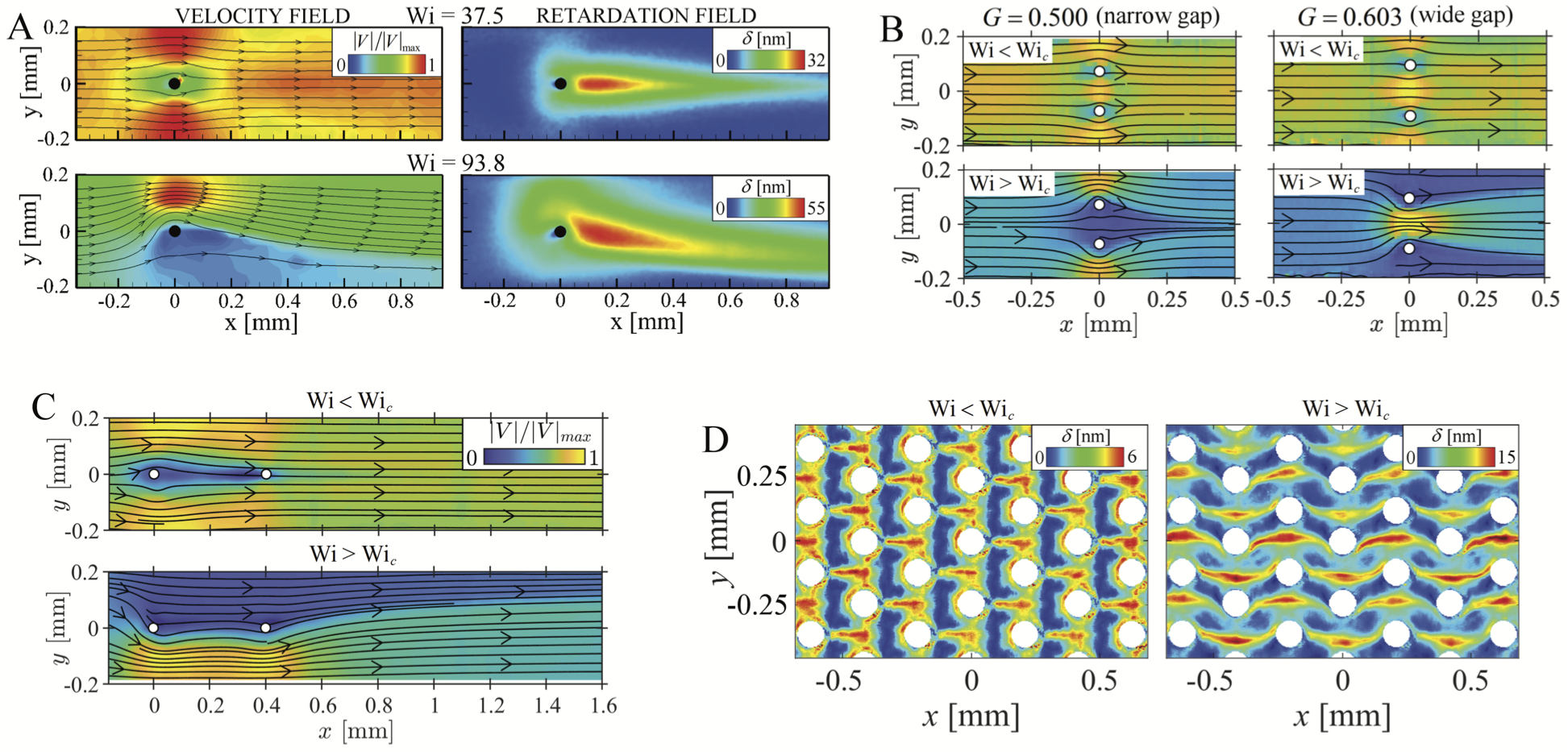}
    \caption{Transitions to steady asymmetric flow states in various geometries constructed from microscale cylinders as the Weissenberg number is increased beyond a critical value $\textrm{Wi}_c$. (A) Flow past a single cylinder positioned on the flow axis. Reproduced from \cite{Haward2019} with permission from the Royal Society of Chemistry. (B) Velocity fields for flow past side-by-side cylinders with different dimensionless intercylinder gap, $G=L_1/(L_1 + L_2)$, where $L_1$ and $L_2$ are the cylinder-cylinder, and cylinder-wall gaps, respectively. Reproduced from \cite{Hopkins2021} with permission. (C) Velocity fields for flow past two axially-aligned cylinders. Reproduced from \cite{Hopkins2020} with permission from John Wiley \& Sons, Inc. (D) Retardation fields for flow through a hexagonal array of cylinders (unpublished data, S. Haward). All cases show the flow from left to right of a shear-thinning viscoelastic WLM solution. }
    \label{Fig1_SH}
\end{figure*}
\subsection{Flow Past Cylinders}

A circular cylinder is arguably the most fundamental shape of an object that can be used for studying flows around obstacles. In viscoelastic fluid flows, circular cylinders are used frequently as building blocks to create complex geometries such as regular or random arrays that model aspects of porous media flows~\cite{Walkama20,Anbari2018,Muller1998,Eberhard2020,De2017,Kawale2017}. Recently, there has also been interest in viscoelastic fluid-structure interactions, where elastic instabilities at high Wi (but negligible inertia, or small enough Re) drive the motion of flexible or cantilevered circular cylinders~\cite{Dey2018,Hopkins2020}.

\begin{figure*}[!t]
    \centering
    \includegraphics[width=\textwidth]{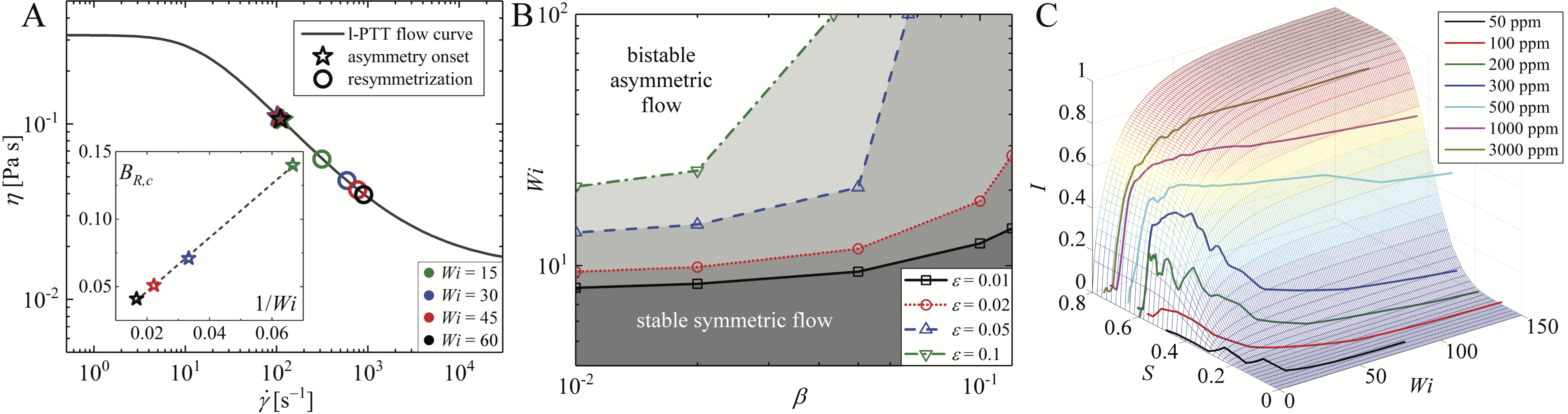}
    \caption{Influence of shear-thinning and elasticity on the onset and development of asymmetric flow states around a single cylinder. (A) Flow asymmetry only occurs when characteristic shear rates near the cylinder correspond to the shear-thinning region of the flow curve. (Inset) The onset of instability is consistent with the scaling predicted by McKinley \textit{et al.}, indicating that elasticity and curved streamlines in the downstream wake provide the initial perturbation to destabilize the flow~\cite{McKinley1996,Varchanis2020}. (B) Stability diagram constructed from simulation results with the l-PTT model examining the interplay between shear-thinning and strain-hardening ($\varepsilon$) ~\cite{Varchanis2020};  in this literature shear-thinning in the l-PTT model is denoted $\beta$ (the label of the horizontal axis). (A-B) are reprinted from \cite{Varchanis2020} with the permission of AIP Publishing. (C) Experimental measurements with polymer solutions over a range of concentration also show that the asymmetric flow around a cylinder (characterized by the magnitude of $I$ plotted on the ordinate axis) requires both shear-thinning and elastic effects in the fluid. Reproduced from \cite{Haward2020} with permission.  }
    \label{Fig2_SH}
\end{figure*}

Using a model viscoelastic wormlike micellar (WLM) solution consisting of 100 mM cetylpyridinium chloride (CPyCl) and 60 mM sodium salicylate (NaSal)~\cite{Rehage1988,Rehage1991}, Haward, Shen, and co-workers examined flows past several different configurations of slender circular cylinders confined within microfluidic channels (see Fig.~\ref{Fig1_SH}), with much larger depth aspect ratios than explored in previous studies. At $24^{\circ}$C (ambient laboratory temperature), the entangled WLM solution has a zero shear viscosity $\eta_{0} \approx 47$~{Pa.s}, exhibits a stress-plateau (shear-banding region~\cite{Fielding2016}), and in small-amplitude oscillation is well-described by a single-mode Maxwell model with relaxation time $\lambda \approx 1.7$ s. The dimensions of the microfabricated glass geometries (channel height, $H = O\text{(1 mm)} \gg$ width, $W \gg$ cylinder radius, $R = O(10~\mu$m)) ensure that inertia is always negligible, and that the flows are approximately uniform (or two-dimensional, 2D) along the length of the cylinder. The Weissenberg number of the flow is defined by $\textrm{Wi} = \lambda U/R$, where the average flow velocity in the channel  $U$ is controlled by a syringe pump.  

For flow around a single rigid cylinder located in the center of the microchannel (Fig.~\ref{Fig1_SH}A), a flow bifurcation occurs as the Weissenberg number exceeds a critical value $\textrm{Wi}_c \approx 60$~\cite{Haward2019}. For $\textrm{Wi}=37.5<\textrm{Wi}_c$, the fluid passes the cylinder symmetrically, with the same flow velocity profile on either side of the cylinder, and a straight elastic wake is observed along the flow axis downstream of the cylinder (as seen in the retardation field). However, for $\textrm{Wi}=93.8>\textrm{Wi}_c$, the fluid selects a preferred path around the cylinder, with a higher average velocity on one side than the other, and the elastic wake becomes correspondingly distorted downstream. This symmetry-breaking transition has been characterized as a supercritical pitchfork bifurcation~\cite{Haward2019}. 

The bifurcation at one cylinder influences (and is influenced by) the bifurcation occurring at neighboring cylinders positioned adjacently (Fig.~\ref{Fig1_SH}B)~\cite{Hopkins2021} or downstream (Fig.~\ref{Fig1_SH}C) in the channel~\cite{Hopkins2020}. In a hexagonal array of circular cylinders, the bifurcation at each obstacle results in a regular pattern of asymmetric wakes where the handedness of the asymmetry alternates between rows (Fig.~\ref{Fig1_SH}D). 

We note that in all the cases illustrated in Fig.~\ref{Fig1_SH}, the flow becomes time-dependent and apparently chaotic as $\textrm{Wi}$ becomes sufficiently large. However, instability progresses from an initial transition to a steady asymmetric flow around each cylinder. These flows all appear to be governed primarily by the bifurcation that occurs at each obstacle for $\textrm{Wi}>\textrm{Wi}_c$. Therefore, to correctly interpret phenomena observed in more complex flows, e.g., path selection through arrays of cylinders representing porous media, we consider it crucial to first understand how instability develops around a single cylinder.

Accordingly, Haward, Shen and colleagues have invested significant efforts in this direction, employing rheologically diverse fluids and a combination of experiments and numerical simulations~\cite{Haward2018,Haward2019,Haward2020,Varchanis2020}. The comprehensive studies indicate that the instability is initiated by random fluctuations in the downstream wake due to a combination of high elasticity and streamline curvature close to the downstream stagnation point, i.e., a purely-elastic instability of the type described by Pakdel and McKinley~\cite{Pakdel1996,McKinley1996,Varchanis2020} (Eq. \ref{mparameter}). As shown in the inset to Fig.~\ref{Fig2_SH}A, from numerical simulations with the simplified linear Phan-Thien and Tanner (l-PTT) model (Table \ref{table1} with $\xi=0$),  the onset Weissenberg number for asymmetric flow scales with the blockage ratio, $B_R = 2R/W$, in excellent agreement with the prediction of McKinley et al.~\cite{McKinley1996,Varchanis2020}. However, from the same set of simulations, performed by varying $B_R$ at fixed $\textrm{Wi}$, asymmetric flows are only supported when the characteristic shear-rate near the cylinder lies on the shear-thinning region of the flow curve (Fig.~\ref{Fig2_SH}A). As the shear rate approaches the high-shear-rate plateau region, symmetry is recovered. 

By fixing the blockage ratio $B_R = 0.1$ and varying the degrees of strain-hardening, $\varepsilon$, and shear-thinning, denoted $\beta$, in the l-PTT model, a stability diagram is obtained in $\textrm{Wi}$--$\beta$ parameter space, where the boundaries marking the onset of asymmetric flows can be followed along lines of constant $\varepsilon$ (Fig.~\ref{Fig2_SH}B). The instability is clearly affected by an interplay between the shear-thinning and the elasticity of the fluid: if strain-hardening is reduced, more shear-thinning is required to induce the asymmetric flow state (and \textit{vice-versa})~\cite{Varchanis2020}. 

These observations are paralleled in experimental measurements using polymer solutions with a range of rheological characteristics (i.e., by varying the shear-thinning and elasticity, see Fig.~\ref{Fig2_SH}C). Here, to understand the role of shear-thinning, Haward, Shen and colleagues employ the ``shear-thinning parameter" defined in \S IIC, $\mathcal{S}=1-(d \ln \tau / d \ln \dot\gamma)$, which is evaluated from the flow curve measured in steady shear~\cite{Sharma2012,Haward2012a,Haward2020}. The quantity ``$I$" reported in Fig.~\ref{Fig2_SH}C  is a measure of the degree of asymmetry in the flow obtained from the difference in flow velocity on either side of the cylinder~\cite{Haward2019,Haward2020}. Elasticity in the wake of the cylinder is considered to depend on the magnitude of $\textrm{Wi}$. Note that both $\mathcal{S}$ and $\textrm{Wi}$ depend on the imposed flow velocity through the microchannel. 

The colored lines in Fig.~\ref{Fig2_SH}C show the trajectories of fluids with different polymer concentrations through the three-dimensional space, while the fitted surface is formed from a combination of sigmoidal curves in $S$ and $\textrm{Wi}$~\cite{Haward2020}. From Fig.~\ref{Fig2_SH}C, it can be observed that fluids with low polymer concentrations (e.g., 50 or 100~ppm) never show significant flow asymmetry ($I \cong 0$); shear-thinning is high only when elasticity is low. Fluids with higher polymer concentrations (e.g., 200 or 300~ppm) show the onset of asymmetry as $\textrm{Wi}$ is initially increased, but the flow recovers symmetry at very high $\textrm{Wi}$ due to the loss of shear-thinning in the high shear-rate plateau. Fluids of very high polymer concentration (e.g., 1000 or 3000~ppm) develop strong flow asymmetries ($I \rightarrow 1$), which can persist up to high $\textrm{Wi}$ since the degree of shear thinning $\mathcal{S}$ remains significant.

The development of the steady flow asymmetry in the cylinder geometry depends on both the degree of shear-thinning and the elasticity of the fluid in question, and the two rheological properties share an interplay whereby strong shear-thinning can compensate for weak elasticity (and \textit{vice-versa}). In light of these results, it may be worthwhile revisiting the role of shear-thinning in other instances of steady viscoelastic flow asymmetries (for instance in the cross-slot geometry~\cite{Arratia2006,Poole2007,Rocha2009,Haward2012a}), where the initial onset of instability gives rise to regions in the flow field with disparate shear rates.

\begin{figure}[!ht]
 \centering
 \includegraphics[width=\columnwidth]{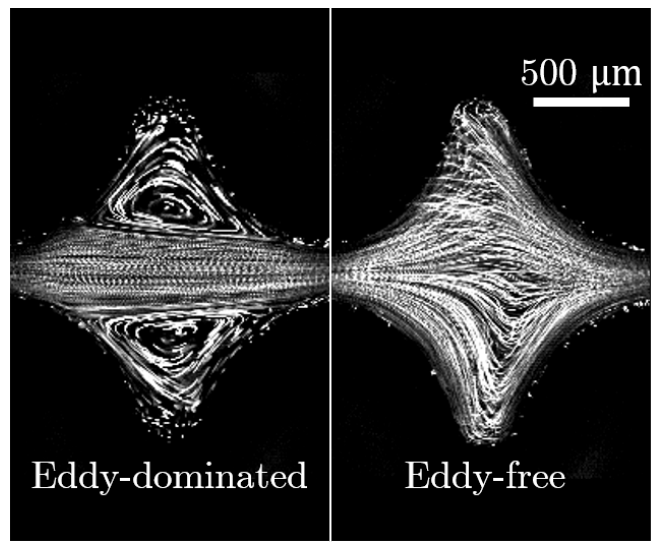}
 \caption{\label{brownebistability} 
 Experimental images of two distinct unstable flow states observed for elastic polymer solution flow through ordered one-dimensional arrays of pore constrictions. Images show fluid pathlines and are adapted from \cite{Browne2020} with permission.}
 \label{bistabilitybrowne}
 \end{figure}

\begin{figure*}[!ht]
 \centering
 \includegraphics[width=\textwidth]{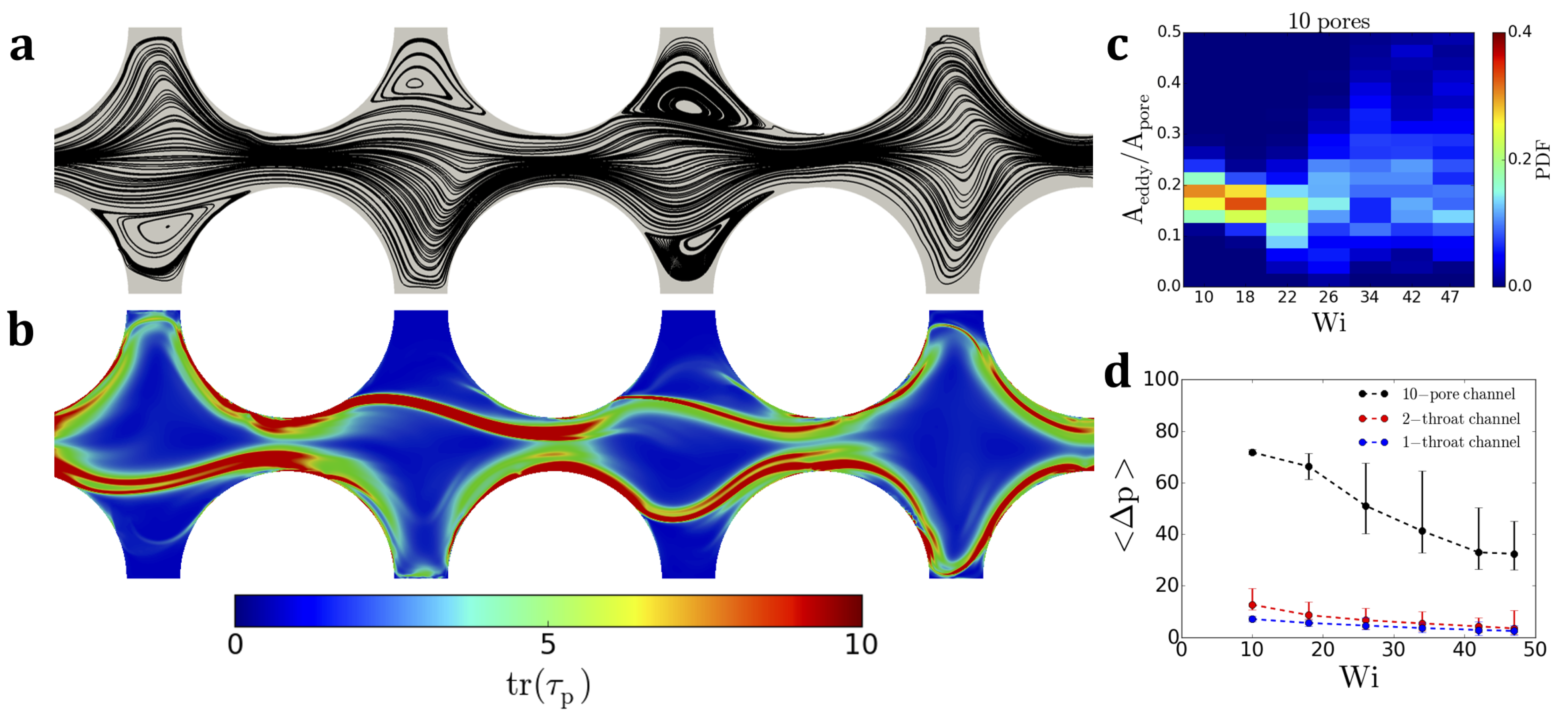}
 \caption{\label{multistability_v2} 
 (a) Multistability of the unstable flow of polymeric fluid through the pores of a converging-diverging channel. (b) Trace of polymeric stress tensor inside the pores. (c) Probability density function (PDF) of the ratio of eddies to pore area ($\rm{A_{eddy}/A_{pore}}$) at different Wi for a channel of 10 closely located pores. $\rm{A_{eddy}}$ represents the total area occupied by eddies in an individual pore and $\rm{A_{pore}}$ is the total area of the pore. Above a threshold Wi, multistability occurs, and the eddy areas take on a broad range of values. (d) Time- averaged pressure drop ($\rm{\langle\Delta p\rangle}$)  across the channels at different Wi. Images are reproduced from~\cite{Kumar2020}.}
 \end{figure*}

\begin{figure*}[!ht]
 \centering
 \includegraphics[width=0.9\textwidth]{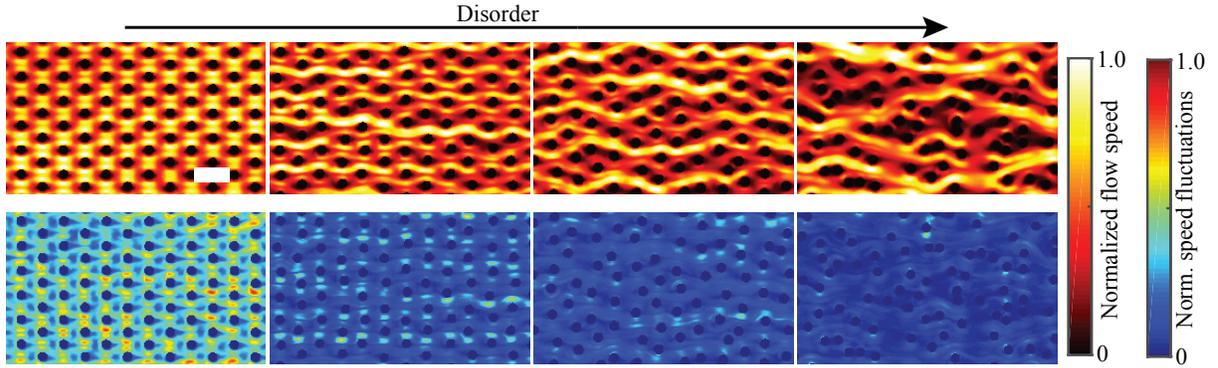}
 \caption{\label{Fig_Guasto} 
Experiments in which disorder reduces chaotic fluctuations in viscoelastic flows through porous media. 
(Top row) Normalized, time-averaged speed field in a microfluidic pillar array for a range of geometric disorders ($\textrm{Wi} \approx 4$). Scale bar, 150~$\mu$m. 
(Bottom row) Local, normalized speed fluctuations as a function of increasing disorder, corresponding to speed fields above. Images are reproduced from \cite{Walkama20}.}
 \end{figure*}

\subsection{Flow in Porous Media}
Studies of the flow of a polymer solution through an isolated constriction or across a single cylinder give some insight to the pore-scale flow dynamics in porous media~\cite{Lanzaro2011,Rodd2005, Haward2020,Varchanis2020}.  However, the higher connectivity and elevated disorder  inherent in natural porous media introduce new complexities to such flows~\cite{Aramideh2018,Datta2013}. Being able to predict and control viscoelastic fluid flow through porous media has several important industrial applications, as reviewed previously in \cite{Browne2019}. Notable examples are enhanced oil recovery (EOR) \cite{sorbie2013} and groundwater remediation \cite{roote1998,smith2008}, in which addition of polymers to a displacing fluid leads to enhanced recovery of a trapped non-wetting fluid \cite{Naik2019computational,Naik2019petroleum, Aramideh2018fuel}. Several mechanisms for this phenomenon have been proposed: adding polymers is thought to (i) increase the viscous drag on trapped immiscible fluid droplets \cite{Datta2014FluidBreakup,Datta2014Mobilization}; (ii) suppress viscous fingering instabilities during fluid displacement \cite{ Aramideh2019a}; (iii) impart strong spatial and temporal velocity fluctuations induced by elastic instabilities \cite{Howe2015,Mitchell2016,Clarke2016a,Aramideh2019}; (iv) reduce the permeability of the medium locally due to polymer retention at solid surfaces, leading to large and heterogeneous local changes in flow \cite{Parsa2020}. However, systematic studies in porous media of varying geometries are needed to parse the influence of these different possible instability mechanisms. 


The accumulation of stresses as polymers traverse successive pores can produce spatial variation in the dominant flow features~\cite{kenney2013,shi2015,shi2016,kawale2017a,kawale2017b,Varchanis2020,Walkama20,Kumar2020}. For example, when flowing around closely-separated obstacles, polymer chains can be advected from an upstream to a downstream obstacle faster than they can relax. This interaction leads to a bifurcation of the unstable polymeric fluid flow into two coexisting flow states in between the two obstacles~\cite{Varshney2017}. More recent work has shown that in tightly-ordered (lower porosity) one-dimensional (1D) arrays of multiple pores, with resemblance to natural porous media, this interaction can produce an unexpected bistability in the unstable flow in which the flow in each pore switches stochastically between two distinct primary structures: an eddy-dominated structure, and an eddy-free structure~\cite{Browne2020} (Fig.~\ref{bistabilitybrowne}). 

Numerical simulations have corroborated these experimental results, showing that even more patterns (i.e., multi-stability) can arise above a critical Weissenberg number: (i) eddy on both the top and bottom of the pore, (ii) eddy-free pore, (iii) eddy-free top of the pore, and (iv) eddy-free bottom of the pore~\cite{Kumar2020} (Fig.~\ref{multistability_v2}a-c). This multi-stability reflects the formation of different regions of high polymeric stress in the pores (Fig.~\ref{multistability_v2}b): the accumulation of stresses as the polymeric chains cross successive pores creates streaks of high polymeric stress that are closely coupled to the flow structures inside the pores. Polymeric chains are highly stretched in the regions of high polymeric stresses, preventing the flow crossing these streaks and inducing eddy formation in different parts of the pore. Intriguingly, the simulations suggest that this multistability can actually reduce the pressure drop across the channel~\cite{Kumar2020} (Fig.~\ref{multistability_v2}d) because the eddies do not contribute to the net volumetric flow through the channel; therefore, an eddy-free pore has a larger apparent width to allow the net volumetric flow than the pore with eddies, which leads to a smaller pressure drop across the eddy-free pore. Further experimental tests of this behavior will be an interesting direction for future work.

Beyond porosity, recent experiments have shown that disorder may also play a fundamental role in the stability of viscoelastic flows through porous media~\cite{Walkama20}.
Similar to single obstacles, viscoelastic flow through an ordered 2D array of cylinders readily transitions to chaos at a critical $\textrm{Wi}_\textrm{cr} = O(1)$. However, the introduction of small deviations from crystalline order in the porous medium can delay the transition to higher $\textrm{Wi}_\textrm{cr}$, and strongly disordered media can have largely suppressed random velocity fluctuations (Fig.~\ref{Fig_Guasto}). 
The mechanism by which disorder may promote stability is by causing a shift in the flow type~\cite{Astarita1979} from extension- to shear-dominated flow.
In the work of Walkama et al.~\cite{Walkama20}, as geometrical disorder increases, stable preferential flow paths emerge and promote shear, which weakly stretches polymers in comparison to extensional flow~\cite{Smith1999,FullerLeal}.
This work also emphasizes the importance of Lagrangian stretch that is accrued along a polymer's flow path in triggering viscoelastic instability.

Exploring how insights developed in 1D and 2D systems relate to flows in more complex 3D porous media \cite{sorbie2013,roote1998,smith2008, Aramideh2018} is an active frontier of current research. In the stable creeping flow regime considered here, spatial correlations of velocity and pore-space have been shown to be almost identical between 2D and 3D media \cite{Aramideh2018}. However, pore scale flow instabilities in 3D geometries can exhibit different patterns than 2D instabilities~\cite{Qin2019,Burshtein2019, McKinley1993}, inducing differences in macroscopic flow and transport as well \cite{Maier2000, De2018, Marafini2020}.

\begin{figure}[!ht]
    \centering
    \includegraphics[width=\columnwidth]{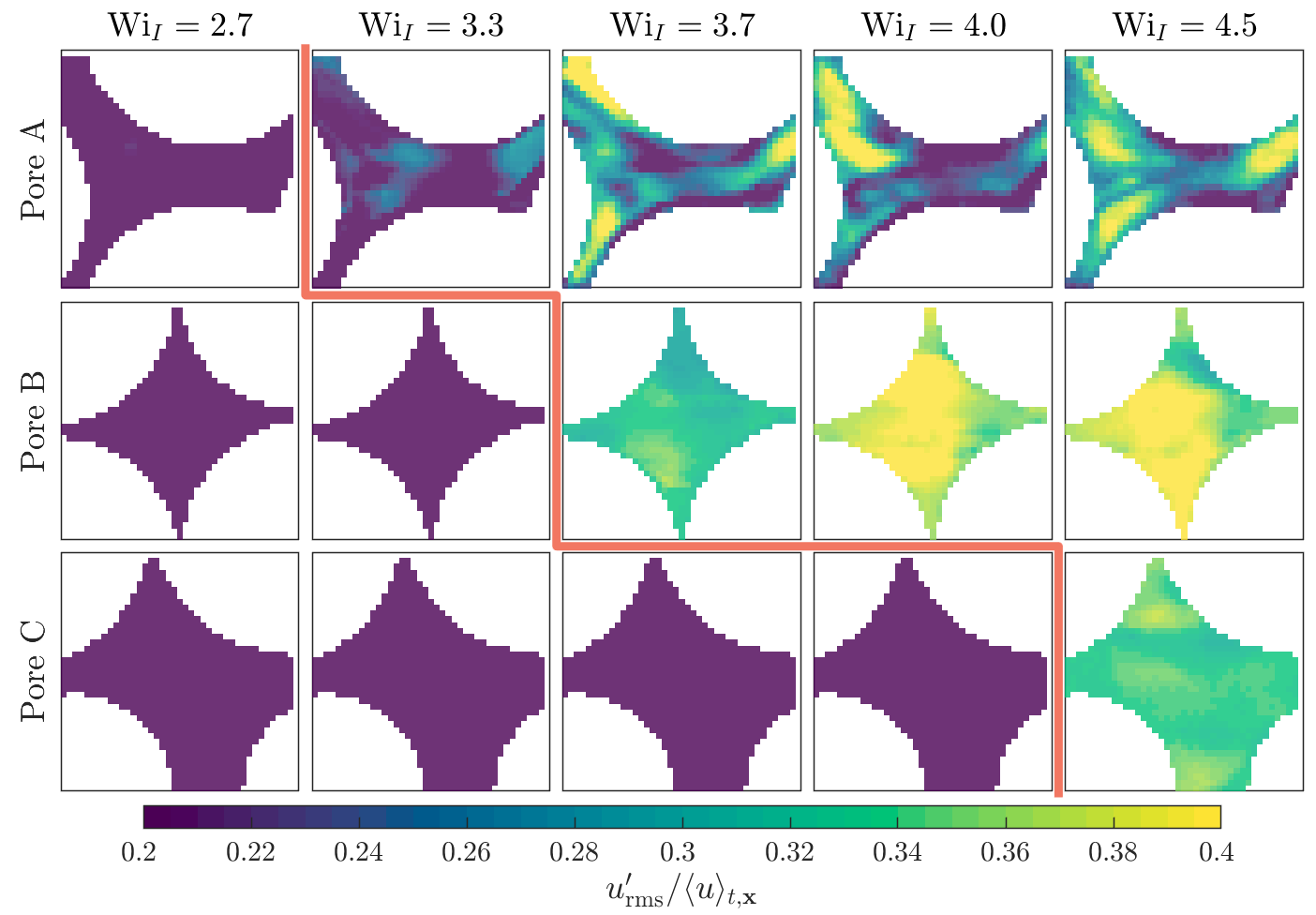}
    \caption{The occurrence of elastic turbulence is spatially heterogeneous throughout a porous medium, reflecting ``porous individualism." Images show the normalized magnitude of root mean square flow fluctuations in different pores and at different flow rates, parameterized by a characteristic Weissenberg number $\mathrm{Wi}_I$. Applied flow is from left to right. Pore A becomes unstable at the lowest flow rate, as shown by the red line in the first row. Pore B becomes unstable at the next highest flow rate, shown by the red line in the second row. Pore C becomes unstable only at even higher flow rates. Note that flow velocity magnitude is denoted by $u$ instead of $v$ as in the rest of the text. Reproduced from \cite{Browne2021}.}
    \label{fig:Patches}
\end{figure}

Indeed, given the observation that disorder can suppress the transition to elastic turbulence in 2D porous media \cite{Walkama20}, it has been unclear whether and how this transition manifests in disordered 3D media --- though elastic turbulence has been speculated to underlie the long-standing observation that the macroscopic flow resistance of an injected polymer solution can abruptly increase above a threshold flow rate in a porous medium, but not in bulk solution~\cite{marshall1967flow,james1975laminar,durst1981,clarke2016,dursthaas1981,kauser,Aramideh2019}. By directly visualizing the flow in a transparent, disordered, 3D porous medium,~\cite{Browne2021} directly verified that elastic turbulence does arise within a disordered 3D porous medium, and used flow velocimetry to link chaotic pore-scale flow fluctuations to the macroscopic flow resistance. In particular, the authors found that the transition to unstable flow in each pore is continuous, arising due to the increased persistence of discrete bursts of instability above a critical value of the characteristic Wi; however, the onset value varies from pore to pore. 

This observation that single pores exposed to the same macroscopic flow rate become unstable in different ways provides a fascinating pore-scale analog of ``molecular individualism" \cite{deGennes1997molecular}, in which single polymers exposed to the same extensional flow elongate in different ways; the authors therefore termed it ``porous individualism", although it is important to note that here, this effect is still at the continuum (not molecular) scale. Thus, unstable flow is spatially heterogeneous across the different pores of the medium, with unstable and laminar regions coexisting (Fig. \ref{fig:Patches}). Guided by these findings, and inspired by the analysis of recent simulations \cite{de2017viscoelastic}, the authors quantitatively established that the energy dissipated by unstable pore-scale fluctuations generates an anomalous increase in flow resistance through the entire medium that agrees well with macroscopic pressure drop measurements.

Thus, by linking the onset of unstable flow at the pore scale to transport at the macroscale, such research is beginning to yield generally-applicable guidelines for predicting and controlling unstable flows of polymer solutions in porous media. Indeed, experimental developments using confocal microscopy in model 3D porous media~\cite{Anbari2018,Datta2013,DoNascimento2019}, defocusing particle tracking velocimetry~\cite{Guo2019}, holographic particle tracking velocimetry~\cite{Qin2020}, and fast synchrotron-based X-ray computed microtomography in real porous rocks~\cite{Berg2013,Pak2015} provide access to flows \textit{in situ} that will likely continue to refine our understanding of these complex systems.


\section{elastoinertial Flow Instabilities}
\label{EITsection}
\newcommand{\Wi}{\textrm{Wi}}


Ever since the iconic experiments of Osborne Reynolds in 1883~\citep{Reynolds1883}, it has been well-known that Newtonian pipe flow undergoes a laminar-turbulent transition when the eponymous dimensionless parameter (the Reynolds number, $\textrm{Re}$) exceeds a threshold. The complexity of this transition was already understood by Reynolds, as evidenced by the following remark in Ref.~\cite{Reynolds1883}:  ``...it was
observed that the critical velocity was very sensitive to disturbance in the water
before entering the tubes." Later experiments have indeed shown that the laminar state in Newtonian pipe flow  can be maintained up to $\textrm{Re} \approx 10^5$~\citep{pfenniger1961boundary}; this behavior is consistent with the current consensus that Newtonian pipe flow is linearly stable at all Reynolds numbers~\citep{meseguer_trefethen_2003}.  The Newtonian pipe flow transition from laminar to turbulent flow is therefore very different from that observed in the Taylor-Couette geometry (with the inner cylinder rotating) discussed earlier in \S~\ref{ViscoelasticIntro}. In the latter case, the  transition is marked by a sequence of reasonably well-defined bifurcations starting from the initial linear instability, and leading to a gradual increase in the spatio-temporal complexity~\cite{andereck_liu_swinney_1986}.  In stark contrast, in Newtonian pipe flow (and indeed in the other canonical shear flows such as plane Couette and Poiseuille flows),  transition is abrupt, and is marked by the appearance of localized structures known as turbulent puffs and slugs (or turbulent spots in the aforementioned plane shear flows) that already exhibit the full spatio-temporal complexity of the ensuing turbulent state~\citep{Eckhardt2007,Barkley2020}. Indeed, the original  paper by G. I. Taylor~\citep{Taylor1923} on the centrifugal instability in the geometry that now (partly) bears his name involved a successful comparison  between theory and experimental observations of the transition from the base-state azimuthal flow. However, as discussed below, more than a century was required after the original  paper of Reynolds for the emergence of a rigorous theoretical understanding of the Newtonian pipe flow transition.
 
Transition to turbulence in canonical rectilinear shearing flows, e.g., plane Couette, and plane- and cylindrical-Poiseuille flows, of a Newtonian fluid, beyond a threshold $\textrm{Re}$ is a complicated process~\citep{eckhardt_etal_2007,avila2011}, largely due to the absence (plane Couette and pipe flow) or irrelevance (plane Poiseuille) of an underlying linear instability~\citep{Drazinreid}.
Note that \textrm{Re} here is the Reynolds number based on the half-height/pipe radius, maximum velocity of the laminar flow, and the total solution viscosity and density. The Weissenberg number, \textrm{Wi},  is  based on the polymer relaxation time, maximum velocity of the laminar flow and the half-height/pipe radius.
 Nevertheless, the Newtonian transition is now regarded as well understood from a dynamical systems perspective, with the eventual transition being presaged by the appearance of non-trivial three-dimensional solutions (the so-called exact coherent states or `ECS' in short) of the Navier-Stokes equations~\cite{Fabian_PRL_1998,waleffe_2001,wedin_kerswell_2004,kerswell_2005}, which are disconnected from the trivial laminar state, and serve as a scaffold, in an appropriate phase space, for the turbulent dynamics after transition~\citep{barkley2016,budanur_etal_2017,GrahamFloryan2021} (also see \S~\ref{subsubsec:TheorylowRe}).  These ECSs contain the basic self-sustaining ingredients of transitional Newtonian turbulence, i.e., quasi-streamwise vortices and streaks. A comprehensive review of ECS can be found in~\cite{Barkley2020,GrahamFloryan2021}.
 
While the transition to inertial turbulence in Newtonian pipe and plane Poiseuille flow is now relatively well understood as described above, recent experimental, theoretical and computational studies have shown that the transition scenario in viscoelastic counterparts of the above flows may be markedly different.
Both linear and nonlinear mechanisms, with no analogues in the Newtonian realm, have been proposed for viscoelastic rectilinear shearing flows.  Thus,  
while the focus in \S~\ref{FlowTransitionSection} was on instabilities of rectilinear shearing flows pertaining to the low-$\textrm{Re}$ regime, this section emphasizes the crucial role played by both fluid inertia \textit{and} elasticity in destabilizing the laminar base state, and the focus is on what may be appropriately referred to as `elastoinertial' instabilities.
 In \S~\ref{subsec:dragreduction} below, we begin with the well known drag-reducing effect of polymers on fully developed Newtonian turbulence, before moving on to the mechanistic underpinnings of turbulent drag reduction in \S~\ref{subsec:linkingtoDR}. We then summarize in \S~\ref{subsec:transitionscenarios} various transition scenarios for viscoelastic pipe  and plane Poiseuille flows for different fixed values of the ratio between solvent and total viscosity, denoted $\beta$.

\subsection{Turbulent Drag Reduction and Elastoinertial Turbulence (EIT)} 
\label{subsec:dragreduction}

The addition of long chain polymer molecules to a fluid has tremendous effects on wall-bounded turbulence, the most dramatic being the substantial reduction of the friction factor~\cite{toms49,tomsremin,Virk}, which is proportional to the pressure drop for a given flow rate (or Reynolds number). This phenomenon has found wide use in various applications that seek energy efficiency in flow processes~\citep{Fink:2012,Burger:1982,King:2012vq}. Not surprisingly, there is also a large literature seeking to understand and/or exploit this phenomenon.

In this section, we now broaden our perspective and focus on situations in which fluid inertia is non-negligible. We focus on high-Reynolds-number channel flow of a dilute solution of high molecular weight polymer, so the ratio between solvent and total viscosity, $\beta$ satisfies $1-\beta\ll 1$, and the Trouton ratio (i.e.,~the ratio between extensional and shear viscosities) $\mathrm{Tr}\gg1$.  For the FENE-P constitutive model with chain length parameter $b (\equiv L^2)$, this requires that $b(1-\beta)\gg 1$.  This is the regime of primary relevance for drag reduction, where as a practical matter it is desired to keep the shear viscosity of the fluid low (i.e., $1-\beta\ll 1$), but the extensional viscosity high (i.e., $b(1-\beta)\gg 1$). The Reynolds number regime considered is $\R\sim 10^3-10^4$, i.e., near transition.



Important features of turbulent flow when the degree of polymer-induced drag reduction is large include a very small Reynolds shear stress and a mean velocity profile that closely approaches the so-called Virk maximum drag reduction (MDR) asymptote \cite{Virk}. It is interesting that this profile is nearly independent of the composition or concentration of the polymer.  

With respect to mechanism, it is well-known that viscoelasticity suppresses the near-wall streamwise vortices that dominate Newtonian turbulence~\citep{Kim:2007dq, White:2008hs}. A number of studies have captured this phenomena by studying the effect of viscoelasticity on the aforementioned ECS solutions~\citep{Stone:2002dj, Stone:2003gq,Stone:2004jk,Roy2006,Li:2005vl,Li:2006gk, Li:2007ii}.
In particular, Li and coworkers \cite{Li:2006gk,Li:2007ii} found that the ECS are so weakened by viscoelasticity that they are no longer self-sustaining and so should fail to exist.  However, recognizing that, in general, viscoelasticity is not experimentally observed to drive relaminarization of the flow, these authors suggested the possibility of new viscoelastic mechanisms for sustaining turbulence and becoming unmasked as the Newtonian structures are suppressed~\citep{Li:2006gk}.

\begin{figure}
\centerline{\includegraphics[width=1.\textwidth]{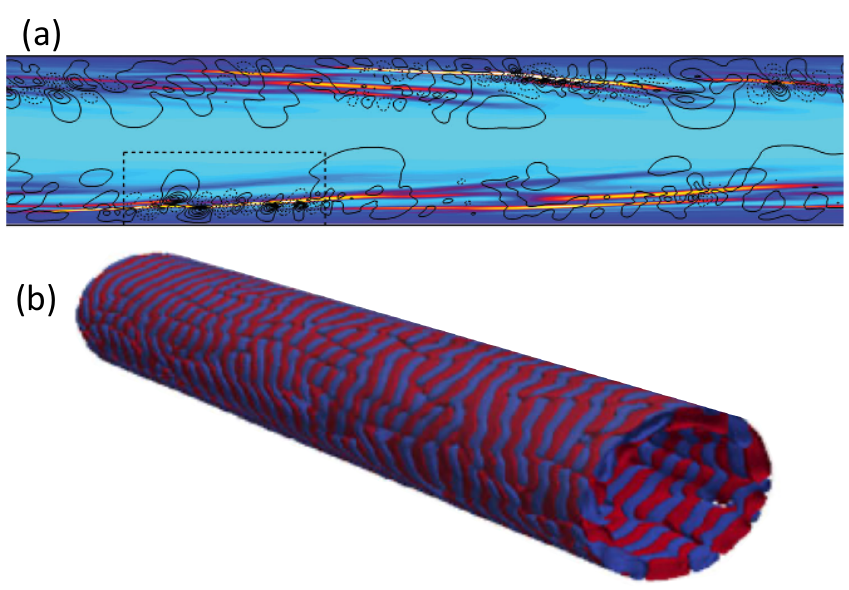}}
	\caption{Snapshots of simulations of EIT in (a) channel flow~\cite{Terrapon:2014kn} and (b) pipe flow~\cite{Lopez:2019ct}. In (a) color contours indicate polymer stretching and lines indicate the magnitude $Q$ of the second invariant of the velocity gradient
tensor; reproduced with permission from \cite{Terrapon:2014kn}. In (b) isosurfaces indicate $Q$; ; licensed under a Creative Commons Attribution (CC BY) license. }\label{fig:EITpix}
\end{figure}

Indeed, instead of complete relaminarization of the flow (except in narrow parameter ranges at transitional $\R$ as detailed later), recent studies have unearthed a polymer-driven chaotic flow state dubbed elastoinertial turbulence (EIT), which dominates high-Reynolds-number flows at high levels of viscoelasticity~\citep{Samanta:2013el}. In this parameter regime EIT displays multilayered sheets of polymer stretch emanating from near the walls (see Fig.~\ref{fig:EITpix}a) and very weak, spanwise-oriented vortices, which is in sharp contrast to the 3D quasi-streamwise vortex structures of  Newtonian wall turbulence. 
Similarly, near-wall localized, nearly-axisymmetric vortex and stress structures (Fig.~\ref{fig:EITpix}b) have been reported in pipe flow simulations of EIT~\cite{Lopez:2019ct}.  

Using computations in channel flow at $\R=1500$, Shekar et al.~\cite{Shekar:2019hq} observed a narrow zone of $\textrm{Wi}$, roughly $10-18$, where the only attractor was the laminar base state. This zone separated drag-reduced Newtonian turbulence at lower $\textrm{Wi}$ and EIT at higher $\textrm{Wi}$, corroborating the experimental observations of~\cite{Choueiri:2018it}. In this case, the laminar flow remains linearly stable in the EIT regime, but only very small (but finite) perturbations are required to drive the flow to EIT. This observation suggests that extreme care must be taken in interpreting experimental observations of a transition to a very weak EIT state: what  appears to be a linear instability may not be. 


When discussing the theory (below) to understand flow structures in viscoelastic shearing flows, it is first useful to recall the known structure, and the associated features, of the Newtonian eigenspectrum for  plane- and pipe-Poiseuille flows. The spectrum has a characteristic `Y'-shaped structure, with the two arms of the Y  comprising the wall modes (the so-called A branch with modal phase speeds and decay rates approaching zero) and the center modes (the so-called P branch with phase speeds approaching the centerline maximum and decay rates approaching zero)  for $\textrm{Re} \sim 1000$ and higher. 
The (lower) stem of the Y-structure corresponds to the S-branch that consists of a denumerable infinity of modes that propagate at two-thirds of the base-flow maximum, and with progressively increasing decay rates down the stem. Note that, with increasing \textrm{Re}, the underlying Y-template remains unchanged,
while there is a progressive increase in the number of modes along each of the three branches. 
The Tollmien-Schlichting  (TS) mode in Newtonian plane-Poiseuille flow corresponds to a wall mode belonging to the A branch that becomes unstable at $\textrm{Re} \approx 5772$; Newtonian pipe-Poiseuille flow, in contrast, is known to be stable for all \textrm{Re}. 

For large \textrm{Re}, the stream-wise velocity eigenfunction for the TS mode displays a sharp localization at wall-normal locations called `critical layers', near the top and bottom walls, where the base-flow velocity equals the phase speed. A balance of  inertial and viscous effects shows that the thickness of the critical layer decreases as Re$^{-1/3}$, consistent with the aforementioned localization.
%
 Critical layers can be thought of as the most favorable positions for energy exchange between the mean flow and the fluctuations, because they are the positions where both the fluctuations and the base flow have the same speed. As we discuss below, recent computational studies on viscoelastic channel flow at $\textrm{Re} = 1500 $ indicate a role for TS-like critical layer mechanisms in EIT. 

EIT in this parameter regime  displays polymer stretch fluctuations localized near the wall. In particular, a  resemblance was noted between the EIT structure and the viscoelastic extension of the classical TS mode, which at the chosen parameters is the slowest decaying mode from linear stability analysis. 
This viscoelastic TS mode displays polymer stretch fluctuations that are sharply localized to critical layers near the top and bottom walls.
Similarly, resolvent analysis predicts strong amplification of this structure in the presence of viscoelasticity. This strong amplification implies, consistent with the fully nonlinear results, that even very weak disturbances may be sufficient to trigger EIT.  We note that Haward  et al.~\cite{Haward:2018he,Haward:2018hs} present experiments and analysis for viscoelastic flow over a wavy wall that illustrate amplification of perturbations in the critical layer.



%

\begin{figure}
\centerline{\includegraphics[width=1.\textwidth]{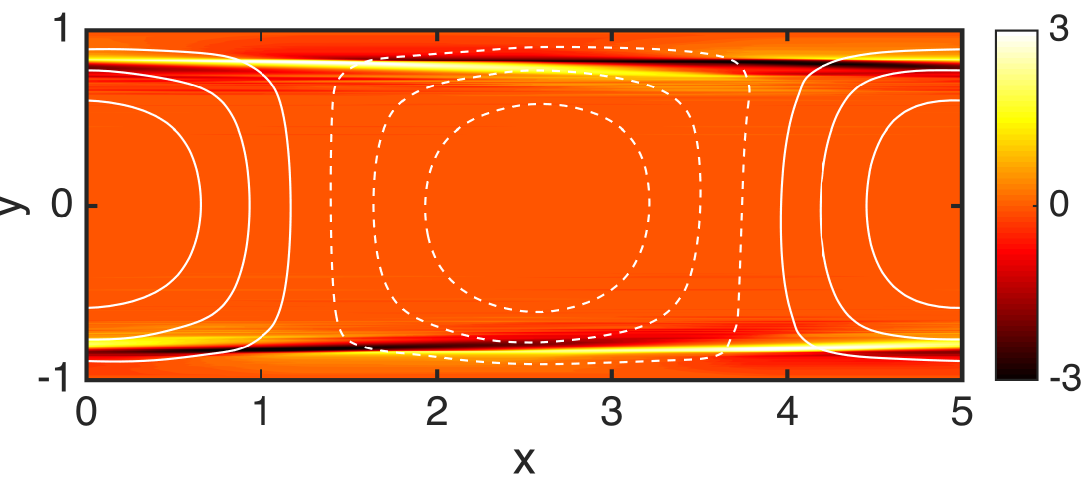}}
\caption[]{Snapshot of the finite amplitude Tollmien-Schlichting wave solution at $\R=3000$, $\rm{Wi}=10$~\cite{Shekar:2020gt}. White contours are wall-normal velocity, colors are deviations of $xx$ polymer stretch from laminar values. Reproduced with permission from \cite{Shekar:2020gt}.}
\label{fig:Re3000VETS}	     
\end{figure}

Building on the above observations, Shekar et al.~\cite{Shekar:2020gt} performed direct simulations of two-dimensional plane channel flow  with the FENE-P constitutive equation at $\R=3000$, revealing the existence of a family of attractors whose structure is virtually identical to the linear TS mode, and in particular exhibits strongly localized stress fluctuations at the critical layer position of the TS mode, as illustrated in Fig.~\ref{fig:Re3000VETS}. At the parameter values chosen, this solution branch is not connected to the nonlinear TS solution branch found for Newtonian flow, and thus represents a solution family that is nonlinearly self-sustained by viscoelasticity: The laminar state remains linearly stable, though again, as in~\cite{Shekar:2019hq}, only an extremely small perturbation is required to drive the solution away from the laminar state. Evidence indicates that this branch is connected through an unstable solution branch to two-dimensional EIT.

\begin{figure}
\centerline{\includegraphics[width=1.\textwidth]{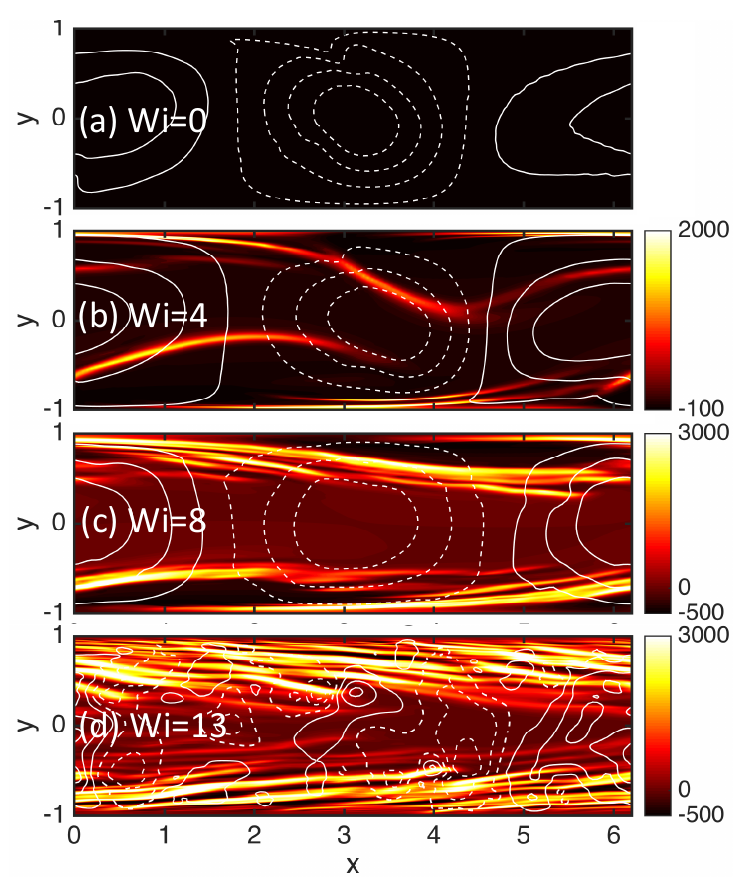}}
\caption[]{Snapshots of the finite amplitude Tollmien-Schlichting wave solution at $\textrm{Re}=10000$ and (a) $\textrm{Wi}=0$, (b) $\textrm{Wi}=4$, (c) $\textrm{Wi}=8$, (d) $\textrm{Wi}=13$~\cite{Shekar:2021vp}. White contours are wall-normal velocity and colors are deviations of $xx$ polymer stretch from laminar values.}
\label{fig:Re10000}	     
\end{figure}

Now we summarize very recent work of Shekar et al.~\cite{Shekar:2021vp} that extends earlier work from the same group \cite{Shekar:2019hq,Shekar:2020gt} to higher $\textrm{Re}$. 
At $\textrm{Re}=10000$, unlike $3000$, the Newtonian TS attractor evolves continuously and without hysteresis into  EIT as $\textrm{Wi}$ is increased from zero to about $13$ -- the two flows are part of the same solution family. Figure \ref{fig:Re10000} illustrates the evolution of the flow and stress fields as $\textrm{Wi}$ increases.  Note the resemblance between Fig.~\ref{fig:Re10000}d and Fig.~\ref{fig:EITpix}a. The simple sheet structures that originate with the TS critical layer structure evolve into the multilayered structure of EIT through a process that has been denoted ``sheet-shedding": Individual sheets associated with the critical layer structure break up, with the fragments further sheared as they travel downstream.

The linear instability to Tollmien-Schlichting waves does not arise for pipe or plane Couette flow, so the scenario described here does not directly apply to those geometries.  On the other hand, in these geometries elastoinertial turbulence with very similar features does arise in simulations in the same general parameter regime: namely, fluctuations localized in a layer near the wall, with a sheet-like stress structure and little to no activity in the center of the flow as illustrated in Fig.~\ref{fig:EITpix}b ~\cite{lopez2019dynamics,pereira2019beyond}. Furthermore, while linearly stable, wall modes analogous to the TS wave do exist in these other geometries~\cite{Drazin:1981wx}, and may be subject to nonlinear critical layer excitation, and subsequent evolution into EIT, just as the TS mode is in the channel flow case. Indeed, Zhang~\cite{Zhang:2021ef} performed resolvent analysis for pipe flow in the same parameter regime considered here,  demonstrating that the most amplified mode has strong stress fluctuations localized in a critical layer near the wall, just as is found by Shekar et al.~\cite{Shekar:2019hq}. See \cite{Shekar:2020gt} for further discussion of these issues.

 At the same time, in more strongly viscoelastic regimes, Garg et al.~\cite{garg2018}, Chaudhary et al.~\cite{chaudharyetal_2021}  and Khalid et al.~\cite{Khalid:2021ix} have found a linear center-mode instability for pipe flow and channel flow respectively, as described in \S \ref{EITsection}.C. 
  Choueiri et al.~\cite{choueiri2021experimental} also note the appearance of ``chevron" shaped structures resembling the unstable center mode in pipe flow up to  $\textrm{Re}=O(100)$ before being taken over by near-wall modes at higher $\textrm{Re}$. These results open up the possibility that other states unrelated to the nonlinear excitation of a wall mode may  play a role in elastic and/or elastoinertial turbulence, with Reynolds numbers in the aforementioned range. Nevertheless, the work described here demonstrates a direct connection between a wall mode (the TS mode) and EIT structures. 

	\subsection{Linking Back to Drag Reduction}
	\label{subsec:linkingtoDR}
Next, we return to the issue of the maximum drag reduction phenomenon (MDR). Based on the results above, the following scenario can be hypothesized: In the MDR regime, the flow cannot stay classically turbulent because streamwise vortices are so strongly suppressed by viscoelasticity that they cannot persist, but on the other hand the flow cannot fully laminarize either, because  viscoelastic TS waves (or something else) are nonlinearly excited by small but finite perturbations even when the laminar flow is linearly stable. Nevertheless, weak quasi-streamwise vortex and streak structures are experimentally observed to exist at MDR~\cite{White:2008hs}. Based on these points, MDR may be a marginal state where weak critical layer (or other) excitations keep the flow from laminarizing and provide sufficient perturbations to the flow for the mean shear to generate weak quasistreamwise vortices. 

We provide two further comments on this hypothesis. First, for Newtonian and viscoelastic channel flows, Xi and Graham~\cite{Xi:2012gh} computed ``edge states", which are dynamical trajectories that are marginal in the sense that they lie on the state-space boundary between laminar and turbulent flow.  Near transition, these states display a mean velocity profile very close to the Virk MDR profile. Furthermore, very recent computations by Zhu and Xi \cite{Zhu:2021di} indicate the presence of an intermittent process in viscoelastic channel flow involving quasi-2D structures with near-wall critical layer characteristics and 3D quasi-streamwise structures, again with a mean velocity profile that lies on or above the Virk MDR profile.

\begin{figure*}
\centerline{\includegraphics[width=0.9\textwidth]{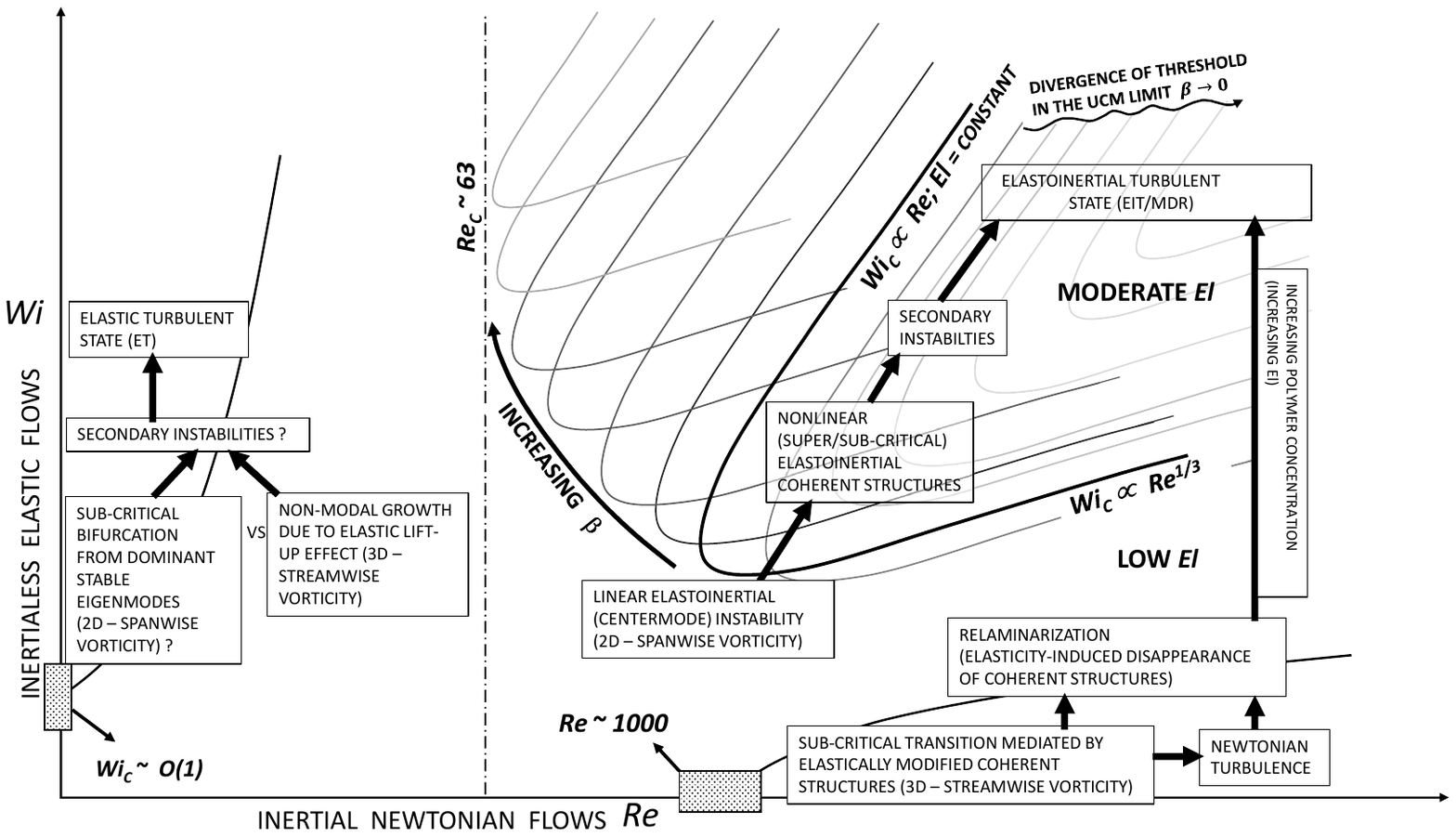}}
\caption{Schematic representation of various  transition scenarios for viscoelastic pipe flow in the \textrm{Re-Wi} plane. The laminar flow is characterized by the Poiseuille velocity profile with rectilinear streamlines. The linearly unstable regions in the interior of the \textrm{Wi-Re} plane, corresponding to the center-mode instability, are marked by a thick black line (solid) for one specific values of $\beta$, while those for other values of $\beta$ are depicted by light gray lines.  
}
\label{fig:bigpicturepipe}
\end{figure*}

\begin{figure*}
\centerline{\includegraphics[width=0.9\textwidth]{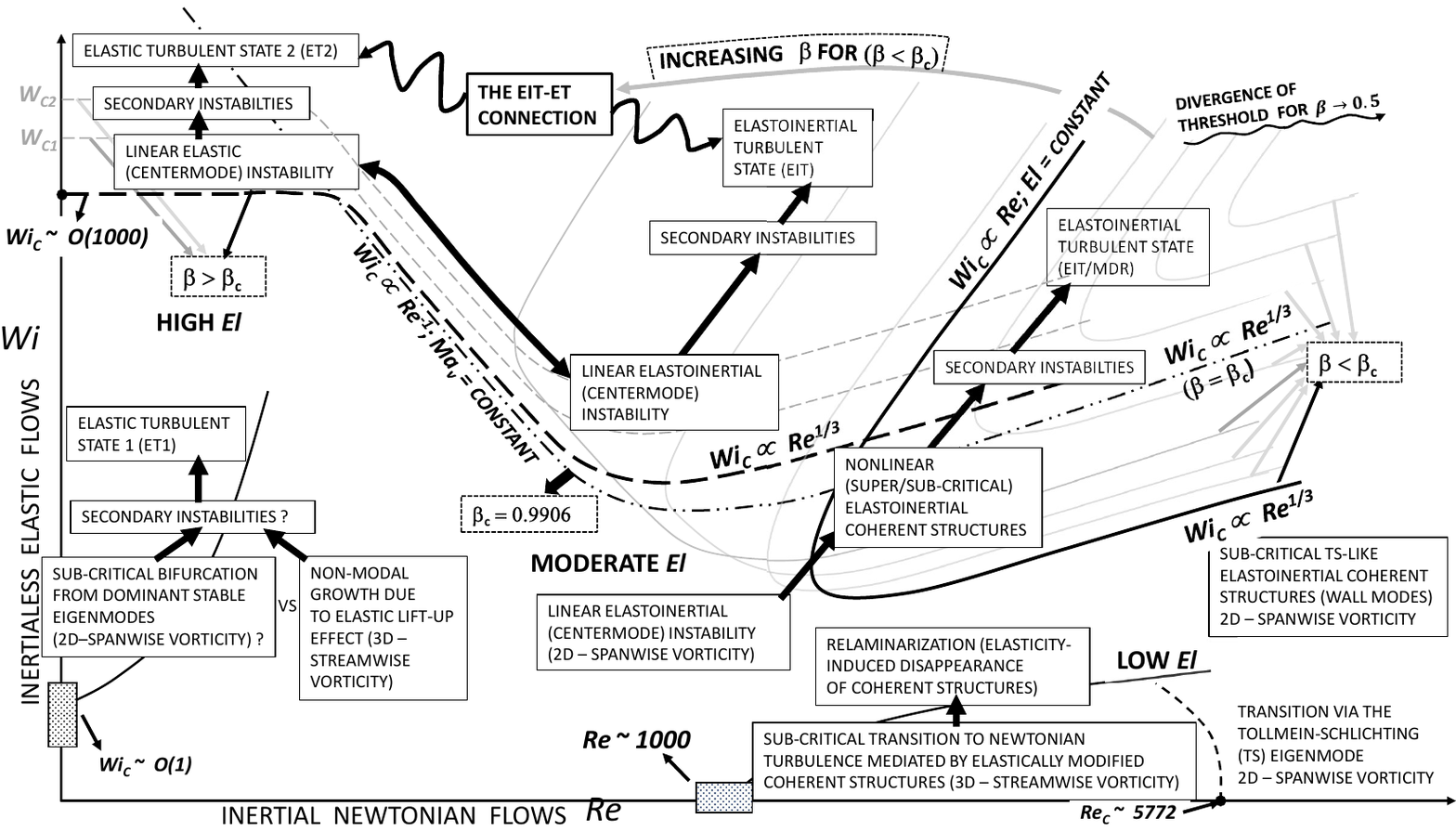}}
\caption{Schematic representation of various  transition scenarios for viscoelastic plane Poiseuille flow in the $\textrm{Re}-\textrm{Wi}$ plane. The linearly unstable regions in the interior of the \textrm{Wi-Re} plane, corresponding to the center-mode instability, are marked by thick black lines (solid, dashed or dash-dotted) for specific values of $\beta$, while those for other values of $\beta$ are depicted by light gray lines. Note that the critical value $\rm{Re}_c\sim5772$ is defined using the laminar equivalent centerline velocity. 
 }
\label{fig:bigpicturechannel}
\end{figure*}

\subsection{Elasticity-induced Transition Scenarios in Re-Wi Space}
\label{subsec:transitionscenarios}

In this section, we focus on the role of polymer on the incipient transition from the laminar state.
Transition for viscoelastic fluids such as polymer solutions, even within the framework of the simplest constitutive equations (the Oldroyd-B equation, for instance), is characterized by at least two additional parameters --- \textrm{Wi} and  $\beta$~\cite{chaudhary_etal_2019,chaudharyetal_2021}. Transition from the steady laminar base state, to states characterized by non-trivial spatiotemporal dynamics, can occur along multiple pathways in the \textrm{Re}-\textrm{Wi}-$\beta$ space; for example, the work on purely elastic instabilities described in \S~\ref{FlowTransitionSection} and the opening of this section, \S~\ref{EITsection}, explored pathways characterized by Re = 0. 

The recent prediction of a linear center-mode instability
for both viscoelastic pipe and channel flows (alluded to above)~\citep{garg2018,chaudharyetal_2021,Khalid:2021ix}
is
qualitatively different from the Newtonian scenario, where pipe flow is linearly stable at all $\textrm{Re}$, while  plane Poiseuille flow becomes unstable to the TS mode at $\textrm{Re} = 5772$, a value that is much higher than the observed threshold for transition. It is important to note that this center mode does not bear a direct relation to the Newtonian center mode, and this in turn is due to the elastoinertial spectrum being very different, and significantly more complicated, than its Newtonian counterpart (which has the APS template described earlier). One of the reasons for this  is  the presence of continuous spectra, which happen to be branch cuts, and discrete eigenmodes can appear and disappear out of the branch cut with variation in the different parameters. The structure of the elastoinertial spectrum in plane- and pipe-Poiseuille flows has been discussed, in some detail, in Refs.~\citep{chaudharyetal_2021,Khalid:2021ix}.

 The discovery of a linear instability in viscoelastic pipe flow, in particular,  marks a radical departure from the earlier literature, which had assumed this flow to be stable in the $\textrm{Re}$-$\textrm{Wi}$-$\beta$ parameter space~\citep{Pan_2012_PRL,MorozovVanSaarloos2005,Sid_2018_PRF}. The existence of a linear pathway to transition was also strongly suggested by the earlier experiments of
Samanta et al.~\citep{Samanta:2013el}, where the threshold Reynolds number was independent of whether the flow was forced at the inlet or not, beyond polymer concentrations of 200 ppm.
Both recent computations of Page et al.~\citep{PageDubiefKerswell2020} and experiments of Choueiri et al.~\citep{choueiri2021experimental} have pointed to the connection of the center-mode eigenfunction to the eventual nonlinear state (a novel EIT coherent structure in the computations) that emerges above threshold. 

We now attempt to bring together the ideas described above, both in this section that deals with elastoinertial transition and turbulence, and the earlier sections that focused on elastic instabilities and transition in rectilinear shearing flows, via
Figs.~\ref{fig:bigpicturepipe} and \ref{fig:bigpicturechannel}, which attempt to summarize the transition scenarios for pipe and plane Poiseuille flows, respectively, in the \textrm{Wi-Re} plane for different fixed values of $\beta$.  The linearly unstable regions in the interior of the \textrm{Wi-Re} plane, corresponding to the center-mode instability, are marked by thick black lines (solid, dashed or dash-dotted) for specific values of $\beta$, while those for other values of $\beta$ are depicted by light gray lines. In both figures, regions adjacent to the \textrm{Re} and \textrm{Wi} axes correspond to the onset of predominantly inertial and elastic instabilities, respectively, with the former underlying the sub-critical Newtonian transition.  
 
 We begin with a brief discussion of the features common to both figures, before going on to describe those unique to Fig.~\ref{fig:bigpicturechannel}, which make the transition in plane Poiseuille flow a potentially richer playground for both linear and nonlinear transition mechanisms. 
The 3D ECS-driven mechanism that triggers the Newtonian transition becomes less relevant for weakly elastic flows on account of the Newtonian ECSs being suppressed by increasing elasticity~\citep{stone_graham2002,stone_graham2003,Stone:2004jk,li_etal_2006,Graham2007}. 
While this suppression has been demonstrated specifically for plane Poiseuille flow, it is reasonable to conjecture  that a similar scenario should prevail for pipe flow on account of the similarity of the underlying ECSs~\citep{chaudharyetal_2021}. 
The suppression and eventual disappearance of the ECSs is thought to be responsible for a delayed transition to, and eventual disappearance of, the Newtonian turbulent state. In both Figs.~\ref{fig:bigpicturepipe} and \ref{fig:bigpicturechannel}, the Newtonian-turbulent-like state is therefore confined to a region between the $\rm{Re}$-axis and a curve that corresponds to a  $\rm{Re}$-dependent critical value of the Weissenberg number $\rm{Wi_c}$. At higher levels of elasticity, the aforementioned linear center-mode instability becomes operative.

Although the extent of the linearly unstable region depends sensitively on flow-type and $\beta$, the unstable regions for both pipe and channel flows, Figs.~\ref{fig:bigpicturepipe} and \ref{fig:bigpicturechannel} respectively, bear a close resemblance in the range $0.5 < \beta < 0.98$, with 
$\textrm{Wi}_c \propto \textrm{Re}^{1/3}$ along the lower branch of the unstable region, while $\textrm{Wi}_c \propto \textrm{Re}$ along the upper 
 branch (the latter corresponds to a constant elasticity number $\textrm{El}$, and represents an experimental path for a given flow geometry and polymer solution). For both geometries, the center-mode eigenfunction likely gives way to supercritical nonlinear structures that, either directly, or \textit{via} secondary instabilities, might underlie the observed EIT dynamics. In this sense, the center-mode instability, for both pipe and channel flows, provides a continuous pathway from the laminar state to the EIT (and the eventual MDR) regime.  Although not shown, the EIT and Newtonian turbulence domains overlap at higher $\rm{Re}$, where the original center mode gives way to a wall mode -- indeed, this overlap has been found by several authors~\cite{pereira2019beyond,ZhuXi2021,ShekarThesis2021} -- implying that the latter transitions in a continuous manner to the former, without an intervening relaminarization. This was believed to always be the case in earlier literature. The vertical path shown on the right in Fig.~\ref{fig:bigpicturepipe} corresponds to the one in Ref.~\citep{Choueiri:2018it}, which first accessed an intermediate quasi-laminar state with increasing Wi at a fixed \textrm{Re} (=3600), thereby contradicting the aforesaid long-held belief. 
 A vertical path at fixed Re in the \textrm{Wi--Re} plane implies an increase in elasticity number $\textrm{El} = \lambda \nu/R^2$, introduced earlier in \S~\ref{subsec:Dimlessparameters}, but defined here with the pipe radius $R$ as the relevant length scale and $\nu\equiv\mu/\rho$.
In this context, it is useful to note that in the dilute limit, strictly speaking, both the relaxation time $\lambda$  and $\nu$ are independent of polymer concentration. Thus, in this regime, an increase in $\textrm{El}$ can be accomplished only by decreasing the pipe radius $R$. However, in the experiments of Choueiri et al. ~\citep{Choueiri:2018it}, the pipe radius $R$ is fixed; instead, the authors increase the polymer concentration in the vicinity of the overlap value, which results in an increase in both $\lambda$ and $\nu$, and thence $\textrm{El}$ (while, presumably, adjusting the flow rate to keep $\textrm{Re}$ fixed).

Despite the above similarities, there remain significant differences between the  instabilities of pipe and plane Poiseuille flows outside of the aforementioned range of $\beta$. The center-mode instability disappears for $\beta < 0.5$ for channel flow, while it persists down to $\beta \approx 10^{-3}$ for pipe flow. The opposite limit of $\beta \rightarrow 1$,
discussed above in the drag reduction context, is also of particular interest from the linear stability viewpoint.  While the center-mode instability
appears to be restricted to $\rm{Re} > 63$ for pipe flow (Fig.~\ref{fig:bigpicturepipe}), remarkably, it morphs into a purely elastic instability for channel flow, continuing to arbitrarily small $\textrm{Re}$ for $\beta > \beta_c \approx 0.9905$ \citep{khalid_creepingflow_2021}. As a result, the `nose' of the original unstable region in Fig.~\ref{fig:bigpicturechannel} begins to broaden for $\beta \rightarrow \beta_c$, eventually opening out into a plateau that extends right up to the Wi-axis for $\beta > \beta_c$. Rather intriguingly, for $\beta$ close to $\beta_c$, the lower branch ($\textrm{Wi}_c \propto \textrm{Re}^{1/3}$) and the small-$\rm{Re}$ plateau are separated by an intermediate asymptotic regime with $\textrm{Wi}_c \propto \textrm{Re}^{-1}$ (this corresponds to a constant viscoelastic Mach number, $\textrm{Ma}_v$ = $V/V_{shear} = O(1-\beta)^{-1}$, with $V_{shear}  = \sqrt{\frac{(1-\beta)\eta}{\rho \lambda}}$ being the shear wave speed). However, the implied shear-wave signature may not be relevant to the recent observation of `elastic waves' in sheared dilute polymer solutions~\citep{Varshney2019,khalid_creepingflow_2021}. 
 Considerations of continuity imply that the crossover from the intermediate scaling regime to the creeping-flow instability must pass through a special $\beta = \beta_c $ for which the scaling $\textrm{Wi}_c \propto \textrm{Re}^{-1}$ should persist down to $\textrm{Re} \rightarrow 0$! (the dash-dotted line in Fig.~\ref{fig:bigpicturechannel}). Importantly, the aforementioned transformation of the original center-mode instability into a purely elastic one (that in turn might give way to a turbulent state) highlights the existence of an EIT-ET connection for channel low (via an underlying modal pathway). This might serve as a novel template in a search for purely elastic coherent structures.

In regions of the \textrm{Re}-\textrm{Wi}-$\beta$ space where the center mode is linearly stable, novel subcritical mechanisms likely dominate the transition process. In this regard,
and as discussed in \S~\ref{subsec:dragreduction}, 
recent work~\citep{Shekar:2019hq,Shekar:2020gt}  has identified a nonlinear mechanism closely related to the stable Newtonian Tollmein-Schlichting mode  (although still disconnected from it in phase space until a $\textrm{Re}$ of $10^4$). The fact that there is no analog of the TS-instability in Newtonian pipe-flow, and no evidence of a corresponding nonlinear solution branch in the Newtonian limit,
suggests that the TS-mode-based subcritical mechanism could be specific to plane Poiseuille flow. On the other hand,  as noted in \S~\ref{EITsection}.B, the direct simulations of EIT by Lopez et al.~\cite{Lopez:2019ct} display strong localization of fluctuations near the wall, and the resolvent analysis of Zhang~\cite{Zhang:2021ef}  demonstrates strong linear amplification of a mode with near-wall critical-layer stress fluctuations. Both of these observations are fully consistent with those described by Shekar et al.~\citep{Shekar:2019hq,Shekar:2020gt} in channel flow, wherein subcritical transition to EIT is driven by the amplification of fluctuations with near-wall critical layer structure, suggesting a similar mechanism for EIT in pipe and channel flows. Indeed, in the work of \citep{Shekar:2019hq}, the Reynolds number is so low that no subcritical TS branch exists in the Newtonian limit.  
Returning to the case of channel flow, the recent subcritical continuation of the unstable center mode  to a nonlinear EIT structure~\citep{PageDubiefKerswell2020} implies that subcritical mechanisms based on the center mode might also be operative in certain regions of Re-Wi-$\beta$ space, and thus the relevance of the center mode might extend outside of the linearly unstable regions indicated in Figs.~\ref{fig:bigpicturepipe} and \ref{fig:bigpicturechannel}. 

In the opposite limit of $\textrm{Re} \ll 1$, viscoelastic pipe and Poiseuille flows are linearly stable for $\textrm{Wi} = O(1)$  and when $\beta$ is not very close to unity~\cite{wilson1999,chaudharyetal_2021}. 
One of the proposed transition scenarios is that of a subcritical 2D nonlinear instability~\cite{MorozovVanSaarloos2005,morozov_saarloos2007}, although this has been demonstrated only for $\textrm{Wi}= O(1)$ and $\beta \rightarrow 0$.  The existence of a linear instability at the other extreme -- $\textrm{Wi} = O(1000)$  and $\beta \rightarrow 1$ ~\citep{khalid_creepingflow_2021} -- implies the possibility of a bifurcation to a distinct elastic turbulent state. It is therefore possible to envisage (at least) two different ET states (labeled ET1 and ET2 in Fig.~\ref{fig:bigpicturechannel}), in inertialess plane Poiseuille flow, depending on Wi.
Even in this limit, however, there is a wide intermediate range of $\beta$ ($0 < \beta < \beta_c$) for which the nature of the subcritical transition remains an open question.

It is worth summarizing, in a succinct manner, the implications of the findings detailed in this section with respect to transition to EIT in pipe and channel flows of polymer solutions. For moderate-to-strongly elastic polymer solutions ($\textrm{El} >0.1$, $\beta \sim 0.5-0.9$), where transition to EIT occurs directly from the laminar state, both experiments and theory point to the relevance of the center mode at onset ~\citep{garg2018,chaudharyetal_2021,Khalid:2021ix,PageDubiefKerswell2020,choueiri2021experimental}.
On the other hand, for weakly elastic dilute polymer solutions of the type investigated in the context of drag reduction ($\textrm{El} < 0.02$, $\beta \rightarrow 1$), 
when the primary transition to turbulence is akin to the Newtonian one, 
the eventual EIT state is dominated by wall modes that appear to be closely related to the nonlinear travelling-wave solutions identified in~\citep{Shekar:2019hq,Shekar:2020gt}.
It is worth noting that there are vast tracts of the viscoelastic parameter space that remain to be understood from the transition perspective. For instance, for dilute solutions (with $\beta = 0.97$) at higher $\textrm{El}$  ($0.02 < \textrm{El} < 0.5$), it is the continuous spectrum that is the least stable (see Fig.~19b of Ref.~\citep{Khalid:2021ix}), and may perhaps be expected to play a dominant role in the (subcritical) transition dynamics. Future research will be necessary to disentangle the roles of wall- and
center-mode-based structures, and perhaps other structures (e.g., modes belonging to the continuous spectrum) as well, for EIT in various geometries and parameter regimes.

\subsection{Non-modal Scenarios}

The above discussion of transition scenarios is restricted to either new modal pathways induced by elasticity, or the elastic modification of essentially Newtonian non-modal pathways. There also exist efforts 
that have highlighted novel non-modal pathways due to elasticity alone~\citep{jovanovic_kumar_2010,jovanovic_kumar_2011}, or due to a non-trivial interplay of elasticity and inertia~\citep{Zaki2013}. The non-modal pathways, in the inertialess limit in particular, point to the importance of spanwise varying disturbances (much like the Newtonian case) that are amplified by an elastic analog of the lift-up effect, and by an amount that increases with increasing $\rm{Wi}$.  The significance of the essentially 3D non-modal pathways~\citep{jovanovic_kumar_2010,jovanovic_kumar_2011} relative to the aforementioned 2D nonlinear modal mechanism \citep{MorozovVanSaarloos2005,morozov_saarloos2007} requires more detailed examination; in light of this, Figs.~\ref{fig:bigpicturepipe} and \ref{fig:bigpicturechannel} indicate both non-modal and modal pathways leading to the ET state (ET1 for channel flow). 

The experiments reported so far~\citep{Arratia_PRL_2017,Arratia_PRL_2019} cannot reliably be used to emphasize either pathway especially because the nonlinear elastic state accessed is for a channel with a cross-sectional aspect ratio of unity; the sensitivity of this state to the precise form (`shape') of the inlet disturbance, including the relative significance of streamwise vis-a-vis spanwise variations, remains to be established. Although the ET state has been reasonably well characterized statistically in the aforementioned experiments, recent experiments \citep{JhaSteinberg1} have, in channels with higher aspect-ratio cross-sections, begun exploring the underlying structural motifs that might help identify the elastic analogs of the Newtonian ECSs. In contrast to the above, the EIT state accessed in both pipe and channel geometries only exhibits minor spanwise variations, and this essential two-dimensionality is consistent with the underlying modal picture~\citep{Piyush_2018,chaudharyetal_2021,Khalid:2021ix}.


\section{Free surface instabilities in polymeric fluids}
\label{SectionFreeSurface}
\begin{figure}
\centering
\includegraphics[width=\hsize]{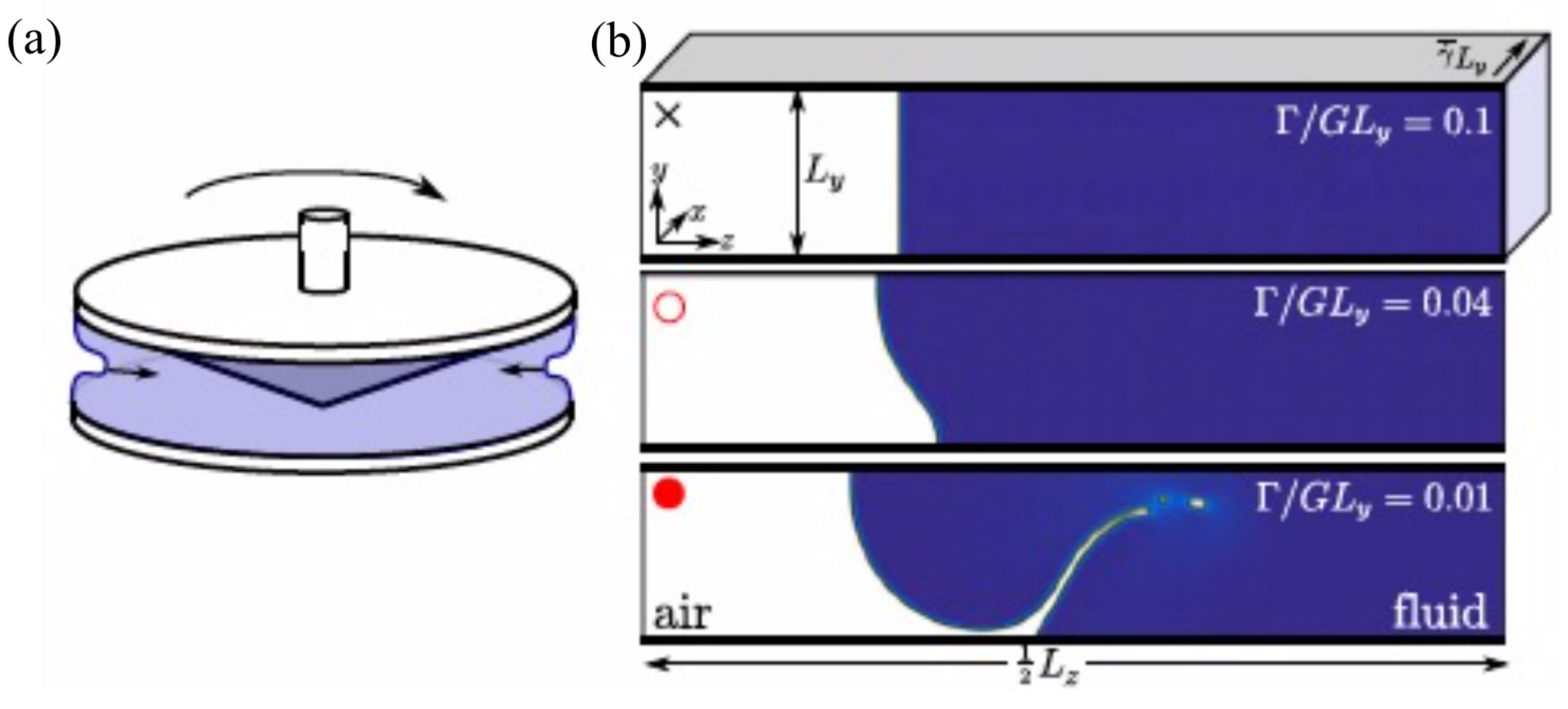}
\caption{Free surface instabilities in edge fracture. (a) Schematic of a cone and plate device where the interface undergoes an instability. (b)
Snapshots from simulations of the Giesekus
model between rigid walls. Note that in this figure, $\Gamma$ is used to denote interfacial tension, while in the corresponding text we use $\sigma$ instead, and $G$ is the elastic modulus~\cite{hemingway2017edge}.  }
\label{FreeSurfaceInstabilities}
\end{figure}

\begin{figure}
\centering
\includegraphics[width=\hsize]{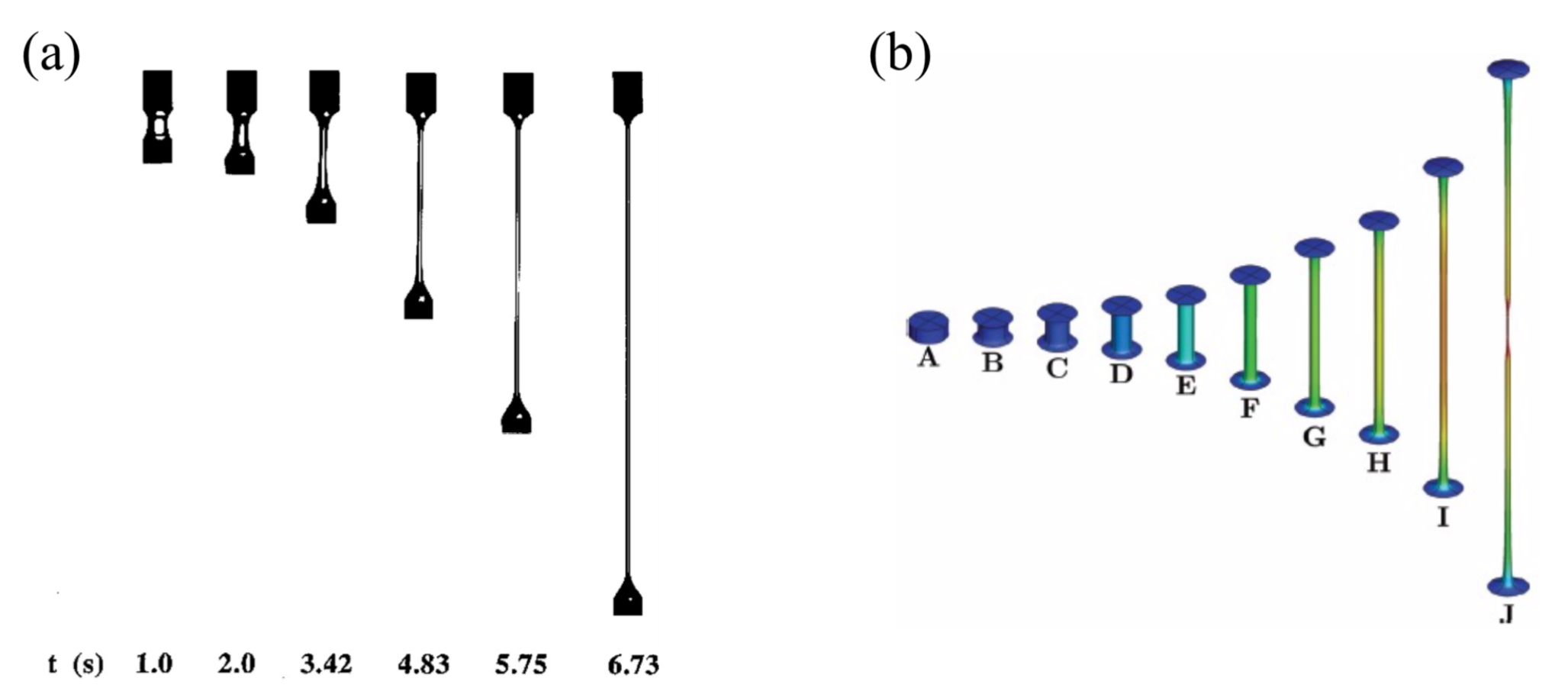}
\caption{Extensional necking. (a) Experiments of exponential elongation of a filament of a viscoelastic fluid 
(0.31 wt \% polyisobutylene
in polybutene)~\cite{McKinley2002b}. Reprinted from \cite{Sridhar}, with permission from Elsevier. (b) Numerical simulations of the so-called pom-pom model for an imposed strain rate $\dot\epsilon$, with the color scale indicative of the tensile stress~\cite{Hoyle1}; see figure 9 of the reference for the time of each image. Reproduced from \cite{Hoyle1}, with the permission of the Society of Rheology. }
\label{FreeSurfaceInstabilities2}
\end{figure}

The previous sections described flow instabilities within bounded domains. In this section, we address flow instabilities that arise at the free
surface between a polymeric fluid and the outside air. We do not
attempt a comprehensive review, but instead focus on three specific
instabilities. The first, often termed `edge fracture', is widely
observed when a highly viscoelastic polymeric fluid is sheared in a
torsional (cone-plate or plate-plate) flow device (Fig.~\ref{FreeSurfaceInstabilities}).  The second
concerns the necking of a filament of viscoelastic polymeric fluid in
which the constituent polymer chains are highly entangled, in the
regime where the bulk viscoelastic stresses dominate surface
tension (Fig.~\ref{FreeSurfaceInstabilities2}). The third concerns the breakup of a thread of high
molecular weight elastic polymer in the regime where surface tension
dominates (Fig.~\ref{complex}). 

\subsection{Edge Fracture in Sheared Complex Fluids}

Measurements of a fluid's shear rheology are commonly performed in a
torsional rheometer, often using either a cone-plate or plate-plate
flow cell.  In the linear viscoelastic regime, measurements are
generally well controlled and reproducible. The measurement of
stronger flows is often hindered by flow instabilities.  For example,
above a critical value of the imposed rate of shear, $\gdot$, the free
surface where the fluid sample meets the outside air can destabilize
towards a more complicated profile, e.g., Fig.~\ref{FreeSurfaceInstabilities}, despite having been neatly trimmed
initially, forming an indentation of the interface that can then invade the bulk.  Part of 
the sample can even be ejected from the flow cell, leading to
unreliable data. This phenomenon is known as
`edge-fracture'~\cite{snijkers2011cone,jensen2008measurements,lee1992does,inn2005effect,sui2007instability,schweizer2008departure,mattes2008analysis,dai2013viscometric}. Several
experimental strategies have been developed aimed  at mitigating its
effects~\cite{mall2002normal,schweizer2003comparing,schweizer2004nonlinear,snijkers2011cone,meissner1989measuring,schweizer2013cone,costanzo2018measuring}.

From a theoretical viewpoint, an early insightful
work~\cite{tanner1983shear,keentok1999edge} argued that edge fracture
must be driven by the second normal stress difference $N_2$ in the
fluid, positing instability to arise above a critical magnitude
$|N_2(\gdot)|>\sigma/R$, where $\sigma$ is the surface tension of
the fluid-air interface and $R$ a pre-assumed surface indentation
radius. Experimental work later confirmed this important role of
$N_2$ in driving edge fracture~\cite{lee1992does,keentok1999edge}.

More recent theoretical studies have revisited this
phenomenon~\cite{hemingway2017edge,hemingway2019edge,hemingway2018edge,hemingway2020interplay}. By
means of linear stability analysis, an updated criterion for the onset
of edge fracture was put forward (note that the notation has been adjusted slightly to be consistent with usage in this paper) ~\cite{hemingway2017edge,hemingway2019edge},
\be
\frac{1}{2}\Delta\tau\frac{d|N_2(\gdot)|}{d\gdot}\bigg/\frac{d\tau}{d\gdot}=\frac{1}{2}\Delta\tau \frac{d|N_2|}{d\tau}>\frac{2\pi\sigma}{L_y},
\label{eqn:criterion}
\ee
where $\tau=\tau(\gdot)$ is the shear stress in the fluid,
$\Delta\tau$ the jump in shear stress between the fluid and outside
air ($\Delta\tau\approx \tau$, given the low viscosity of
air), and $L_y$ the gap size in the rheometer. This updated criterion was
shown to agree with the instability threshold found in direct
nonlinear simulations at low shear rates (which is the regime in which
it was developed)~\cite{hemingway2017edge,hemingway2019edge}.  
%
%

In the limit of low shear rates, in fluids in which the shear stress
scales linearly with shear rate and $N_2$ scales  quadratically, the 
criterion (\ref{eqn:criterion}) predicts the same scaling as that of the earlier
Refs.~\cite{tanner1983shear,keentok1999edge}, if the pre-assumed
indentation radius in the earlier work is now instead identified as
the rheometer gap size. It is however worth noting that the updated
criterion correctly predicts the prefactor and identifies the important role of shear stresses in
contributing to instability. Importantly, the new criterion also departs
markedly from the early ones in stronger shear.

The linear stability analysis of
Refs.~\cite{hemingway2017edge,hemingway2019edge} also elucidated for
the first time the basic physical mechanism of edge fracture, which
can be understood as follows.  Were the interface between the fluid
and air to remain flat, the jump in shear stress across it would be
consistent with force balance.  It is helpful to recognize that, with $x$ as the flow direction, 
 were the interface oriented with its normal in the flow-gradient
direction $y$, then the shear stress $\tau_{xy}$ would have to be continuous across
it. However, this perturbed interface has its normal in the
vorticity direction ($z$) so the shear stress $\tau_{xy}$ can jump across it.
(This structure is actually the same as allowing vorticity bands
with layer normals in the vorticity direction, with a jump in shear stress
$\tau_{xy}$ between the bands,  which has been discussed in the literature \cite{goveas}.) So imagine that a small disturbance from a planar state now
develops in the interfacial profile. This exposes the jump in shear
stress, potentially disturbing the force balance across the interface. To
recover local equilibrium, a perturbation is needed in the shear stress, and so in
the shear rate. This in turn perturbs the second normal stress, which
must be counterbalanced by a perturbation to the extensional stresses
in the vicinity of the interface. The imbalance then requires a perturbation to the
velocity gradient and therefore velocity near the interface, which can
be shown to enhance the original interfacial disturbance, giving the
runaway positive feedback of the edge fracture instability. The
mechanism just described resembles that of other interfacial
instabilities between layered viscoelastic
fluids~\cite{hinch1992instability,wilson1997short,nghe2010interfacially}.

The work of Refs.~\cite{hemingway2017edge,hemingway2019edge} also
suggested a possible route to mitigating edge fracture
experimentally. In particular, the left-hand side
of Eq.~(\ref{eqn:criterion}) contains the term $\Delta\tau$, which is the
jump in shear stress between the fluid and outside medium. By
immersing the flow cell in an immiscible Newtonian fluid with a
viscosity more closely matched to that of the original fluid, the jump
$\Delta\tau$ will be reduced, thereby potentially mitigating the
instability. Another strategy could be to engineer a larger
interfacial tension $\sigma$, again by suitable choice of the
(Newtonian) bathing medium.

In addition, the interplay of edge fracture with the bulk flow instability known as
shear banding has been
considered~\cite{skorski2011loss,hemingway2018edge,hemingway2020interplay}. These
works show that modest edge disturbances that constitute a precursor to edge
fracture can lead to a noticeable apparent shear banding effect that
can penetrate far into the bulk, for a fluid with a relatively flat
underlying constitutive relation of shear stress as a function of
shear rate~\cite{hemingway2018edge}.  Conversely, shear banding can
lead to edge fracture~\cite{skorski2011loss}. More generally, a
complicated interplay is expected to exist between the two
effects~\cite{hemingway2020interplay}, potentially informing the long
standing debate concerning whether bulk shear banding occurs in
entangled
polymers~\cite{sui2007instability,li2015startup,schweizer2008departure,li2013flow,wang2014letter,li2014response,boukany2015shear}.

Notable challenges in understanding edge fracture remain. For example,
the work of
Refs.~\cite{hemingway2017edge,hemingway2019edge,hemingway2018edge,hemingway2020interplay}
considered only fluids with a negative second normal stress
difference; it would be interesting in future studies to consider the
case of a positive $N_2$. Furthermore, these works considered only
fluids with a finite terminal relaxation time, $\lambda$, for which the
shear stress $\tau\sim\gdot\lambda$ and second normal stress $N_2\sim
-(\gdot\lambda)^2$ for low shear rates, $\gdot\lambda\ll 1$. Future work
should consider non-Brownian suspensions~\cite{denn2014rheology}, in
which $N_2$ scales linearly with shear rate. 

The criterion discussed
above also assumes an underlying base flow of steady shear, while edge
fracture is widely seen in transient rheological protocols, the
modelling of which would require a time-dependent underlying base
state. Finally, the phenomenon of wall slip  arises widely in
strongly sheared entangled polymers, and is therefore likely often to
occur alongside edge fracture. The interplay of these two widely
occurring phenomena remains to be considered theoretically.

\subsection{Extensional Necking in Entangled Polymeric Fluids}

Extensional flows provide a key benchmark for the development of
rheological constitutive models of highly entangled polymeric fluids,
with many nonlinear flow features being apparent only in extension.  A
common experimental protocol consists of stretching out in length an
initially cylindrical filament of material in a filament stretching rheometer, e.g., Fig.~\ref{FreeSurfaceInstabilities2}. Such
experiments can be performed by switching on a Hencky strain rate
$\edot$, which is held constant thereafter; or a tensile stress
$\tau_{\rm{E}}$~\cite{Munstedt1975,Munstedt2013,Munstedt2014,Alvarez2013};
or a tensile force $\force$ (which provides a closer model of some
industrial processes, such as fibre
spinning~\cite{Wagner2002,Szabo2012}). In many such experiments, the
region of the filament furthest from the sample ends will often thin
more quickly than the sample as a whole, forming a `necked' region and
finally even causing the filament to
fail~\cite{Barroso2005a,Liu2013,Burghelea2011a,Tripathi2006,Bhardwaj2007,Arciniaga2011,Smith2010}. This
necking instability has been observed at constant tensile stress
\cite{Andrade2011}, constant Hencky strain rate
\cite{Burghelea2011a,Malkin2014}, and during the process of stress relaxation after an initial Hencky strain
ramp~\cite{Wang2007}.

From a theoretical viewpoint,  recently criteria for the onset of necking have
 been
developed~\cite{PhysRevLett.107.258301,ISI:000352990500012,Hoyle1,Hoyle2,hoyle2017necking},
separately for the flow protocols of constant imposed tensile stress,
tensile force, and Hencky strain rate, and considering necking during
stress relaxation after an initial extensional strain ramp.  These
criteria were initially derived analytically within a constitutive
model written in a highly generalised form, then checked to indeed
apply in numerical calculations performed in several different widely
used polymer constitutive models~\cite{Larson1988} (the Oldroyd B,
Giesekus, FENE-CR, Rolie-poly~\cite{likhtmangraham03} and
pom-pom~\cite{McleLars98} models). The focus throughout these studies was on the
case of highly viscoelastic filaments of sufficient radius that bulk
stresses dominate surface tension.

For a filament subject at time $t=0$ to the switch-on of a constant
tensile stress $\tau_{\rm{E}}$, the Hencky strain rate $\edot$ quickly attains its value
prescribed by the underlying steady state extensional constitutive
curve before any appreciable necking develops. The criterion for a
neck subsequently to develop was then found to be~\cite{Hoyle2} 
\be
\frac{d\tau_{\rm{E}}}{d\edot}>0.
\ee
This shows that, in fact, any highly viscoelastic material with a
positively sloping extensional constitutive relation $\tau_{\rm{E}}(\edot)$
must ultimately be unstable to necking in filament stretching.

For a filament subject instead to the switch-on of a constant tensile
force $F$, a filament was predicted to become unstable to necking in
any regime where the time-differential $\edot(t)$ of the extensional
creep curve $\epsilon(t)$ simultaneously has positive slope and
positive curvature~\cite{Hoyle2}:
\be
\label{eqn:creepExt}
\frac{d^2\edot}{d t^2}/\frac{d\edot}{dt}>0.
\ee

A filament subject to the switch-on of a constant Hencky strain rate was shown
to be unstable to necking if  the tensile stress response shows negative
curvature as a function of the accumulating Hencky
strain~\cite{PhysRevLett.107.258301}, 
\be
\label{eqn:creepStrainRate}
\frac{d^2\tau_{\rm{E}}}{d\epsilon^2}<0.
\ee

A full discussion of these criteria can be found in
Refs.~\cite{ISI:000352990500012,PhysRevLett.107.258301,Hoyle1,Hoyle2,hoyle2017necking}. They
were derived within a so-called `slender filament' approximation, in
which the wavelength of necking variations along the filament's length
is assumed long compared with the filament radius. They furthermore
ignore any effects of the endplates, beyond their role in seeding an
initial heterogeneity in the way that the filament starts to
deform. In future work, it would be interesting to perform fully 3D
simulations in microscopically motivated rheological models, move
beyond the slender filament approximation, and incorporate endplate effects.

\subsection{Instabilities in Polymeric Pinching}
The breakup of a solution of high molecular weight, elastic,
polymers, driven by surface tension, is very different from its
Newtonian counterpart, even at concentrations as low as 10 ppm
in weight \cite{McK_review,MAK14}. Polymers are stretched by the extensional
flow leading to breakup and resist it, resulting in a strong increase
of the extensional viscosity $\eta_E$ (the extensional stress
$\tau_{zz}-\tau_{rr}$
divided by the elongation rate $\dot{\epsilon}$). What would have been
two or a whole series of isolated
drops in the Newtonian case, are now connected by thin threads of
highly stretched material of almost uniform radius. This is known as
the ``beads-on-a-string'', (BOAS) structure, characterized by a strong
buildup of stress inside the threads, the extensional viscosity
growing by several orders of magnitude in the process. Since the capillary
pressure inside a thread is much higher than inside a drop, the thread
empties into the drop and thins further \cite{BVER81}, limited by the
buildup of stress. In a regime where inertia is important, it is known from the
Newtonian case that so-called satellite drops of smaller size
are formed between two main drops \cite{E97}. The same is true
in the elastic case, but with threads connecting the main and satellite drops
\cite{LF03,WABE05,BAHPMB10}.

Taking into account stress relaxation, an analysis of the
Oldroyd-B viscoelastic equations \cite{BAH87} with a single relaxation time
$\lambda$ shows that the thinning of the thread (radius $h_{thr}$) is 
exponential \cite{BER90}:
\beq
h_{thr} = h_0 e^{-t/(3\lambda)}.
\label{exp}
\eeq
Although even monodisperse 
polymer solutions are known to exhibit a spectrum of relaxation
times rather than a single $\lambda$ \cite{DoiBook86}, \eqref{exp}
works remarkably well for a wide range of flexible polymer systems, 
in both low- and high viscosity solvents. The reason is that
\eqref{exp} is dominated by the longest relaxation time \cite{EH97},
which can be estimated as the Zimm time \cite{AM01}. Good agreement
between $\lambda$ found from fitting Eq. \eqref{exp} to experimental data,
and the Zimm time is found for dilute solutions \cite{AM01,CPKOMSVM06,DVB18},
but the case of higher concentrations is often more complicated.
For example, \cite{CPKOMSVM06,DHEB20}
found a strong (power-law) dependence of $\lambda$ on polymer concentration,
with an exponent that depends on the quality of the solvent. 

In order to relate the prefactor $h_0$ in \eqref{exp} to the stress,
one needs to match the thread to the drop into which it is emptying
\cite{CEFLM06}, as has recently been done on the basis of
the full three-dimensional, axisymmetric Oldroyd-B equations
\cite{EHS20,DHEB20}. Thus one can determine the extensional
viscosity by measuring the thread radius alone, knowing the surface
tension $\sigma$:
\beq
\eta_E \equiv \frac{\tau_{zz}-\tau_{rr}}{\dot{\epsilon}} = 
\frac{3\sigma \lambda}{h_{thr}} = -\frac{\sigma}{\dot{h}_{thr}}.
\label{eta_E}
\eeq
The remarkable feature of
\eqref{eta_E} is that it is independent of the history of 
the filament (for example any pre-stretch), or the geometry
(a dripping geometry or a free jet). Of course, this is true
only in an asymptotic sense, such that the regime of exponential
thinning is long. However, eventually the polymer reaches full stretch,
and crosses over to a faster thinning law, which is observed to 
be linear instead of exponential, with $\eta_E$ saturating at a
constant value \cite{AM01,DS19,DS20}.

The physical idea is that at full stretch of the polymer, the viscosity
can grow no longer, and the polymeric solution behaves once more like
a Newtonian fluid, but with an elevated value of the elongational viscosity.
Indeed, a theoretical analysis \cite{Ren02,FL04} of the
FENE-P model \cite{BAH87}, which incorporates finite extensibility, 
predicts an {\it instability} of the uniform thread, leading to 
localized pinch solutions of the same self-similar form as for a
Newtonian thread \cite{EV08}, but with an effective viscosity
that grows linearly with the length of the polymer. By observing
the minimum radius of the localized pinch $h_{min}(t)$, the extensional
viscosity can be inferred from $\eta_E = -3\alpha \sigma / \dot{h}_{min}$,
where $\alpha = 0.0709$ for
symmetric pinching (inertia subdominant), and $\alpha = 0.0304$ for
asymmetric pinch solutions (the asymptotic case for small $h_{min}$)
\cite{EV08}. 

The latter case predicts an extensional viscosity more than a factor
of 10 smaller than \eqref{eta_E}, which indeed is for a uniform thread,
which is at best an unstable solution once finite extensibility comes
into play. In order to interpret thread radius data correctly, it
therefore seems important to monitor the radius of the thread in space,
which can no longer be assumed uniform. A theoretical analysis of the
crossover between a uniform thread and localized pinching remains to 
be done. 

However, as first reported in \cite{OM05}, uniform polymeric threads
are frequently subject to a more complicated, delocalized instability,
leading to the sudden growth of many small droplets all along the thread.
The instability can proceed through several generations \cite{CDK99,OM05},
producing drops of different sizes, but the initial instability was observed
to follow an exponential growth law \cite{SWE08}, indicative of a linear
instability. While there is a superficial resemblance to the original
BOAS structure, the physical process is the opposite: a localized relaxation 
of stress, leading to droplets. To differentiate between the two processes,
''blistering instability'' has been proposed as a name for the instability
of a highly stretched polymeric thread.

The blistering instability cannot be understood as the linear (Rayleigh-Plateau)
instability of a Newtonian thread, but with an elevated elongational viscosity:
the growth rate is at least an order of magnitude faster than a Newtonian
instability would predict \cite{SWE08,SGEW12}. Instead, \cite{CDK99} proposed
an instability localized at the end of the thread, resulting in a relaxation
of stress, followed by elastic recoil, and triggering the formation of a
thinner filament. While such localized instabilities have also been
seen by others \cite{SGEW12}, they are distinct from the linear instability
leading to quasi-simultaneous growth along the entire thread. 

To explain the observed linear instability, it has been proposed
\cite{E_pol14} that the coupling between stress and the local polymer
density \cite{HF89,CVFL13} has to be taken into account. Density 
fluctuations are automatically part of the description when deriving
continuum models using kinetic theory \cite{BM94}, but are usually 
neglected in continuum descriptions. The stress-density coupling
results in a flow of polymers toward regions of high stress (this is
true independent of the flow type \cite{HF89}), leading
to further stress relaxation in polymer-poor parts of the thread,
driving an instability. The idea of a non-uniform polymer density is
consistent with the observation that for polymer concentrations of
above 1000 ppm, threads eventually solidify and never break
\cite{FBBP64,SWE08,SGEW12,SKTAM15}. 

Based on the linear stability analysis of a uniform thread in the exponentially
thinning regime \eqref{exp} \cite{E_pol14}, a transition is expected to take
place when the thread radius is smaller than the ``blistering'' radius 
\beq
h_{bl} \approx \sqrt{D \lambda}, 
\label{hbl}
\eeq
where $D = k_BT / (6\pi\eta_s a)$ is the diffusion constant of the
polymer \cite{DoiBook86}.  Here $\eta_s$ is the solvent viscosity and
$a \propto M_w^{1/2}$ the polymer radius, so that $D$ decreases
strongly with molecular weight $M_w$.  
The prediction \eqref{hbl} has been confirmed experimentally 
in \cite{DVB18}, varying $D$ and $\lambda$ independently. This was 
achieved using two different polymers, whose relaxation
times have a different dependence on temperature. 

\begin{figure}
\centering
\includegraphics[width=\hsize]{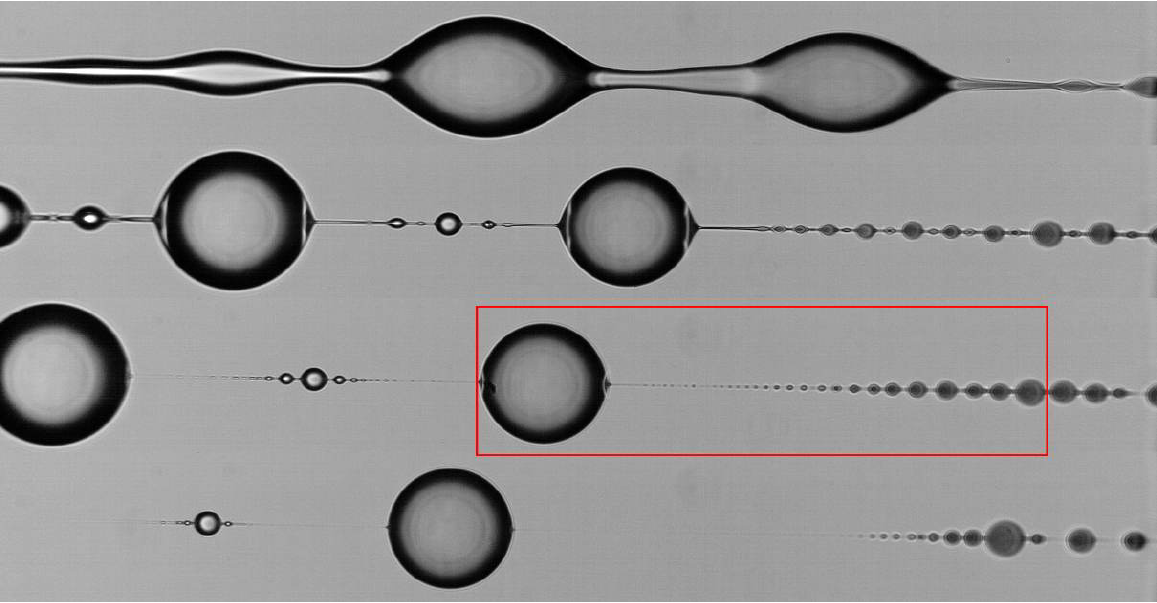}
\caption{The late stages of the blistering process of an aqueous 1000 ppm PEO
solution \cite{SGEW12}. The first image is at $t$ = 250 ms
after the formation of the cylindrical filament; subsequent images are
taken every 30 ms. A thin fiber with the small beads is drawn out of the
large droplet (red box). The width of the image is about 300~$\mu$m. Image is from R. Sattler, S. Gier, J. Eggers, and C. Wagner, \textit{Phys. Fluids} {\bf 24}, 023101, (2012); licensed under a Creative Commons Attribution (CC BY) license.
\label{complex}}
\end{figure}

It is however worth pointing out that as nonlinear effects become more
important, and phase separation progresses, the blistering dynamics can
become remarkably complicated, as illustrated by the sequence shown in
Fig.~\ref{complex}. As the thread evolves, droplets of widely varying size
are created on the thread, in a manner that seems difficult to predict.
However, there are also some organizing features, like the
sequence of smaller and smaller drops highlighted in the red box, which
a partially solidified thread draws out of a drop. Relating such small-scale
features to a fundamental description of polymer solutions appears to
be a daunting yet worthy challenge!

\section{Flow instabilities in non-polymeric systems}
\label{Nonpolymericsystems}
The previous sections focused on flow instabilities of polymeric fluids (with a brief discussion of some instabilities exhibited by viscoelastic wormlike micellar solutions in \S VA). In this section we discuss briefly some features of the flow and instability of three other classes of materials with  complex rheological characteristics: yield stress fluids, wormlike micelle solutions, and liquid crystals.

\subsection{Yield Stress Fluids}
In many industrial and natural settings, we often deal with yield stress fluids. Examples include natural muds, cement pastes, injectable hydrogels, biological fluids, and hygiene products. Yield stress fluids have a threshold in stress, namely the yield stress, above which they flow like a nonlinear viscous fluid. In the past, the main research activities have been focused on yield stress fluids with inelastic properties (e.g., natural muds). Therefore, ideal yield stress models or visco-plastic models, such as Bingham and Herschel-Bulkley, have been used by researchers to address flows involving a yield stress~\cite{BalmforthFrigaard}. In the ideal yield stress models  the shear rate is set to zero when the second invariant of the deviatoric stress tensor falls below the yield stress. In these flow regions, the effective viscosity becomes infinite, the material is unyielded, and the state of stress is undetermined. We refer the reader to the recent book edited by Ovarlez and Hormozi~\cite{Ovarlez}, which includes several lectures on theoretical, computational, and experimental approaches in visco-plastic fluid mechanics. 

However, recent experimental studies show that even a slight elasticity in polymer-based yield stress test fluids has an essential role in the flow dynamics. For example, in the absence of inertia, the loss of the fore-aft symmetry and the formation of the negative wake are observed when a sphere settles in a yield stress fluid (see e.g.,~\cite{Holenberg,Ahonguio}). Also, Firouznia et al. showed an asymmetric disturbance velocity field around a neutrally buoyant sphere in Carbopol gels (i.e., an accepted model yield stress test fluid) subjected to linear shear flows. Additionally, the authors showed that the trajectories of two spheres are asymmetric in the absence of contact~\cite{Firouznia}. None of the aforementioned observations can be explained \textit{via} ideal yield stress models that produce symmetric flow solutions. However, the loss of symmetry and reversibility can be explained by including viscoelastic effects in modelling flows of yield stress fluids around obstacles~\cite{Fraggedakis,Sarabian}. In addition, applications in industry frequently utilize yield stress fluids for rheological innovation, e.g., designing injectable hydrogels and engineering inks for additive manufacturing, which are advantaged as these  materials are self-supporting and hence preserve their shape. These new polymer-based gels have substantial elasticity as well as yield stress.

It is only relatively recently that soft matter scientists have begun to incorporate elastic effects into constitutive equations for yield stress fluids. For example, Saramito proposed a constitutive law in which the material behaves as a nonlinear viscoelastic fluid above the yield stress and as a nonlinear viscoelastic solid below the yield stress~\cite{Saramito}. Modeling practical yield stress fluids \textit{via} this constitutive description results in an unrealistic zero loss modulus below the yield stress since an ideal Hookean solid remains in-phase with the imposed strain. This issue has been resolved in recent models where McKinley and co-workers have developed a class of elastoviscoplastic constitutive models adapted from ideas in the nonlinear plasticity literature, collectively known as isotropic and kinematic hardening (IKH)~\cite{DimitriouEwoldt,DimitriouMcKinley}. The evolution of the yield stress is captured through an internal tensorial back stress, which describes the residual stresses that develop in the microstructure as it is deformed elastoplastically prior to yield. This framework results in a set of Oldroyd-type evolution equations that contain up to nine material constants, which can be determined using a sequence of steady and time-varying viscometric flows. The aforementioned constitutive laws can be used in numerical simulations to predict non-viscometric flow fields. The comparison of the results with experimental observations then provides a basis for further improvement of such constitutive laws. 

To our the best of our knowledge, the stability analysis of elastoviscoplastic fluids has not yet been performed. As far as the yield stress property is concerned, the first study of the hydrodynamic stability of a Bingham fluids came more than a century after the Newtonian equivalent, and even simple plane channel flow has been studied only recently~\cite{Nouar}. The primary assumption in these efforts is that yielded surfaces remain invariant as instabilities develop, which is a crippling approximation that leads to mathematical anomalies~\cite{Metivier2005}. There is a dearth of literature in this area, with only a few weakly nonlinear and nonlinear (energy) stability results~\cite{HormoziFrigaard,Metivier2010,MoyersGonzalez,NouarFrigaard}. The difficulty arises because, for yield stress fluids, the nonlinearity of the problem is not only in the inertial terms, if the Reynolds number is finite, but also in the shear stress/shear rate relationship and in the existence of unyielded plug regions, which are defined in a non-local fashion even for simple flows. Therefore, the gap between linear and nonlinear theories is much broader and more complex than with Newtonian fluids. Therefore, it is essential to study how the knowledge of the stability of the ideal yield stress models can be extended to practical elastoviscoplastic fluids. To further our ability to design, predict and optimize flows of elastoviscoplastic fluids, we must first build new scientific knowledge regarding the behavior of these fluids.

\subsection{Wormlike Micellar Solutions}

\begin{figure*}[ht]
\centering
\includegraphics[scale=0.59]{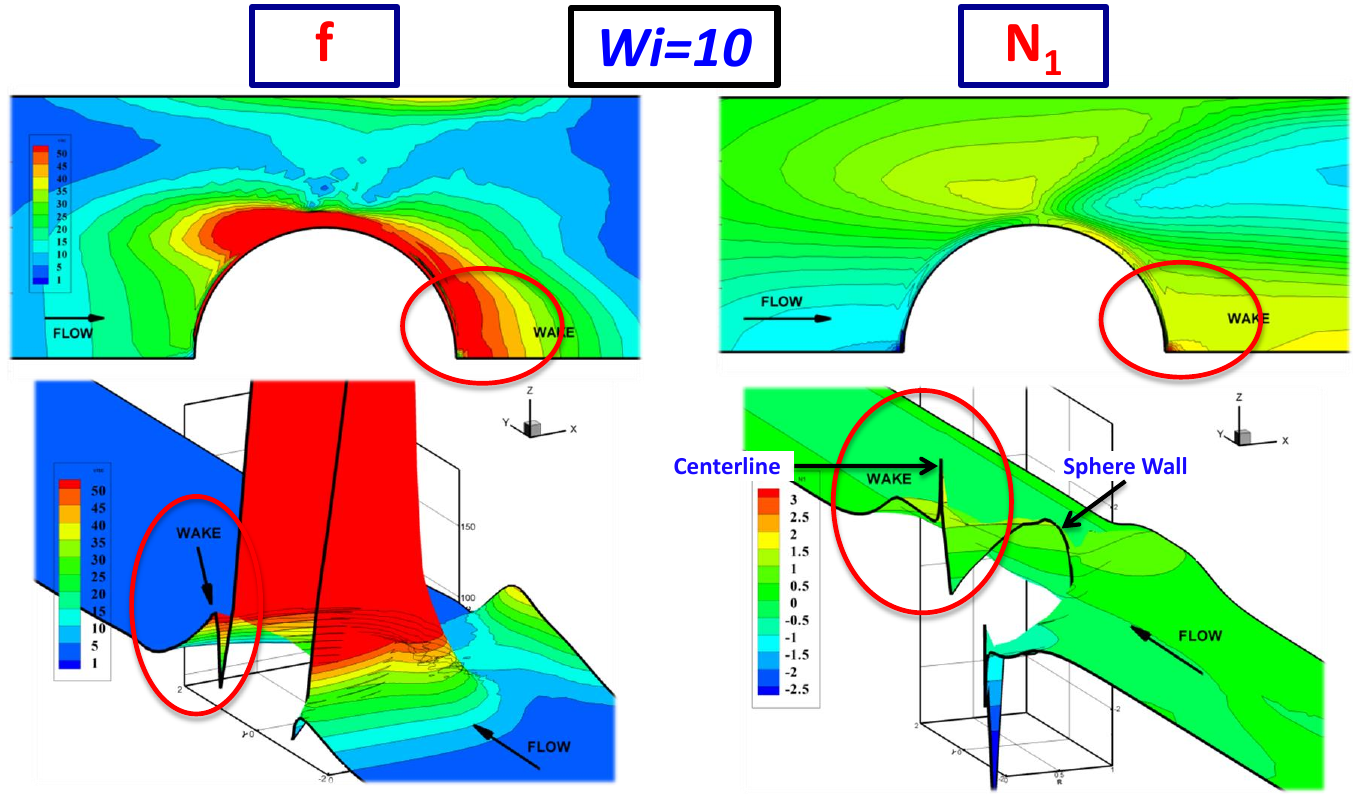}
\caption{Dimensionless fluidity and first normal-stress difference $ N_{1} $ in a flow past a sphere (sphere-to-tube aspect ratio 1:2) of a WLM solution; solvent-fraction $ \beta = 5 $x$ 10^{-3} $ and moderate hardening features, Wi = 10. See L{\'o}pez-Aguilar et al.~\cite{WLMJNNFM2018} for further details in the WLMs rheology used.}
\label{Figure17}
\end{figure*}

Wormlike micellar solutions (WLMs)  are materials of wide use in industrial and technological applications due to their versatile rheological properties, i.e., they may be thixotropic, shear-thinning, strain hardening and softening viscoelastic fluids, capable of  displaying shear-banding and plasticity \cite{RothsteinReviewWLMJNNFM2020,DreissReviewWLMSoftMatter2007,YangReviewWLM2002, OlmstedReviewSB2008}. WLMs are utilized in home-care and health-care products, such as shampoos, soaps, detergents, pharmaceuticals, biocompatible drug-delivery systems; in the oil industry, as drilling and reservoir stimulation fluids; drag-reducing agents in pipelines and lubricants; and as emulsifiers, among other numerous uses \cite{RothsteinReviewWLMJNNFM2020,DreissReviewWLMSoftMatter2007,YangReviewWLM2002}. The list of rheological properties and applications of WLMs are promoted by their time-dependent internal structure, which may be broken-down and rebuilt under deformation. Accordingly, WLMs are also referred to as `living polymers', since they can relieve stress, in addition to undergoing reptation, with a mechanism of dynamic construction-destruction of their internal-structure  \cite{YangReviewWLM2002}.

One constitutive framework to characterize the rheology of WLMs  is the Bautista-Manero-Puig (BMP) family of constitutive laws, whose novel BMP+\_\texttau \textsubscript{p} model-variant has been proposed recently \cite{WLMJNNFM2018}. The BMP+\_\texttau \textsubscript{p} model is embodied in a generalized differential equation-of-state of the Oldroyd-B type in stress-split form, i.e., $ \boldsymbol{ \tau } = \boldsymbol{ \tau }_{p} + 2 \eta_{s} \boldsymbol{ \dot{\gamma} }$, with a solvent of Newtonian nature with viscosity $ \eta_{s} $ and $ \boldsymbol{ \dot{\gamma} } = \frac{1}{2} [\nabla \boldsymbol{v} + \nabla (\boldsymbol{v})^T ]$, which feeds the thixotropic internal-structure into the polymeric viscosity $ \eta_{p} $ \textit{via} an evolution equation for a dimensionless fluidity $ f = \frac {\eta_{p0}} {\eta_{p}} $; here, $ \eta_{p0} $ represents the polymeric first Newtonian-plateau level. 

An equation for the microstructure considers rates of internal-structure construction and destruction of wormlike micelles, involving  viscoelasticity in the dynamics as follows: $ ( \frac{\partial} {\partial t} + \boldsymbol{v} \cdot \nabla ) f = \frac{1} {\lambda_{s}} (1-f) + k (\frac{G_{0} \lambda_{1}} {\eta_{\infty}+\delta}) \lvert \boldsymbol{ \tau }_{p} : \boldsymbol{ \dot{\gamma} } \rvert $, where internal structure destruction is promoted by the energy dissipated by the solute in flow. Here, $ \lambda_{s} $ is a structural construction time, $ k $ is a  destruction parameter whose inverse corresponds to a structure-destruction stress, $ G_{0} $ is the elastic modulus, $ \lambda_{1} $ stands for the viscoelastic relaxation-time, and $ ({\eta_{\infty}+\delta}) $ represents the polymeric viscosity at high deformation rates. This constitutive equation reflects a rheological response of shear-thinning, finite strain hardening and softening, alongside non-linear first normal-stress difference in shear \cite{WLMJNNFM2018}, which are all common features of WLM systems~\cite{RothsteinReviewWLMJNNFM2020,DreissReviewWLMSoftMatter2007,YangReviewWLM2002}. In addition, the BMP+\_\texttau \textsubscript{p} model predicts flow-segregation, such as shear-banding and apparent yield stress~\cite{WLMJNNFM2018,SBBMP+2017}.

In the workshop, L{\'o}pez-Aguilar presented numerical solutions of complex flows of WLMs in  axisymmetric contraction-expansion flow and flow past a sphere, produced with their in-house hybrid finite element/volume algorithm (\cite{WLMJNNFM2018} and references there in). Particular attention was paid to the flow-structure relation and its correlation with the rheological properties of WLMs characterized by the BMP+\_\texttau \textsubscript{p} model \cite{WLMJNNFM2018}. In an axisymmetric abrupt 10:1 contraction-expansion, distinct flow transitions are observed for WLMs (studying extension-hardening and solute-concentration variations \cite{WLMJNNFM2018}). Strong correlation is recorded between vortex evolution and the normal-stress distribution in the recess-zones. Here, for solutions with solvent-fraction $ \beta \leq 1/9 $ and high flow-rates, strong-hardening WLM flow-structures evolve to have upstream elastic-corner vortex phases: a step preceding strong time-dependency of the  numerical solutions with further flow-rate increase (see  \cite{BogerJNNFM2019} and references therein). In contrast, for dilute solutions, upstream lip-vortices are captured at intermediate deformation-rates.

For flow past a sphere, the BMP+\_\texttau \textsubscript{p} dimensionless fluidity $ f $ is used to analyze the internal-structural changes of extremely concentrated WLMs in the wake of a sphere. In the three-dimensional plots of Fig.~\ref{Figure17}, for $ \beta = 5 \times 10^{-3} $, moderate hardening features and \textrm{Wi} = 10, coinciding maxima are captured in $ N_{1} $ and $ f $, located on the symmetry-line downstream of the sphere. In the companion 2D-representation, a highly unstructured fluid (red fringe level of high $ f $-values) is recorded on the sphere wall, reflecting a shear-thinned material with relatively small $ N_{1} $-values. Such a red-fringe of fluid connects to the centerline downstream of the sphere, where extensional deformation prevails and the fluid develops a $ N_{1} $-hardening peak. This $ N_{1} $-peak coincides in location with a fluidity maximum, which, under extensional deformation and the BMP+\_\texttau \textsubscript{p}  formalism, may reflect a growth of extensional viscosity~\cite{WLMJNNFM2018}. The complex interplay between changes in the localized material properties and mixed deformations may play a role in the further understanding of instabilities observed experimentally when a sphere settles in WLMs~\cite{RothsteinReviewWLMJNNFM2020,McKinleyRheolActa1998}.

\subsection{Liquid Crystals}
Nematic liquid crystals are a class of fluids with intrinsic orientational order. In equilibrium, nematics tend to uniformly align their anisotropic constituents as a means to minimize energy, which annihilates topological defects. When driven away from equilibrium by an externally applied flow, the continual injection of energy can destabilize the defect-free alignment. The emergence of shear-induced structures has attracted significant attention in studies of nematic thermotropic liquid crystals and liquid crystal polymers \cite{beris1994thermodynamics,Larson1999,oswald2005nematic,mather1996flow,larson1992development}. 

Most nematic thermotropic liquid crystals are shear-aligned nematics, in which the director evolves towards an equilibrium polar angle. A variety of different defect topologies nucleate beyond a critical Erickson number Er~=~$\frac{{\eta}{\dot{\gamma}}L^2}{{K}}$, which denotes the condition when the viscous torques become dominant over the elastic ones. Here, $\eta$ is the dynamic viscosity, $\dot{\gamma}$ the shear rate, $L$ the typical scale of deformation (often the thickness of the liquid crystal layer), and $K$ the Frank elastic constant. The shear-alignment in the bulk flow leads to an irreconcilable alignment of the directors with those in the surface-anchored region. The high elastic stresses of the director gradient at the boundary between the two regions are released through the formation of defects. Liquid crystal polymers, by contrast, are typically tumbling nematics characterized by a nonzero viscous torque for any orientation of the director. Their tumbling characteristics facilitate the nucleation of singular topological defects. Recently, topological structures and their dynamics have been described in three-dimensional active nematics, where disclination loops undergoing complex dynamics and recombination events are identified as the primary unstable structure \cite{duclos2020topological}.

The flow behavior of lyotropic chromonic liquid crystals has so far remained largely unstudied. Lyotropic chromonic liquid crystals~(LCLCs) are aqueous dispersions of organic, disk-like molecules that self-assemble into cylindrical aggregates, which form nematic and columnar phases beyond a certain concentration \cite{collings2015nature}. Due to their bio-compatibility, they have opened paths for controlling assembly and dynamics of biological systems when used as a replacement for isotropic fluids in microfluidic devices \cite{zhou2014living,peng2016command}. At rest, LCLCs exhibit unique properties distinct from those of thermotropic liquid crystals and liquid crystal polymers. In particular, LCLCs have significant elastic anisotropy and their twist elastic constant, $K_2$, is much smaller than the bend and splay elastic constants, $K_{1}$ and $K_{3}$ \cite{zhou2012elasticity}. The resulting relative ease with which twist deformations occur can lead to a spontaneous symmetry breaking and the emergence of chiral structures in static LCLCs under spatial confinement, despite the achiral nature of the molecules \cite{tortora2011chiral,park2020periodic}.

A recent study by Baza et al. has revealed a variety of complex textures that emerge in Couette flow in the nematic LCLC disodium cromoglycate~(DSCG) \cite{baza2020shear}. The liquid crystal tends to tumble, which leads to a high sensitivity to shear rate; with increasing shear rate the materials transitions from a log-rolling state, where the director realigns perpendicular to the flow direction, to polydomain textures and finally to periodic stripes in which the director is predominantly along the flow direction.

For pressure-driven flow of nematic DSCG, during the workshop Bischofberger discussed the emergence of pure-twist disclination loops for a certain range of shear rates, which form as a consequence of the smallness of the twist elastic constant $K_2$. Their characteristic size is governed by the balance between the nucleation force and the annihilation force acting on the loop. Remarkably, chiral domains spontaneously form at lower shear rates, suggesting that not only topological confinement, but also weak shear can induce chiral structuring in achiral materials. These observations hint towards the wealth of phenomena that are still awaiting to be discovered in flows of lyotropic chromonic liquid crystals.

\section{Outlook and open questions}
\label{outlook}

A reader who has even read only one or two sections of this wide-ranging perspective will note that, in spite of substantial progress in the field over the past thirty years, including advances in experimental measurements, theory, and detailed numerical simulations by computational rheologists, where each brings their own insights, there remain important areas where the integration of ideas is needed and new discoveries remain to be made. This is true for the flow  transition that occurs in channel and pipe flows of polymeric fluids (\S\ref{FlowTransitionSection}, \S\ref{EITsection}) and worm-like micelle solutions (\S\ref{Nonpolymericsystems}). It is also the case for more complex configurations, such as the cross-slot geometry~\cite{Arratia2006,Poole2007,Rocha2009,Haward2012a}), where shear-thinning apparently triggers flow asymmetries so that the initial onset of instability gives rise to regions in the flow field with disparate shear rates. Similar issues are at the heart 
of unstable flows at the pore scale for a wide range of heterogeneous and porous media;  recent experimental developments seeking to  provide \textit{in situ} access to these flows are mentioned in \S\ref{ElasticInstability}. Not surprisingly, observations of other microstructurally complex fluids \S\ref{Nonpolymericsystems}, such as  lyotropic chromonic liquid crystals, or free-surface flows of complex fluids that are impacted by surface tension \S\ref{SectionFreeSurface}, hint at a wealth of discoveries that are yet to be made. 

\textit{Models of viscoelastic flows.} Throughout the workshop there was discussion surrounding the fact that elastic instabilities are typically discovered at ``high Weissenberg number" in polymer solutions -- where the veracity of constitutive equations at the requisites shear rates always comes into serious question. Indeed, when modeling highly elastic viscoelastic flows, the Oldroyd-B model is almost never quantitative at high Weissenberg number either in shear flows or extensional flows of polymer solutions. This can also be said of nonlinear extensions of this model, including the FENE models as well as the Giesekus and PTT (see \S\ref{ModelsAndNumericalSimulations}). In particular, detailed molecular studies have demonstrated that internal degrees of freedom cannot be neglected in these flows if one wants to capture the extra polymer stress. Thus, progress in understanding elastic instabilities must almost, by the definition of the phenomena, be made hand-in-hand with advancements in the rheological modelling of elastic fluids. 

\textit{Spectral properties of elastic turbulence.} There was also significant discussion surrounding the fact that the eigenspectrum of elastic two-dimensional Taylor-Couette flow was essentially unstudied. Many of the attendees thought that such a study deserved attention in an effort to further understand the work described in \S\ref{FlowTransitionSection}, especially since the beginnings of instability at values of Wi $\sim 10$, albeit at large gap ratio, had been found.

\textit{Importance of a Lagrangian perspective.} Remarkably, shear plays an important role in all detailed mechanisms worked out for purely elastic instabilities so far. This is despite the fact that purely elongational flows can have a dramatic effect on single molecules, with a strong dependence on individual histories, as exemplified by the coil-stretch transition of a single polymer, which can be sharp and hysteretic. Moreover, a dependence on the history does not play any role in the currently understood instability mechanisms, as the flows considered have all been viscometric, e.g., the flows have been such that each infinitesimal fluid element sees effectively a steady shear. 

Consequently, the history of the flow and stress experienced by a fluid element should enter considerations for flows that are Lagrangian unsteady.  In particular, even the notion of extension- or shear-dominated flow may depend on Lagrangian properties such as the residence time of fluid elements.  Beyond classification, it would be interesting to see if new, purely Lagrangian, instability mechanisms could be found, e.g., using insights from bead-spring models to inform the expected stresses and the coupling to the base flow. A Lagrangian perspective might be useful also at the next level of complexity, when conceptually the flow is made up of many coupled flow units, e.g., see \S\ref{ElasticInstability} on flows in different porous systems. Indeed, stress correlations over the path of a fluid element can play a nontrivial role in coupling the flow units and determining the overall stability of the flow, which largely remain to be understood.   

{\textit{Another perspective on elastic waves.} Because the topic of elastic waves was actively discussed at the workshop, here we include another perspective put forward by V. Steinberg. Three main features characterize the elastic waves observed by Varshney and Steinberg \cite{Varshney2019}, as described by the group of Steinberg: (i) The waves are transverse and manifest as a peak in the power spectrum of span-wise velocity fluctuations; (ii) The velocity of wave propagation depends on $(\rm{Wi}-\rm{Wi}_c)^{\beta}$  with $\beta=0.73$; (iii) The measured wave dispersion relation is linear \cite{ShnappSteinberg}. Given these three features, and given the apparent agreement with the predictions of \cite{balkovsky}, the authors propose that such waves are indeed elastic waves. Moreover, they note that they observed these waves exclusively in random flows: either in chaotic flows above the non-normal mode bifurcation and further in ET in a straight channel with and without strong perturbations \cite{JhaSteinberg1,ShnappSteinberg} or only above the transition to ET in a flow past an obstacle or between two obstacles hindering a channel flow \cite{Varshney2019,kumar21}. In addition, the elastic waves were not found in flow geometries with curvilinear streamlines including ET.

Because of these apparent similarities as the elastic waves predicted in \cite{balkovsky}, they authors use the expression for the wave velocity $c_{el}^2=\tau/\rho$ to estimate an elastic stress, $\tau$, which depends on the flow, in the direction of wave propagation. In the experiments of \cite{Varshney2019}, $c_{el}$ varied from 2 to 17 mm/s, yielding an estimate for $\tau$ ranging from $\sim5\times10^{-3}$ to $0.375$ Pa, and a corresponding Mach number Ma = u/$c_{el}$ of the order $0.3 < 1$. In a later paper~\cite{JhaSteinberg1}, the Steinberg group reported reaching $c_{el}\approx45$ mm/s corresponding to $\tau\approx10$ Pa, whereas in \cite{ShnappSteinberg}, much smaller wave velocities close to the onset were measured from 0.05 to 0.2 mm/s corresponding to $\tau$ between $3\times10^{-6}$ and $5\times10^{-5}$ Pa. Thus, summarizing these experiments, the range of the elastic stress derived from $c_{el}$ is $\sim3\times10^{-6}$ up to 10 Pa --- considerably smaller than the values noted in \S IIC, where the shear wave speed is tied to a flow-independent material property. Clarifying the underlying physics will be an important direction for future research. 

\textit{Open questions regarding EIT.} Substantial progress has been made, both experimentally and computationally, in understanding the nature of the turbulent drag reduction phenomenon, and more broadly in nonlinear viscoelastic dynamics in straight pipes and channels. The mechanism by which viscoelasticity suppresses near-wall coherent structure are understood, and the discovery of elastoinertial turbulence helps explain why flows remain turbulent even when the Newtonian near-wall structures are strongly suppressed. Many questions about EIT remain, however. In channel flow, it is directly connected to the Tollmien-Schlichting instability mode and corresponding near-wall critical layer, and is known to subcritically excite this mode, driving flow away from the laminar state even when that state is linearly stable. A simple mechanistic explanation of this viscoelastic excitation of the TS mode does not yet exist. Simulations reveal similar near-wall critical layer fluctuations in plane Couette and pipe flows, even though they do not display a linear instability of Tollmien-Schlichting type -- perhaps, in analogy to channel flow, wall modes are highly susceptible to amplification, driving bypass transition. 

Additionally, a new mode of linear instability, with a critical layer near the centerline, arises in both pipe and channel flows. This mode also appears to be strongly subcritical, leading to flows with substantial polymer stretch fluctuations near the centerline.  This may be the dominant mode at low Reynolds numbers and high Weissenberg numbers, helping explain experimental observations in this regime. 

In addition, the maximum drag reduction asymptote may be a marginal flow regime in which Newtonian and elastoinertial flow structures coexist, perhaps in an intermittent fashion. These issues are ripe for future study.

\textit{Controlling and using flow instabilities.} Finally, 
an interesting research direction for the future will be to connect this emerging understanding of the physics underlying elastic instabilities and turbulence to applications of viscoelastic fluids. For example, one of the most fundamental descriptors of such flows in applications is the relationship between the overall pressure drop and the volumetric flow rate, often described using an ``apparent viscosity" $\eta_{\rm{app}}$ that quantifies macroscopic flow resistance; however, prediction and control of this quantity remains challenging, despite its central importance in applications. Recent work has made progress in doing so for flows in porous media \cite{Browne2021}, and further research along these lines will be important in the rational application of viscoelastic flows in diverse settings. 

Another important direction for future work will be in exploring ways to \textit{control} elastic instabilities and turbulence as well as \textit{engineer} spatio-temporal flow patterns in viscoelastic fluids. So far, mostly geometric control has been employed such as the flow between two cylinders \cite{Varshney2017}, in a cross-slot geometry \cite{Davoodi19}, in disordered obstacle arrays \cite{Walkama20}, or in designed porous media \cite{Browne2020,Browne2021}. An attractive alternative is to employ time-dependent or modulated shear rates for active open-loop control of viscoelastic fluid flow. As demonstrated in Ref. \cite{Buel20}, this allows  tuning or control of the occurrence of elastic turbulence and therefore, for example, the mixing of complex fluids. Exploring this and other approaches (e.g., employing deformable structures \cite{RothsteinCantileverWLMJNNFM2020}) will be a useful direction for future work. 

In terms of direct applications of elastic turbulence, a number of different research groups have shown that, in addition to enhancing passive scaling mixing as was originally demonstrated in the earlier works of Steinberg and collaborators \cite{Burghelea,Steinberg_AFM}, these kinds of viscoelastic chaotic flows can also be used to effectively enhance heat transfer at low Reynolds numbers both in macro-sized ``von Karman flow" \cite{Traore} and also at the microscale in serpentine and wavy channel geometries \cite{Abed,LiMicrofluid2017,YangThermal}. Finally, elastic turbulence has been used to create emulsions from immiscible viscous liquids in a simple shear mixing device \cite{PooleEmulsion}. In the absence of elasticity, but at identical viscosity ratio, Reynolds and capillary numbers, no mixing was observed and the immiscible liquids remain separated. Elastic turbulence thus offers a unique pathway to create dispersions in viscous liquids at low Reynolds numbers in nominally shear-dominated flows. Further studies of these phenomena, and other potential applications, would be a fruitful avenue for additional research.

\textit{In conclusion.} Any reader that has gotten this far will hopefully agree that the subject is fascinating, both from the standpoint of fundamentals, but also because the materials that make up the complex fluids field find a wide range of applications. Thus, we hope this article serves future researchers as a basis for next steps in advancing research and understanding of the flows and instabilities of complex fluids. \\

\section{acknowledgements}
We thank the Princeton Center for Theoretical Science at Princeton University for their support of this virtual workshop. In particular, we are grateful for the help provided by Charlene Borsack, both leading up to and during the workshop. We also thank Manuel Alves for helpful feedback on this manuscript. 

SSD acknowledges the Donors of the American Chemical Society Petroleum Research Fund for partial support of this research through grant PRF 59026-DNI9. 

SSD and HAS acknowledge the NSF for work that was partially supported  through Princeton University'€™s Materials Research Science and Engineering Center DMR-2011750.

GHM acknowledges the National Science Foundation for funding support under Grant No. CBET-2027870.

JELA acknowledges the support from Consejo Nacional de Ciencia y Tecnolog{\'i}a (CONACYT, Mexico), from Universidad Nacional Aut{\'o}noma de M{\'e}xico UNAM (grant numbers PAPIIT IA105620 and PAIP 5000-9172 Facultad de Qu{\'i}mica); and from Laboratorio Nacional de C{\'o}mputo de Alto Desempeo UNAM (grant number LANCAD-UNAM-DGTIC-388), for the computational time provided on the \textit{Miztli} Supercomputer.

The research leading to the results of SMF et al. summarized in \S\ref{SectionFreeSurface} A-B has received funding from the European Research Council under the EU's 7th Framework Programme (FP7/2007-2013) / ERC grant number 279365.

MDG acknowledges support from  NSF  CBET-1510291, AFOSR  FA9550-18-1-0174, ONR N00014-18-1-2865 and N00014-17-1-3022.

JSG acknowledges support by NSF Awards CBET-1701392 and CAREER-1554095. 

SJH and AQS gratefully acknowledge the support of the Okinawa Institute of Science and Technology Graduate University (OIST) with subsidy funding from the Cabinet Office, Government of Japan, and also funding from the Japan Society for the Promotion of Science (JSPS, Grant Nos. 18K03958 and 18H01135) and the Joint Research Projects (JRPs) supported by the JSPS and the Swiss National Science Foundation (SNSF). 

RJP would like to acknowledge funding from the UK's Engineering and Physical Sciences Research Council (EPSRC) under grant number EP/M025187/1.

HS acknowledges support from the Deutsche Forschungsgemeinschaft in the framework of the Collaborative Research Center SFB 910.


\end{document}